\documentclass[12pt]{article}
\pdfoutput =1
\usepackage{graphics}
\usepackage{graphicx}
\DeclareGraphicsExtensions{.pdf}
\usepackage{float} %Include figure filesusepackage{graphicx} %Include figure files
\usepackage{subfloat}
\usepackage{amsmath}
\usepackage{slashed}
\usepackage{amsfonts}  
\usepackage{amssymb}
\usepackage{anyfontsize}

\def\b{\beta}

\def\s{\sigma}
\def\om{\omega}
\def\th{\theta}
\def\ph{\phi}

\def\Om{\Omega}

\usepackage{color}

\usepackage{subcaption}
\begin{document}
\date{}
%%%%%%%%%%%%%%%%%%%%
\title{{\bf{\large  Hall transports from Taub-NUT AdS black holes}}}
%%%%%%%%%%%%%%%%%%%%
\author{
{\bf { ~Mohd Aariyan Khan}$
$\thanks{E-mail::  aariyan\_km@ph.iitr.ac.in}}\\
 {\normalsize  Department of Physics, Indian Institute of Technology Roorkee,}\\
  {\normalsize Roorkee 247667, Uttarakhand, India}
\\[0.3cm]
{\bf { Hemant Rathi}$
$\thanks{E-mail:  hemant.rathi@saha.ac.in}}\\
 {\normalsize  Saha Institute of Nuclear Physics, 1/AF Bidhannagar, }\\
  {\normalsize Kolkata 700064, India }
  \\[0.3cm]
 {\bf { ~Dibakar Roychowdhury}$
$\thanks{E-mail:  dibakar.roychowdhury@ph.iitr.ac.in}}\\
 {\normalsize  Department of Physics, Indian Institute of Technology Roorkee,}\\
  {\normalsize Roorkee 247667, Uttarakhand, India}
\\[0.3cm]
}
\maketitle
\abstract{We compute Hall transport coefficients associated with Taub-NUT AdS black holes in four space-time dimensions using the probe D-brane approach. In particular, we examine the effects due to the NUT parameter ($n$), or equivalently, the novel frame-dragging on the holographic charge transport properties. In our analysis, we treat the external electric field as a constant background, while varying the magnetic field ($B$) from small to  finite. Within this framework, we analyze conductivities in both low and high temperature regions, focusing on locations that are both near and far from the Misner string. Our calculations show that frame-dragging effects are significant primarily at lower temperatures and near the Misner string, while a small magnetic field is maintained. However, these effects become negligibly small at a ``finite" magnetic field and even at lower temperatures. Our analysis reveals the existence of finite Hall transport, that has its origin in the novel frame-dragging.}

%We compute holographic DC conductivity associated with the Taub-NUT-AdS4 black holes following the probe D-brane approach. In particular, we examine the effects of frame dragging on charge transport in both low and high temperature regimes. Our analysis reveals that in the low temperature regime, the conductivity is sensitive to the presence of the Misner string that causes frame dragging. Notably, the increase in the conductivity near the Misner string is sharper as compared to points farther away from it. On the other hand, in the high temperature regime, the effects due to frame dragging are significantly suppressed, and the thermal contribution to the charge transport takes over that due to the U(1) charge carriers.

\section{Introduction and motivation\label{secint}}
The $AdS_{d+1}/CFT_d$ duality or commonly known as the gauge/gravity duality \cite{Witten:1998qj}-\cite{Gubser:1998bc} is a powerful tool in theoretical physics to study the strongly correlated quantum dynamics in condensed matter system. In this context, $AdS_{d+1}$ stands for the Anti-de-Sitter space-time in $d+1$ dimensions that relates the Conformal field theory ($CFT_d$) in one lower dimension.

The authors in \cite{A.~O'Bannon}-\cite{Karch:2014mba} employ the tools of $AdS/CFT$ to investigate the Ohmic and Hall conductivities in %four dimensional
boundary theory using the probe D-brane approach. These conductivities are fundamental quantities observed in condensed-matter systems when electric and magnetic fields are applied orthogonally \cite{Nagaosa:2009ycg}-\cite{Anderson:1991ixg}. The Ohmic conductivity is associated with the free charge carriers that flow parallel to the electric field, while the Hall conductivity corresponds to the electric current that flows orthogonal to the electric field in the presence of a magnetic field (due to the Lorentz force).

Motivated by the work of \cite{A.~O'Bannon}-\cite{Karch:2014mba}, we explore the Hall and Ohmic conductivities associated with the Taub-NUT (Newman, Unti and Tamburino) AdS black holes in four dimensional space-time\footnote{For detailed discussion on Taub-NUT AdS black holes, see \cite{Khan:2025fne}. } \cite{Durka:2019ajz}. These black holes are the straightforward generalization of the 4D Schwarzschild space-time metric characterized by the NUT parameter ($n$) \cite{Newman:1963yy}-\cite{Taub:1950ez}. The Taub-NUT AdS black holes are of significant interest due to their several
interesting properties. For instance, these black holes are not globally AdS, but rather, they are asymptotically locally AdS and contain the line singularities known as the Misner string \cite{Misner:1963fr}, which leads to the novel frame-dragging effects \cite{Durka:2019ajz}, \cite{Zhang:2016gzk}-\cite{Ong:2016cbo}. The thermodynamic properties of these black holes were examined by the authors in \cite{Bordo:2019tyh}-\cite{Rodriguez:2021hks}, where various formulations of the first law of black hole thermodynamics were discussed. This work was further extended by the authors of \cite{Jiang:2019yzs}-\cite{Chen:2023eio}, where they compute the holographic complexity associated with the charged TN-AdS black holes. Additionally, Taub-NUT black holes have been explored in the context of condensed matter systems \cite{Kalamakis:2020aaj}-\cite{Kalamakis:2025daq}.

Given the above state of art, in the present work, we explore the effects of the NUT parameter ($n$) on the Hall transports (Ohmic and Hall conductivities collectively) associated with the Taub-NUT $AdS_4$ black holes in $(t,z,\th,\ph)$ coordinates using the probe D-brane approach \cite{A.~O'Bannon}, \cite{Karch:2007pd}. The effects due to frame-dragging (or the NUT parameter ($n$)) on the Ohmic (or DC) conductivity in the presence of a small external electric field ($E<<1$) have been investigated in \cite{Khan:2025fne}. Here, we extend the previous work by applying a finite magnetic field ($B$) on top of the electric field. We apply the electric field in the direction of $\hat{\phi}$ and the magnetic field in the direction of $\hat{z}$. This causes the charge carriers to drift along the $\hat{\phi}$ and $\hat{\th}$ directions, due to the Lorentz force. This results in a Ohmic conductivity along $\hat \ph$ direction and Hall conductivity along $\hat \th$ direction.

Our calculations reveal that both the Ohmic and Hall conductivities receive contributions due to $U (1)$ charge carriers as well as the thermally produced charge pairs. The current due to thermally produced pairs are sourced due to novel ``frame-dragging".  This marks a notable deviation from previous findings reported by the authors in \cite{A.~O'Bannon}-\cite{Lee:2010uy}, where the thermal contribution to the Hall conductivity vanishes identically. We examine the conductivities both at low and high temperature limits. We conducted our analysis at a small $(B<<1)$ and a finite magnetic field $(B>1>E)$, while considering an electric field ($E$) in the background.

In the presence of a small magnetic field (i.e., $B<<1$), we observe that the Ohmic conductivity is greater than the Hall conductivity for the respective charge carriers ($U(1)$  and thermally produced charge pairs). At a low temperature regime ($T\sim T_{min}$), the frame-dragging effects are significantly predominant near the Misner string. As a result, both Hall and Ohmic conductivity of $U(1)$ and thermally produced charge carriers are greater near the Misner string as compared to the points farther from the Misner string. This increase in conductivity results from the additional drift of charge carriers near the Misner string. Furthermore, we found that at low temperatures $(T\sim T_{min})$ and in the presence of a small magnetic field, both Ohmic and Hall conductivities are dominated by $U(1)$ charge carriers.

On the other hand, at higher temperatures ($T>>T_{min}$), frame-dragging effects are negligible. As a result, the behaviors of Ohmic and Hall conductivities, both near and far from the Misner string, are nearly identical. At high temperatures with a small magnetic field, the Ohmic conductivity is dominated by thermal charge carriers. On the other hand, the Hall conductivity remains dominated by $U(1)$ charge carriers because the Hall conductivity due to thermal charge carriers arises as a result of frame-dragging, which becomes negligible at high temperatures.

%As a result, the behavior of conductivities (Ohmic and Hall), both near and far from the Misner string, are nearly equal.

 Our analysis reveals that at finite magnetic field $(B>1>E)$, the frame-dragging effects are negligible even at low temperatures. As a result, the Hall conductivity due to thermally produced charge pairs vanishes and it is predominantly influenced by $U(1)$ charge carriers at both low and high temperatures. On the other hand,  the Ohmic conductivity is primarily governed by thermally generated charge carriers at high temperatures and at low temperatures, the behavior is contingent upon the density of $U(1)$ carriers and the strength of the magnetic field. Specifically, when $J_t > B$, the Ohmic conductivity due to $U(1)$ carriers surpasses that of thermal pairs. Conversely, when $J_t < B$, the Ohmic conductivity is predominantly dominated by the thermally produced charge pairs, irrespective of the position of the Misner string. We also observe that the Hall conductivity due to  $U(1)$ carriers (in a finite magnetic field) exceeds the Ohmic conductivity due to the Lorentz force at both low and high temperatures.

Finally, our calculations reveal a nice relation between Ohmic and Hall conductivity while they are shown to be related up to an overall factor that determines the position of the Misner string. This observation holds at both low and high temperatures and should be valid at all magnetic field strength.

The organization for the rest of the article is as follows.

$\bullet$ In Section 2, we briefly review the transport properties of the system and provide the definitions of Ohmic and Hall Conductivities. 

$\bullet$ In Section 3, we derive the holographic Ohmic and Hall conductivities associated with Taub-NUT $AdS_4$ black holes by employing the probe D-brane method. In this analysis, we treat the external electric field $(E)$ as a constant background, while we change the strength of the magnetic field $(B)$.

$\bullet$ In Section 4, we examine the effects of the NUT parameter ($n$) on the Hall transports at low magnetic field ($B<<1$). We perform our analysis in both low and high temperature regimes.

$\bullet$ In Section 5, we explore the frame-dragging effects on the Hall transports at finite magnetic field ($B>1>E$). We perform our analysis in both low and high temperature regimes.

$\bullet$ In Section 6, we finally conclude our discussion with some interesting future projects.

 %where we previously only turned on $A_t, A_\phi$.

%****paper summary and its organization**** 

\section{A quick review of Ohmic and Hall Conductivities}\label{secint1}
In this Section, we provide a brief review of the transport properties in QFT, in particular, focusing on the definitions of Ohmic and Hall conductivities \cite{Tong:2016kpv}. Transport properties describe how a system responds to an external electromagnetic field. When both electric and magnetic fields are present, the induced current ($J^i$) can be expressed in terms of the conductivity tensor ($\s^{ij}$) as follows: $J^i=\s^{ij}E_j$. 

For an isotropic system, the conductivity tensor \cite{A.~O'Bannon} can be described as 
\begin{align}
  \s^{ij} =\left(\begin{matrix}
    \s_{xx} & \s_{xy}\\-\s_{xy}&\s_{xx}
\end{matrix}\right),
\end{align}
where $\s_{xx}$ represents Ohmic conductivity that arises when $U(1)$ charge carriers are drifted in the direction of an applied electric field. On the other hand, in the presence of a magnetic field, a few charge carriers move transverse to the direction of the electric field due to the Lorentz force. This phenomenon gives rise to the Hall conductivity, represented as $\sigma_{xy}$.

A straightforward microscopic understanding of these transport coefficients can be described by the Drude model \cite{Tong:2016kpv}. In this model, the charge carriers are treated as classical particles that scatter with a characteristic relaxation time, $\tau$. To be specific, let us consider charged particles carrying a charge $q$, moving under the influence of an electric field, given by $\vec{E} = E_x \hat{i} + E_y \hat{j}$, and a magnetic field, represented by $\vec{B} = B \hat{k}$, where the fields are orthogonal to each other. The equation of motion for the charged particle is given below \cite{Tong:2016kpv}
\begin{align}
    \frac{d\vec{p}}{dt}=q \vec{E}+q\left(\vec{v}\times \vec{B}\right)-\frac{\vec{p}}{\tau}. \label{DE}
\end{align}

The steady state (or the equilibrium) solution can be found by setting $\frac{d\vec{p}}{dt}=0$ in the above equation \eqref{DE}, which yields
 \begin{equation}
     q \vec{E}+q\left(\vec{v}\times \vec{B}\right)-\frac{M\vec{v}}{\tau}=0. \label{DEc}
 \end{equation}

The above expression (\ref{DEc}) can be expanded in terms of the components as follows 
 \begin{align}
&qE_x=\frac{M v_x}{\tau}-qv_yB, \label{twoeqn0}\\&
qE_y=\frac{M v_y}{\tau}+qv_x B. \label{twoeqn}
\end{align}

Next, we multiply equations \eqref{twoeqn0}-\eqref{twoeqn} by  the charge density $(J^t)$ and substitute $J^x=J^tv_x,J^y=J^tv_y$. This recast \eqref{twoeqn0}-\eqref{twoeqn} in the matrix form as follows
%-e\left(\begin{matrix}E_x  \\E_y  \end{matrix}\right)=\left(\begin{matrix}M\m & eB \\-eB & M\mu\end{matrix}\right)\left(\begin{matrix}v_x  \\v_y  \end{matrix}\right)
\begin{align}
qJ^t \left(\begin{matrix}
E_x  \\
E_y  
\end{matrix}\right)=
\frac{M}{\tau}\left(\begin{matrix}
1 & -\frac{qB\tau}{M} \\
\frac{qB\tau}{M} & 1
\end{matrix}\right)\left(\begin{matrix}
J^x  \\
J^y  
\end{matrix}\right). \label{Erhoj}\end{align}
%Where we multiply by $J^t$ in \eqref{twoeqn} and substitute $J^x=J^tv_x,J^y=J^tv_y$ .now we can invert above equation \eqref{Erhoj} to get $J=\s E$ in matrix form.

One can express \eqref{Erhoj} in the form $J=\s E$ after inverting the matrix as follows
\begin{align}
   \left(\begin{matrix}
J^x  \\
J^y  
\end{matrix}\right) =\frac{qJ_t\tau/M}{\left({qB\tau}/{M}\right)^2+1}\left(\begin{matrix}
    1 & \frac{qB\tau}{M}\\ -\frac{qB\tau}{M}&1
\end{matrix}\right)\left(\begin{matrix}
E_x  \\
E_y  
\end{matrix}\right),
\end{align}
where we identify the Ohmic and Hall conductivities as follows 
%\begin{align}
%\left(\begin{matrix}
%J^x  \\
%J^y  
%\end{matrix}\right) =\left(\begin{matrix}
 %   \s^{xx} & \s^{xy}\\\s^{yx}&\s^{yy}
%\end{matrix}\right)\left(\begin{matrix}
%E_x  \\
%E_y  
%\end{matrix}\right)\end{align}  
\begin{align}
    \s_{xx}=\s_{yy}=\left(\frac{qJ_t\tau}{M}\right)\frac{1}{\left({B\tau}/{M}\right)^2+1}, \s_{xy}=-\s_{yx}=\left(\frac{qJ_t\tau}{M}\right)\frac{{B\tau}/{M}}{\left({B\tau}/{M}\right)^2+1}.\label{co2}
\end{align}

The above expression (\ref{co2}) shows that the magnetic field bends the trajectories of charge carriers. As a result, longitudinal transport is suppressed, leading to the formation of a transverse Hall current.  In particular, the Ohmic conductivity ($\s_{xx}$) decreases with increasing magnetic field, whereas the Hall conductivity ($\s_{xy}$) rises and dominates in the strong magnetic field regime  \cite{Tong:2016kpv}-\cite{ashcroft2011solid}.

Notice that the Drude model provides valuable insights into electrical conductivities, but it is limited to classical systems characterized by a single relaxation time. A more comprehensive approach is offered by quantum field theory through the concept of linear response. 

The conductivity tensor can be described by the Kubo formula \cite{Ramallo:2013bua}-\cite{Hartnoll:2009sz}
\begin{align} \s_{ij}=-\frac{i}{\omega}G^R_{J_i,J_j}(\omega,k=0),\end{align}
where $G^R_{J_i,J_j}$ represents the retarded current-current correlation function. Here $\om$ and $k$ denote the frequency and momentum, respectively. This formulation expresses the transport coefficients in terms of equilibrium fluctuations and does not rely on the existence of well-defined quasi-particles. In cases where quasi-particles have an extended lifetime, the Kubo formalism resembles the Drude model, with the relaxation time corresponding to the decay timescale of the current correlation function.

However, linear response theory is applicable only to infinitesimal external electric fields. Notably, the holographic constructions based on the probe D-branes \cite{Karch:2007pd} naturally extend transport theory to finite electric fields. In the framework developed by the authors in \cite{A.~O'Bannon}, \cite{Karch:2007pd}, the electric current is derived by varying the Dirac–Born–Infeld (DBI) action \cite{Karch:2007pd}
\begin{equation} 
J^i=\frac{\delta S_{DBI}}{\delta A_i},\hspace{2mm} J^i=\s^{ij}E_j,\label{condcfromDBI}
\end{equation}
which is intrinsically nonlinear in the field strength. As a result, the conductivity exhibits a nontrivial dependency on both electric and magnetic fields, providing a nonlinear extension of transport.  In the weak-field limit, the DBI action reduces to Maxwell theory, thus recovering the standard linear-response (Kubo/Drude) results.

\section{Holographic Ohmic and Hall conductivity \label{hall conductivity}}  
In this section, we calculate the holographic Ohmic conductivity \cite{Karch:2007pd} and Hall conductivity  \cite{A.~O'Bannon} for the Taub-NUT Anti-de Sitter (TN-AdS) black holes in four-dimensional space-time \cite{Newman:1963yy}. In particular, we investigate how the NUT parameter $(n)$ affects these conductivities. 

To begin with, we express the Dirac-Born-Infeld (DBI) action in four dimensions \cite{A.~O'Bannon}
\begin{equation}
    S_{DBI}=-T_p\int dt d\phi dz d \theta \sqrt{- det(g_{ab}+2\pi \alpha'F_{ab})} \label{S indp of t,phi},
\end{equation} 
where, $T_P$ represents the D-brane tension, and $F_{ab}$ denotes the world-volume $U(1)$ gauge field strength tensor.  Furthermore, here, $\alpha'$ is a constant parameter and $g_{ab}$ denotes the induced metric on the world-volume of D-brane.

Next, we embed the D-brane in the Taub NUT AdS background \cite{Khan:2025fne}. The space-time metric for TN-AdS black holes in four dimensions \cite{Durka:2019ajz} is given below\footnote{For a detailed review on TN-AdS space-time see \cite{Khan:2025fne}.}
 \begin{align}
         ds^{2}=\hspace{1mm}&-f(z)\left[dt^2+4n^2(\cos\theta+\beta)^2d\phi^2 -4n(\cos\theta+\beta)d\phi dt\right]+\frac{L^4}{z^4f(z)}dz^{2}+\nonumber\\
         &\frac{(L^4+n^{2}z^2)}{z^2}\left(d\th^2+\sin^2\th d\phi^2\right), \label{TN metric}
    \end{align}
where the function $f(z)$ is given below
\begin{equation}
    f(z)=\frac{\left(z_h-z\right) \left(L^8 z z_h+n^2 z^3 z_h^3 \left(L^2+3 n^2\right)+z_h^2 \left(L^8+L^6 z^2+6 L^4 n^2 z^2\right)+L^8 z^2\right)}{L^2 z^2 z_h^3 \left(L^4+n^2 z^2\right)}.
\end{equation}
Here, $L$ is the AdS length scale, $z_h$ denotes the location of the horizon, and $n$ is the NUT parameter. Furthermore, here $\b=0,\pm1$ is a parameter that denotes the location of the Misner string \cite{Durka:2019ajz}. 

The Hawking temperature for the TN-AdS black holes is given below \cite{Khan:2025fne}
\begin{align}
    T=\frac{1}{4\pi z_hL^4}\left(z_h^2\left(L^2+3n^2\right)+3L^4\right).\label{htn}
\end{align}
The authors in \cite{Abdusattar:2023fdm}-\cite{Li:2008xw} further demonstrate that black holes can exist only beyond a certain minimum temperature, $T_{min}=\sqrt{3}\sqrt{L^2+3n^2}/2\pi L^2$. For temperatures $T<T_{min}$, only a thermal gas configuration is present. 

%\textcolor{red}{By inverting \eqref{htn} and using definition of $T_{min}$, we can express $z_h$ in terms of $T$ and $T_{min}$ as,
%\begin{align}
 %   z_h=\frac{3 }{2 \pi   T \pm 2 \pi \sqrt{ T^2-T_{min}^2}}\hspace{1mm}\label{zhT}
%\end{align}}
In the present analysis, we adopt the following choice for the $U(1)$ gauge field, 
\begin{align}\label{ansatzu1g}
     A_{t}=A_t(z),  A_\phi=A_\phi(z,t)=-Et+H(z),A_{\theta}=B \phi +K(z),
\end{align}
where $E$ and $B$ are {respectively the world-volume }electric and magnetic fields, oriented in the $\phi$ and $z$ directions, respectively. Here, $H$ and $K$ are functions of the $z$ coordinate only.

Substituting (\ref{TN metric})-(\ref{ansatzu1g}) into (\ref{S indp of t,phi}) and after integrating over the variables $t$ and $\phi$, we obtain
\begin{align}
  S_{DBI}=\int dz d \theta L_{DBI}, 
\end{align}
where the Lagrangian density ($L_{DBI}$) is given\footnote{Here, we absorb the overall factor $\int dt d\phi T_P= 2\pi t_0 T_P$ (where $t_0$ denotes the  time interval) into the definition of DBI action.} by
\begin{align}
    L_{DBI}=-\Bigg(g_{\theta\theta}\bigg[|g_{tt}|g_{zz}g_{\phi \phi}+|g_{tt}|H^{'2}-g_{\phi \phi}A_t^{'2}-2|g_{t \phi}|A_t'H'+g_{zz}(|g_{t \phi}|^2-E^2)\bigg]\nonumber\\ -A_t'^2 B^2+2 A_t' B E K'+B^2 g_{zz}|g_{tt}|+K'^2 \left(-E^2+|g_{tt}| g_{\phi \phi}+|g_{t\phi}|^2\right)\Bigg)^{1/2}.\label{DBI indp of A,H} 
\end{align}
Here, $'$ represents the derivative with respect to the coordinate $z$.

%{\footnotesize \begin{align} S_{DBI}=-\int dz d \theta\sqrt{\begin{aligned}
%g_{\theta\theta}\Bigg[|g_{tt}|g_{zz}g_{\phi \phi}+|g_{tt}|H^{'2}-g_{\phi \phi}A_t^{'2}-2|g_{t \phi}|A_t'H'+g_{zz}(|g_{t \phi}|^2-E^2)\bigg]\\ -A_t'^2 B^2+2 A_t' B E K'+B^2 g_{zz}|g_{tt}|+K'^2 \left(-E^2+|g_{tt}| g_{\phi \phi}+|g_{t\phi}|^2\right)     
%\end{aligned}}\label{DBI indp of A,H}
%\end{align}}

It is important to note that the DBI Lagrangian density (\ref{DBI indp of A,H}) contains only the derivatives of the functions $K(z)$, $A_t(z)$, and $H(z)$. This implies that there are three conserved charges ($b$, $c$ and $d$ {respectively}) associated with these functions, as given below
\begin{align}
b=&\hspace{1mm}\frac{\partial L_{DBI}}{\partial K'}=-\frac{A'_t B E+K' \left(|g_{tt} |g_{\phi \phi}+|g_{t\phi}|^2-E^2\right)}{L_{DBI}},\\
c=&\hspace{1mm}\frac{\partial L_{DBI}}{\partial A_t'}=\frac{A'_t \left(B^2+g_{\theta\theta}g_{\phi \phi}\right)-B. E. K'+g_{\theta\theta}|g_{t\phi}|H'}{L_{DBI}},\\
d=&\hspace{1mm}\frac{\partial L_{DBI}}{\partial H'}  =\frac{{g_{\theta \theta}}(A_t'|g_{t\phi}|-|g_{tt}|H')}{L_{DBI}}.
\end{align}

A straightforward calculation shows that the fields $K(z)$, $A_t(z)$, and $H(z)$ can be expressed in terms of the conserved charges $b$, $c$, and $d$ and the components of the TN-AdS metric, as shown explicitly in Appendix \ref{A,H,K expr}. Finally, by using these expressions, one can write the on-shell DBI Lagrangian density (\ref{DBI indp of A,H}) as follows\footnote{When we substitute the fields $K(z)$, $A_t(z)$, and $H(z)$ into the Lagrangian (\ref{DBI indp of A,H}), an overall factor of $\sqrt{g_{\theta \theta} g_{zz} \left(|g_{tt}|g_{\phi \phi }+|g_{t\phi }|^2\right)}$ appears. Upon inserting the TN-AdS metric components \eqref{TN metric}, this expression simplifies to $\frac{(L^4+n^{2}z^2)L^2\sin\theta}{z^4}$, which appears in the on-shell Lagrangian (\ref{onshell action}).} 

\begin{equation}
    L_{DBI}^{on-shell}=-\frac{(L^4+n^{2}z^2)L^2\sin\theta}{z^4} {\frac{\xi}{\sqrt{\xi \chi-\alpha^2}}}, \label{onshell action}
\end{equation}
where we define the functions $\chi$, $\xi$ and $\alpha$ as
\begin{align}
    \chi=&\hspace{1mm}c^2 g_{tt}-{d}^2 g_{\phi \phi}+g_{\theta\theta}|g_{tt}| g_{\phi \phi}+g_{\theta\theta}|g_{t\phi}|^2+2 {c} {d} |g_{t\phi}|-c^2 g_{\theta\theta},\label{c}\\
    \xi=&\hspace{1mm} B^2 |g_{tt}|+g_{\theta\theta} \left(|g_{tt}|g_{\phi \phi} +|g_{t\phi}|^2-E^2\right),\label{x}\\
    \alpha=&\hspace{1mm}B c| g_{tt}|+B d  |g_{t\phi}|+b E g_{\theta\theta}.\label{a}
    \end{align}

Notice that, as we approach the horizon ($z=z_h$), the functions $\chi$ (\ref{c}) and $\xi$ (\ref{x}) are both negative, while these functions are both positive near the boundary of space-time ($z=0$). On the other hand, the function $\alpha$ (\ref{a}) remains positive both at the horizon and at the boundary. Consequently, for the DBI Lagrangian density (\ref{onshell action}) to be real, the functions $\chi$ and $\xi$ must change sign at some point $z_*$, such that $z_h < z_* < 0$. Furthermore, $\alpha$ should also vanish at the turning point $z_*$. The turning point can be determined by requiring that the equations (\ref{c})–(\ref{a}) vanish at $z_*$, as demonstrated below
 {\begin{align}
   &  \chi(z_*)=0= c^2 |g_{tt}(z_*)|-d^2g_{\phi \phi}(z_*)+|g_{tt}(z_*)|g_{\phi \phi}(z_*)g_{\theta \theta}(z_*) +|g_{t\phi}(z_*)|^2g_{\theta \theta}(z_*)+2dc|g_{t \phi}(z_*)|\nonumber\\&\hspace{19mm}-b^2g_{\theta\theta}(z_*), \label{2nd constent to find d^2}\\
     & \xi(z_*)=0= B^2 |g_{tt}(z_*)|+g_{\theta\theta}(z_*) \left(|g_{tt}(z_*)|g_{\phi \phi}(z_*) +|g_{t\phi}(z_*)|^2-E^2\right), \label{find z_* constrent}\\
     & \alpha(z_*)= 0= B c |g_{tt}(z_*)|+B d  |g_{t\phi}(z_*)|+b E g_{\theta\theta}(z_*)=0.\label{3d cost}
\end{align}}

Using \eqref{2nd constent to find d^2} and \eqref{3d cost}, one can rewrite the conserved charges $b$ and $d$ in terms of $c$ and the component of TN-AdS metric as 
\begin{align}
 b=&\hspace{1mm}\Bigg[Bg_{\phi\phi} \Bigg(E\Omega\sqrt{g_{\phi \phi } g_{\theta \theta } \left(g_{tt} +g_{\phi\phi}\Omega^2\right) \left[B^2 \left(g_{\theta \theta} g_{\phi\phi}^2\Omega^2-c^2 g_{tt}\right)+E^2 g_{\theta \theta } \left(c^2+g_{\theta \theta }g_{\phi \phi}\right)\right]}+\nonumber\\
    &c E^2 g_{\theta \theta} \left(g_{tt} +g_{\phi\phi}\Omega^2\right)\Bigg)\Bigg]/\left(B^2 g_{\phi\phi}^2\Omega^2+E^2 g_{\theta \theta } g_{\phi \phi}\right)Eg_{\theta \theta},
   \label{b^2 without comp}\\
    d=&\hspace{1mm}\Bigg[E\sqrt{g_{\phi \phi } g_{\theta \theta } \left(g_{tt} +g_{\phi\phi}\Omega^2\right) \left[B^2 \left(g_{\theta \theta} g_{\phi\phi}^2\Omega^2-c^2 g_{tt}\right)+E^2 g_{\theta \theta } \left(c^2+g_{\theta \theta }g_{\phi \phi}\right)\right]}+\nonumber\\
    &\hspace{1mm}c g_{\ph\ph}\Omega\left( E^2 g_{\theta \theta}-B^2  g_{tt}\right) \Bigg]/ \left(B^2 g_{\phi\phi}^2\Omega^2+E^2 g_{\theta \theta } g_{\phi \phi}\right),
  \label{d^2 without comp form}
\end{align}
where we introduce the rotational angular velocity, $\Omega = |g_{t \phi }|/g_{\phi \phi}$, which arises due to frame dragging \cite{Durka:2019ajz}, \cite{Zhang:2016gzk}-\cite{Ong:2016cbo}. It is worth noting that these conserved charges $b$ (\ref{b^2 without comp}) and $d$ (\ref{d^2 without comp form}) are consistent with the results of \cite{Khan:2025fne}, in the absence of an external magnetic field $B$.

%Note: If we take \(B=0\) then \(b\) reduces to zero, and \(d^2\) reduces to \(d^2\) of the Ohmic Conductivity.
%\begin{align}
 %   b^2|_{(B=0)}=0
%\end{align}
%\begin{align}
 %   d^2|_{(B=0)}=&\hspace{1mm} g_{\theta \theta}|g_{tt}|+c^2 \frac{|g_{tt}|}{g_{\phi \phi}}+2c^2\Omega^2+\Omega^2 g_{\phi \phi}g_{\theta\theta}\pm2c\Omega\sqrt{g_{\theta \theta}|g_{tt}|+c^2 \frac{|g_{tt}|}{g_{\phi \phi}}+c^2\Omega^2+\Omega^2 g_{\phi \phi}g_{\theta\theta}}.
%\end{align}
%\\and \(b^2\) for \(g_{t\phi=0}\) reduce to following form.
%\begin{align}
 %   b^2|_{(g_{t\phi}=0)}=\frac{B^2 {c}^2 E^2}{\left(B^2+g_{\theta \theta} g_{\phi \phi }\right)^2}
%\end{align}
%This matches the result of \cite{A.~O'Bannon} in which they found \(b=\frac{(2\pi \alpha')^2B {c} }{\left(B^2+g_{xx}^2 \right)}E\) for \(Ads_5\) planer black hole.
%\\\\

Our next goal is to determine the turning point ($z_*$) at which the aforementioned expressions (\ref{b^2 without comp}) and (\ref{d^2 without comp form}) are valid. To calculate  $z_*$, we substitute the components of TN-AdS  (\ref{TN metric}) into (\ref{find z_* constrent}), which yields 
\begin{align}
    &\left(z_h-z_*\right)( B^2z_*^2 +\left({L^4}+n^2{z_*^2}\right)^2 \sin^2\theta) \Bigg(L^8 z_* z_h+n^2 z_*^3 z_h^3 \left(L^2+3 n^2\right)+L^8 z_*^2+\nonumber\\
    &+z_h^2 \left(L^8+L^6 z_*^2+6 L^4 n^2 z_*^2\right)\Bigg)-\left({L^4}+n^2{z_*^2}\right)^2E^2 L^2 z_*^4 z_h^3=0. \label{exact eqn of z_*}
\end{align}

It is evident from the above expression (\ref{exact eqn of z_*}) that solving this equation for an arbitrary value of the external electric field $E$ is quite complicated. However, we can obtain an approximate solution for small values of the electric field, $E<<1$, while keeping the magnetic field $B$ arbitrary. To achieve this,  we expand  $z_*$ as follows \cite{Khan:2025fne}
\begin{align}\label{expansion}
    z_*=z_*^{(0)}-E^2 z_*^{(1)}-E^4z_*^{(2)}\hspace{2mm},\hspace{1mm} E^2<<1.
\end{align}
Here, $z_*^{(0)}$ is the solution of (\ref{exact eqn of z_*}) in the absence of an external electric field $E$. Furthermore, $z_*^{(1)}$ and $z_*^{(2)}$ represent the leading order and next-to-leading order (NLO) corrections in the presence of $E$.
Substituting (\ref{expansion}) into (\ref{exact eqn of z_*}), we obtain $z_*$ up to NLO,
\begin{align}
    z_*=&\hspace{1mm}z_h-\frac{E^2 L^2 z_h^5 \left(n^2 z_h^2+L^4\right)}{\left(L^2 z_h^2+3 n^2 z_h^2+3 L^4\right) \left(B^2 z_h^4+2 L^4 n^2 z_h^2 \sin ^2\theta +n^4 z_h^4 \sin ^2\theta +L^8 \sin ^2\theta \right)}\nonumber\\& +E^4 L^4 z_h^9 \left(n^2 z_h^2+L^4\right)\left[
    \frac{ \left(B^2 z_h^4 \left(-2 L^6 z_h^2+n^2 z_h^4 \left(L^2+3 n^2\right)-3 L^8\right)\right)}{\left(z_h^2 \left(L^2+3 n^2\right)+3 L^4\right){}^3 \left(B^2 z_h^4+\sin ^2\theta  \left(n^2 z_h^2+L^4\right){}^2\right){}^3}\right.\nonumber\\&+\left.\frac{\sin ^2\theta  \left(n^2 z_h^2+L^4\right){}^2 \left(n^2 z_h^4 \left(L^2+3 n^2\right)+2 z_h^2 \left(L^6+6 L^4 n^2\right)+9 L^8\right)}{\left(z_h^2 \left(L^2+3 n^2\right)+3 L^4\right){}^3 \left(B^2 z_h^4+\sin ^2\theta  \left(n^2 z_h^2+L^4\right){}^2\right){}^3}\right].\label{z_* upto E^4}
\end{align}

Next, we substitute the above expression for $z_*$ (\ref{z_* upto E^4}) into the conserved charges \eqref{b^2 without comp} and (\ref{d^2 without comp form}). After comparing the conserved charges $b= <J_\th>\sin\th$ and $d= <J_{\ph}>\sin\th$ with the respective current densities, along with $c=<J_t>\sin\th$ as the charge density,  we obtain the following expressions

\begin{align}&
 <J_\phi>^2=\frac{ \left(L^4+n^2 z_{h}^2\right)^2}{B^2 z_{h}^4 \csc^2\th+\left(L^4+n^2 z_{h}^2\right)^2}\left[\frac{4 B^2 E^4 L^6 z_{h}^8 \csc ^8(\theta ) }{\left(3 L^4+z_h^2 \left(L^2+3 n^2\right)\right) \left(B^2 z_h^4 \csc^2\th+\left(L^4+n^2 z_h^2\right)^2\right)^2}\right.\nonumber\\& \hspace{20mm}\left.+\frac{E^2}{\sin^4\th}+\frac{\Omega ^2  \left(L^4+n^2 z_h^2\right)^2}{z_h^4}\right]+
\frac{ J^2_t \left(L^4+n^2 z_h^2\right)^2}{\left(B^2 z_h^4 \csc^2\th+\left(L^4+n^2 z_h^2\right)^2\right)^2}\Bigg[{3  \Omega ^2 \left(L^4+n^2 z_h^2\right)^2}\nonumber\\&\hspace{22mm}\left.+\frac{E^2  z_h^4}{ \sin^4\th}+\frac{2 E  \Omega  \left(L^4+n^2 z_h^2\right)^2}{J_t\sin^2\th} \sqrt{1+\frac{z_h^4 (J_t^2+B^2 \csc^2\th)}{\left(z_h^2 n^2+L^4\right)^2}}\right. \nonumber
\\&\hspace{22mm}\left.-\frac{4 E^4  L^6 z_h^8\csc^6(\th)  \left(\left(L^4+n^2 z_h^2\right)^2-B^2 z_h^4 \csc^2\th\right)}{\left(3 L^4+z_h^2 \left(L^2+3 n^2\right)\right) \left(B^2 z_h^4 \csc^2\th+\left(L^4+n^2 z_h^2\right)^2\right)^2}
\right],\label{current phi}
\end{align}
\begin{align}&
    <J_\theta>^2=\frac{J_t^2 z_h^8 B^2}{\left(B^2\csc^2\th+\left(z_h^2 n^2+L^4\right)^2\right)^2}\left[\frac{ E^2}{\sin^4\theta}
 +\frac{2  E  \Omega   \left(L^4+n^2 z_h^2\right)^2 }{ J_tz_h^4\sin^2\th}\sqrt{1+\frac{z_h^4 \left(B^2 \csc^2\th+J_t^2\right)}{\left(L^4+n^2 z_h^2\right)^2}} \right. \nonumber\\&\hspace{20mm}  \left.+\frac{3 \Omega ^2 \left(z_h^2 n^2+L^4\right)^2}{z_h^4}-\frac{8 z_h^{4}  E^4 L^6 \csc ^6(\theta ) \left(z_h^2 n^2+L^4\right)^2}{\left(z_h^2 \left(L^2+3 n^2\right)+3 L^4\right) \left(z_h^4 B^2 \csc^2\th+\left(z_h^2 n^2+L^4\right)^2\right)^2}\right]\nonumber\\&\hspace{22mm}+\frac{B^2 \Omega ^2 \left(z_h^2 n^2+L^4\right)^2}{(z_h^4 B^2 \csc^2\th+\left(z_h^2 n^2+L^4\right)^2)},\label{current th}
\end{align}
where the frame-dragging angular velocity can be expressed as\footnote{Notice that the frame-dragging angular velocity is directly proportional to the square of  the electric field $(E)$, which implies that in the absence of the electric field, no charge carriers will be accelerated, hence no frame-dragging effects will be observed.}
\begin{align}
    | \Omega(z_*)|=\frac{E^2}{\sin^4\theta}\frac{2n z_h^4|\left(\cos \theta+\beta\right)|}{\left(n^2 z_h^2+L^4\right){}^2+z_h^4 B^2 \csc^2\th} +O(E^4).\label{omega at z*}
\end{align}

Notice that it is sufficient to consider the angular velocity, $\Omega$ (\ref{omega at z*}), up to $O(E^2)$ as it appears in \eqref{current phi} and \eqref{current th} either as a quadratic term or multiplied with an external electric field ($E$) or magnetic field ($B$). In other words, quartic order corrections to $\Omega$ contribute %negligibly to $J_{\th}$ and $J_{\phi}$.%
higher order in the current densities $J_{\theta}$ and $J_{\phi}$.

Following our discussion in Section \ref{secint1}, we notice that the conductivities are related to current densities and the external electric field through the expression $<J_i>=\sigma_{ij} E_j$ \cite{A.~O'Bannon}. After comparing equations \eqref{current phi} and \eqref{current th} with the standard form \eqref{condcfromDBI}, we obtain the following expressions for the Ohmic conductivity ($\sigma_{\phi\phi}$) and Hall conductivity ($\sigma_{\theta\phi}$),
\begin{align}
<J_\phi>=\s_{\ph\ph}E_{\phi}\hspace{1mm},\hspace{2mm}<J_\th>=\s_{\th\ph}E_{\ph},
\end{align}
where we identify, $E_\ph=E$ and
\begin{align}
    \sigma_{\ph\phi}=&\hspace{1mm} \sqrt{\s_{\phi\phi,\text{thermal}}^2+\s_{\phi\ph, U(1)}^2}\hspace{1mm}, \hspace{2mm}  \sigma_{\th\ph}=\hspace{1mm} \sqrt{\s_{\th\phi,\text{thermal}}^2+\s_{\th\ph ,U(1)}^2}.\label{FC}
\end{align}

The subscript ``$U(1)$" denotes the contribution due to the externally added electric charge carriers. On the other hand, the subscript ``thermal" represents the contribution due to thermally produced pairs. The detailed expressions of these conductivities are given below\footnote{Here, we rescaled the conductivities $(\s_{\phi\phi}$ and $\s_{\th\ph})$ by a factor of $\sin^2\th$.}
{\begin{align} &
    \sigma_{\phi\phi ,U(1)}^{2}=\frac{ J_t^2 \left(L^4+n^2 z_h^2\right)^2}{\left(B^2 z_h^4 \csc^2\th+\left(L^4+n^2 z_h^2\right)^2\right)^2}\left[z_h^4+\frac{2 ez_h  \tilde\Omega  \left(L^4+n^2 z_h^2\right)^2 }{J_t \sin\th}\sqrt{1+\frac{z_h^4 (J_t^2+B^2 \csc^2\th)}{\left(z_h^2 n^2+L^4\right)^2}}\right.\nonumber\\&\hspace{18mm}\left.+\frac{3 \tilde \Omega ^2 e^2z_h^2\left(L^4+n^2 z_h^2\right)^2}{\sin^2\th }-\frac{4 e^2 J_t L^6 z_h^{10}  \left(\left(L^4+n^2 z_h^2\right)^2-B^2 z_h^4 \csc^2\th\right)\csc^2\th}{\left(3 L^4+z_h^2 \left(L^2+3 n^2\right)\right) \left(B^2 z_h^4 \csc^2\th+\left(L^4+n^2 z_h^2\right)^2\right)^2}\right],\label{condu1dc}
\end{align}

\begin{align} &
    \sigma_{\phi\phi ,\text{thermal}}^{2}=\frac{ \left(L^4+n^2 z_{h}^2\right)^2}{B^2 z_{h}^4 \csc^2\th+\left(L^4+n^2 z_{h}^2\right)^2}\left[\frac{4 B^2 E^2L^6 z_{h}^{10} \csc^4\th }{\left(3 L^4+z_h^2 \left(L^2+3 n^2\right)\right) \left(B^2 z_h^4 \csc^2\th+\left(L^4+n^2 z_h^2\right)^2\right)^2}\right.\nonumber\\&\left.\hspace{22mm}+1+\frac{\tilde\Omega ^2e^2  \left(L^4+n^2 z_h^2\right)^2}{z_h^2\sin^2\th}\right],\label{condthdc}\end{align} }

   { \begin{align}
\sigma_{\th\ph,U(1)}^2=&\frac{J_t^2 z_h^8 B^2}{\left(B^2\csc^2\th+\left(z_h^2 n^2+L^4\right)^2\right)^2}\left[1
 +\frac{2  e z_h  \tilde\Omega   \left(L^4+n^2 z_h^2\right)^2 }{ J_tz_h^4\sin\th}\sqrt{1+\frac{z_h^4 \left(B^2 \csc^2\th+J_t^2\right)}{\left(L^4+n^2 z_h^2\right)^2}} \right. \nonumber\\&\hspace{10mm}  \left.+\frac{3\tilde \Omega ^2 e^2\left(z_h^2 n^2+L^4\right)^2}{z_h^2\sin^2\th}-\frac{8 z_h^{6}  e^2 L^6 \csc^2\th \left(z_h^2 n^2+L^4\right)^2}{\left(z_h^2 \left(L^2+3 n^2\right)+3 L^4\right) \left(z_h^4 B^2 \csc^2\th+\left(z_h^2 n^2+L^4\right)^2\right)^2}\right],\label{hallu1} \\
 \s^2_{\th\ph,\text{thermal}}=&\frac{B^2 \tilde\Omega ^2 e^2 z_h^2 \left(z_h^2 n^2+L^4\right)^2}{\sin^2\th\left(z_h^4 B^2 \csc^2\th+\left(z_h^2 n^2+L^4\right)^2\right)},\label{hall conductivity thermal}
  \end{align}}
where we denote $e=E/z_h$ and $\tilde\Omega=\Omega\sin^3\theta/E^2$.

The above expressions  (\ref{FC})-(\ref{hall conductivity thermal}) for the holographic Ohmic and Hall conductivities are the main results of our paper. It is important to notice that these conductivities are affected due to the presence of an external magnetic field ($B$) and the NUT parameter ($n$), which induces frame dragging. In the absence of the external magnetic field ($B$), the above expressions (\ref{FC})-(\ref{hall conductivity thermal}) reduce to those found in the previous paper \cite{Khan:2025fne}.

It is interesting to notice that our results (\ref{FC})-(\ref{hall conductivity thermal}) differ significantly from those reported in the earlier work \cite{A.~O'Bannon}-\cite{Lee:2010uy}. In particular, the authors in \cite{A.~O'Bannon} examined the conductivity of $AdS_5$, the planar black holes, both in the presence of the electric and magnetic fields. They found that only the Ohmic conductivity receives contributions due to thermal and $U(1)$ charge carriers, whereas the Hall conductivity receives only the $U(1)$ contribution. In contrast, our results reveal contributions due to both $U(1)$ and thermally produced charge carriers to Ohmic and Hall conductivities as shown in (\ref{FC})-(\ref{hall conductivity thermal}). The most significant is the novel frame-dragging effects produce an additional drift of the charge carriers, resulting in a non-vanishing thermal contribution to the holographic Hall conductivity (\ref{hall conductivity thermal}).

Ideally, in the absence of frame-dragging and under the influence of an electric field (say $\vec{E}=E_x \hat x$), the particles and their antiparticles in a thermal plasma move with the identical speed but in opposite directions. When we turn on the magnetic field (say $\vec{B}=B_z\hat z$), both particles and antiparticles exhibit motion in the transverse $\hat{y}$ plane due to the Lorentz force ($\vec{F}_L=q(\vec{v}\times \vec{B})$), where $q$ is the charge of the particle. Consequently, this results in a zero Hall current for the thermally produced charge carriers as discussed by the authors in \cite{Hartnoll:2009sz}, \cite{Rath:2021ryd}. In other words, the Hall conductivity due to thermally produced pairs must be much less than its corresponding Ohmic counterpart.

On the other hand, our findings indicate that both $U(1)$ and thermally produced charge carriers contribute to the Ohmic and Hall conductivities (\ref{FC})-(\ref{hall conductivity thermal}). This contribution arises from the anisotropic nature of space-time, which results in a frame-dragging effect. When an electric field $\vec{E} = E_{\phi} \hat{\phi}$ is applied, a particle moving in the longitudinal direction, $+\hat{\phi}$, experiences additional drift. In contrast, negatively charged particles moving in the $-\hat{\phi}$ direction undergo deceleration due to the rotational velocity in the $\hat{\phi}$ direction, which is a consequence of the frame-dragging effect. As a result, particles and antiparticles move at different velocities due to the Lorentz force in the $-\hat\theta$ direction, generating a net non-zero Hall current within the thermal plasma as illustrated in \eqref{hall conductivity thermal}. To summarize, frame-dragging produces an effective (or net) velocity along the  direction of dragging $(\hat\phi)$, which in turn produces a (net) non-zero Hall current. This appears to be one of the major differences in the presence of the frame-dragging which we further elaborate in our subsequent analysis.

  \section{Hall transports at low magnetic field\label{low b 3}}
    In this Section, we examine how does the NUT parameter $(n)$ and the associated frame dragging $(\Om)$ influence the holographic Ohmic and Hall conductivities (\ref{FC})-(\ref{hall conductivity thermal}) associated with the TN-AdS black hole. We study the behavior of these conductivities in the low and high temperature regions, both near and far from the Misner string, while considering a small magnetic field ($B<<1$) limit.

    To start with, we first re-write the angular velocity (\ref{omega at z*}) and the conductivities (\ref{FC})-(\ref{hall conductivity thermal})  in terms of the Hawking temperature (\ref{htn}) and the minimum temperature ($T_{min}$). The angular velocity can be expressed as\footnote{
  The AdS length scale is set to unity (i.e., $L=1$) for the rest of the computation.}
  %\footnote{Since $g_{t\phi}=2nf(z)(\cos\th+\b)$ , $\Omega=\frac{g_{t\phi}}{g_{\ph\ph}}\propto n$ as given at $z_*$ \eqref{omega at z*}and by using \eqref{zhT} into \eqref{omega at z*},we write   $\tilde\Omega(z_*)=\frac{\Omega(z_*)\sin^3\th}{E^2}$ at $T=T_{min}$ and expend in small n(since $T_{min}=\sqrt{1+3n^2}/2\pi$ and for low T regime,  n must be small)  and take leading oreder so $\tilde\Omega\propto n$} 
 \begin{align}
   \big| \tilde\Omega\big|
   =&\hspace{1mm}\frac{162  n }{\left(\left(4 \pi ^2  \left(2 T \tilde{T}-T_{min }^2+2 T^2\right)+9 n^2\right)^2+81B^2\csc^2\th\right)}\cot \left(\frac{\theta }{2}\right),\label{angb1}
 \end{align}
  %\begin{align}
   %\big| \tilde\Omega\big|_{\b=0}=&\hspace{1mm}\frac{162  n }{\left(\left(4 \pi ^2  \left(2 T \tilde{T}-T_{min }^2+2 T^2\right)+9 n^2\right)^2+81B^2\csc^2\th\right)}\cot \left(\theta\right),\label{angb2}
 %\end{align}
 %\begin{align}
  % \big| \tilde\Omega\big|_{\b=-1}=&\hspace{1mm}\frac{162  n }{\left(\left(4 \pi ^2  \left(2 T \tilde{T}-T_{min }^2+2 T^2\right)+9 n^2\right)^2+81B^2\csc^2\th\right)}\tan \left(\frac{\theta }{2}\right),\label{angb3}
 %\end{align}\footnote{For $\b=1$, the Misner string is located at $\th=0$.}
where we define $\tilde{T}=\sqrt{T^2-T_{min }^2}$ and set the location of Misner string\footnote{We discuss conductivities for $\b=0$ and $\b=-1$ in the Appendix \ref{Appc}. As frame-dragging is significant at low magnetic field, therefore we refrain ourselves from discussing the finite magnetic field case.} at $\th=0$ i.e. $\b=1$ . The detailed expressions of the holographic Ohmic and Hall conductivities have been provided in Appendix \ref{exect condT}. Notice that the effects due to $E$ has been absorbed into the definition of $\tilde\Omega$ (as defined below (\ref{hall conductivity thermal})).

\subsection{Low temperature regime}
In this Section, we investigate the Hall and Ohmic conductivities (see Appendix B (\ref{condu1dcT})-\eqref{hallthermal T}) at low temperatures. The low temperature region is defined as the range where the temperature ($T$) is close to the minimum temperature ($T_{min}$), i.e., $T\sim T_{min}$. It is important to note that when the temperature approaches the minimum value ($T\sim T_{min}$) and in the small magnetic field limit ($B<<1$), the angular velocity (\ref{angb1}) can be approximated as

%in the rest of our analysis, we assume that both the magnetic field and the NUT parameter are small, i.e., $B<1$ and $n<1$. Under these conditions, one can approximate the angular velocity (\ref{angb1}) near the minimum temperature as

%\textcolor{red}{We primly conducted our analysis in the small \( B \) limit, as performed by the authors in \cite{Lee:2010ii}, \cite{Hartnoll:2009ns}. They observed that \(\sigma_{xx}\) simplifies to their previously calculated \(\sigma_{xx}\) for \( B=0 \). The authors also determined that the ratio of the Hall and DC conductivities scales inversely with the DC conductivity, \(\frac{\sigma_{xx}}{\sigma_{xy}} \sim (\sigma_{xx})^{-1}\). We will also examine the ratio of the Hall and DC conductivities, both near and far from Misner's string.\\
%First, we write $\tilde\Omega$ at low T in small B by writing \eqref{angb1}  at $T=T_{min}$ and expanding in B up to $B^2$ , then expanding in small n up to $n^2$ to ensure $T_{min}=\sqrt{1+3n^2}/2\pi$ to low enough.
\begin{align}
    |\tilde\Omega|=\frac{81 n \cot \left(\frac{\theta }{2}\right)}{8 \pi ^4  T_{min }^4}\left(1-\frac{81B^2\csc^2\th}{16\pi^4T_{min}^4}\right)+O(\sqrt{T-T_{min}}).\label{am}
\end{align}
%}
Notice that the angular velocity (\ref{am}) is proportional to the NUT parameter ($n$).  In other words, the parameter $n$ actually contributes to the frame-dragging effect, which will become more evident in the subsequent expressions.
\subsubsection{Hall conductivity}
In the present Section, we study the Hall conductivity $(\s_{\th\ph})$ near the minimum temperature, i.e., $T\sim T_{min}$, and consider smaller values of the external magnetic field ($B<<1$) (see Appendix B \eqref{hallu1T}-\eqref{hallthermal T}). In this context, the $\s_{\th\ph}$ can be approximately expressed as
\begin{align}
 &   \s_{\th\phi,U(1)}\bigg|^{fM}_{T \sim T_{min}}=\frac{81BJ_t}{16\pi^4T_{min}^4}+O\left(\sqrt{T-T_{min}}\right)\label{HallUafmb1},\\&\s_{\th\ph,U(1)}\bigg|^{nM}_{T \sim T_{min}}=\frac{729 \sqrt{3} B E n J_t\cot\left(\frac{\th}{2}\right)}{32 \pi ^6\sin\th   T_{min}^6}+O\left(\sqrt{T-T_{min}}\right)\label{HallU1nMb1},%\\&\s_{\th\phi,\text{thermal}}\bigg|^{fM}_{T \sim T_{min}}=\frac{81 B E n }{16 \pi ^4 T_{\min }^4}+O\left(\sqrt{T-T_{min}}\right),
 \\&{\s_{\th\phi,\text{thermal}}\bigg|^{fM}_{T \sim T_{min}}=O\left(\sqrt{T-T_{min}}\right)},\label{hallthfMb1}\\&\s_{\th\phi,\text{thermal}}\bigg|^{nM}_{T \sim T_{min}}=\frac{81 B E n\cot\left(\frac{\th}{2}\right) }{8 \pi ^4\sin\th T_{min }^4}+O\left(\sqrt{T-T_{min}}\right).\label{hallthnMb1}
\end{align}
Here, the superscripts $fM$ and $nM$ denote the points far from the Misner string and close to the Misner string, respectively. Additionally, the parameter $0<\th<<\pi$ measures the proximity to the Misner string.

It is interesting to note that frame dragging appears beyond the quadratic order in the electric field ($E$) as demonstrated in \eqref{current phi}-\eqref{omega at z*}. The effects due to frame-dragging is significant near the Misner string, where $\th\sim0$ as illustrated in equations \eqref{HallU1nMb1} and \eqref{hallthnMb1}. Furthermore, the thermal conductivity ($\s_{\th\phi,\text{thermal}}$) arises from frame-dragging effects. However, as we move away from the Misner string (\ref{hallthfMb1}), its contribution vanishes, provided $BEn<T_{min}^4$. In other words, far from the Misner string, the analysis goes parallel to the previous results \cite{A.~O'Bannon}-\cite{Lee:2010ii}, while near the Misner string it differs significantly.

\begin{figure}[H]
     \centering
     \begin{subfigure}[b]{0.495\textwidth}
         \centering
         \includegraphics[width=\textwidth]{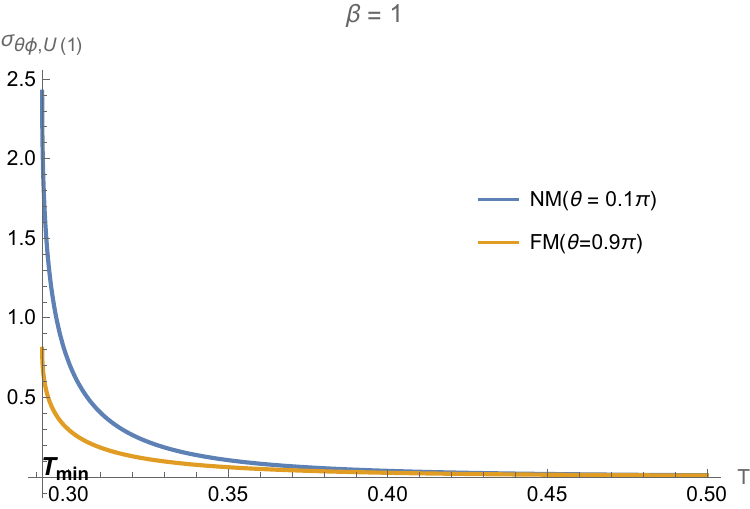}
         \caption{$\s_{\th\ph,U(1)}$ vs $T$ for $B=0.01$ }\label{fig1a}
              \end{subfigure}
              \hfill
     \begin{subfigure}[b]{0.495\textwidth}
         \centering
         \includegraphics[width=\textwidth]{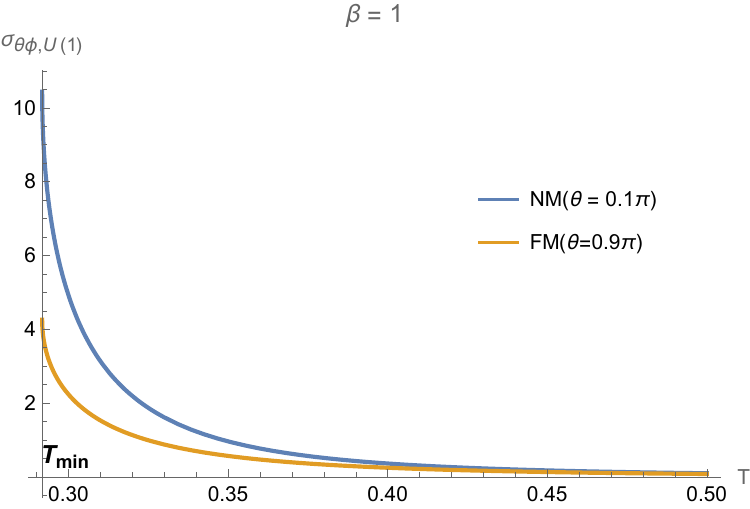}
         \caption{$\s_{\th\ph ,U(1)}$ vs $T$ for $B=0.1$}\label{fig1b}
             \end{subfigure}
        
             \caption{$\s_{\th\ph,U(1)}$ vs temperature plot. Here we set $E=0.1$, $J_t=10$ and $n=0.2$.}
        \label{figure1}
\end{figure}

\begin{figure}[H]
     \centering
     \begin{subfigure}[b]{0.495\textwidth}
         \centering
         \includegraphics[width=\textwidth]{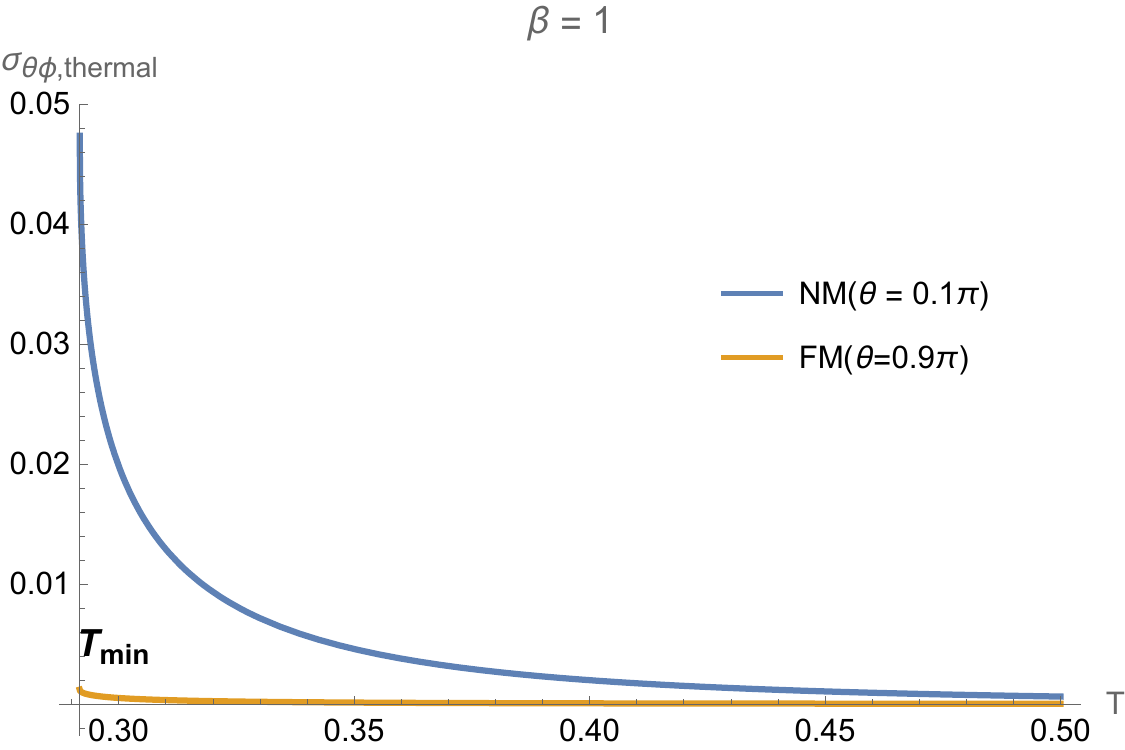}
         \caption{$\s_{\th\ph,\text{thermal}}$ vs $T$ for $B=0.01$ }\label{fig2a}
              \end{subfigure}
              \hfill
     \begin{subfigure}[b]{0.495\textwidth}
         \centering
         \includegraphics[width=\textwidth]{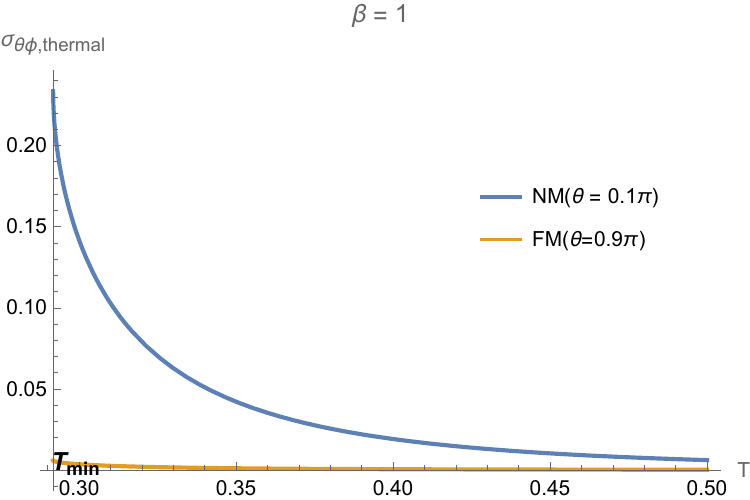}
         \caption{$\s_{\th\ph ,\text{thermal}}$ vs $T$ for $B=0.1$}\label{fig2b}
             \end{subfigure}
        
             \caption{$\s_{\th\ph,\text{thermal}}$ vs temperature plot. Here we set $E=0.1$, $J_t=10$ and $n=0.2$.}
        \label{figure2}
\end{figure}

To provide further clarity, we plot the conductivities ($\s_{\th\ph,U(1)}$ and $\s_{\th\phi,\text{thermal}}$) with temperature (see Figures \ref{figure1} and \ref{figure2}) both near and far from the Misner string. We set $\th=0.1\pi$ for the region near the Misner string and $\th=0.9\pi$ for the region away from it. 

Figures \ref{figure1} and \ref{figure2} show the behavior of $U(1)$ Hall conductivity $\s_{\th\ph,U(1)}$, (which arise from the externally added $U(1)$ charge carriers) and the thermal Hall conductivity ($\s_{\th\ph,\text{thermal}}$) in the low-temperature regime. Notably, $\s_{\th\ph,U(1)}$ \eqref{HallUafmb1}-\eqref{HallU1nMb1} increases as we approach the minimum temperature ($T \sim T_{\text{min}}$), with a particularly steep rise near the Misner string, where $\theta = 0.1\pi$. This increase is due to the additional drift of the $U(1)$ charge carriers as a result of the novel frame-dragging effect $(\tilde{\Omega}\propto n)$ (\ref{am}). Furthermore, as we increase the magnetic field, these charge carriers are further accelerated, contributing to even higher conductivity near the minimum temperature, regardless of the location of the Misner string (see Figure  \ref{fig1b}).

Next, we discuss the contribution due to thermally produced charge carriers to Hall conductivity. Notice that like the $U(1)$ Hall conductivities, $\sigma_{\th\ph, \text{thermal}}$ \eqref{hallthfMb1}-\eqref{hallthnMb1} also increases as we move towards the threshold temperature, $T\sim T_{min}$, and the rise is particularly more significant near the Misner string. On the other hand, the conductivity is negligibly small for the region away from it (see Figure \ref{figure2}). Physically, far from the Misner string, the frame-dragging effect is negligible, resulting in an equal number of particles and antiparticles moving in same direction with identical speed(s) which causes a zero Hall transport as observed in \cite{Hartnoll:2009sz}. Finally, the increase in the magnetic field leads to an increase in the conductivity, as explained above.

To make a comparative analysis,  we take the ratio of \(\sigma_{\th\ph,U(1)}\) near and far from the Misner string \eqref{HallUafmb1}-\eqref{HallU1nMb1}. We apply a similar approach for \(\sigma_{\th\ph,\text{thermal}}\) \eqref{hallthfMb1}-\eqref{hallthnMb1}, which yield the following ratio
\begin{align}
     \frac{\sigma_{\th\ph,U(1)}\bigg|_{T \sim T_{min}}^{nM}}{\sigma_{\th\ph,U(1)}\bigg|_{T \sim T_{min}}^{fM}}  =\frac{18 \sqrt{3} E n\cot\left(\frac{\th}{2}\right)}{\pi ^2\sin\theta T^2_{min }}>>1\hspace{1mm},\hspace{2mm}\frac{\sigma_{\th\ph,thermal}\bigg|_{T \sim T_{min}}^{nM}}{\sigma_{\th\ph,thermal}\bigg|_{T \sim T_{min}}^{fM}}>>1.\label{ratio1}
\end{align}
It clearly shows that both $U(1)$ and thermal contributions (\(\sigma_{\th\ph,U(1)}\) and \(\sigma_{\th\ph,\text{thermal}}\) ) dominate near the Misner string, due to the novel frame-dragging effect $(\tilde{\Omega}\propto n)$ (\ref{am}).

Finally, it is worth noting that in the low temperature regime, the number of externally added $U(1)$ charge carriers exceeds the number of thermally produced charge pairs. As a result, the conductivity $\s_{\th\ph,U(1)}$ is significantly higher than $\s_{\th\ph,\text{thermal}}$, regardless of the position of the Misner string, as illustrated below
\begin{align}
  \frac{\s_{\th\ph,U(1)}}{\s_{\th\ph,thermal}}\Bigg|^{nM}_{T\sim T_{min}} = \frac{9\sqrt{3}J_t}{4\pi^2\ T_{min}^2}>>1\hspace{1mm},\hspace{2mm} \frac{\s_{\th\ph,U(1)}}{\s_{\th\ph,thermal}}\Bigg|^{fM}_{T\sim T_{min}}>>1.\label{ratio2}
\end{align}
\subsubsection{Ohmic conductivity}
Our next goal is to study the holographic Ohmic conductivity ($\s_{\ph\ph}$) associated with the TN-AdS black holes. In the low temperature regime ($T\sim T_{min}$) and for small values of magnetic field ($B<<1$), the conductivity $\s_{\ph\ph}$ (see Appendix \eqref{condu1dcT}-\eqref{condthT}) can be approximately expressed as  
% & \s_{\phi\ph,U(1)}\bigg|^{fM}_{T \sim T_{min}}=\frac{9J_t}{4\pi^2T_{min}^2}\left(1-\frac{81B^2 \csc^2\th }{16 \pi ^4 T^2_{min}}\right)+O\left(\sqrt{T-T_{min}}\right),\label{d1}\\&\s_{\phi\ph,U(1)}\bigg|^{nM}_{T \sim T_{min}}=\frac{81 \sqrt{3} E n J_t\cot\left(\frac{\th}{2}\right)}{4 \pi ^4 \sin\th T_{\min }^4}\left(1-\frac{81B^2 \csc^2\th }{8 \pi ^4 T^2_{min}}\right)+O\left(\sqrt{T-T_{min}}\right), \label{d2}\\& \s_{\phi\ph,\text{thermal}}\bigg|^{fM}_{T \sim T_{min}}=1-\frac{81B^2\csc^2\th }{32  \pi ^4  T_{min}^4}+O\left(\sqrt{T-T_{min}}\right),\label{d3}\\&\s_{\phi\ph,\text{thermal}}\bigg|^{nM}_{T \sim T_{min}}=\frac{9 En\cot\left(\frac{\th}{2}\right)}{2\pi ^2 \sin\th T_{min}^2}\left(1-\frac{243B^2\csc^2\th }{32  \pi ^4  T_{min}^4}\right)+O\left(\sqrt{T-T_{min}}\right),\label{d4}

\begin{align}
   & \s_{\phi\ph,U(1)}\bigg|^{fM}_{T \sim T_{min}}=\frac{9J_t}{4\pi^2T_{min}^2}\left(1-\frac{81B^2 \csc^2\th }{16 \pi ^4 T^4_{min}}\right)+O\left(\sqrt{T-T_{min}}\right),\label{d1}\\&\s_{\phi\ph,U(1)}\bigg|^{nM}_{T \sim T_{min}}=\frac{81 \sqrt{3} E n J_t\cot\left(\frac{\th}{2}\right)}{8 \pi ^4 \sin\th T_{\min }^4}\left(1-\frac{81B^2 \csc^2\th }{8 \pi ^4 T^4_{min}}\right)+O\left(\sqrt{T-T_{min}}\right), \label{d2}\\& \s_{\phi\ph,\text{thermal}}\bigg|^{fM}_{T \sim T_{min}}=1-\frac{81B^2\csc^2\th }{32  \pi ^4  T_{min}^4}+O\left(\sqrt{T-T_{min}}\right),\label{d3}\\&\s_{\phi\ph,\text{thermal}}\bigg|^{nM}_{T \sim T_{min}}=\frac{9 En\cot\left(\frac{\th}{2}\right)}{2\pi ^2 \sin\th T_{min}^2}\left(1-\frac{243B^2\csc^2\th }{32  \pi ^4  T_{min}^4}\right)+O\left(\sqrt{T-T_{min}}\right).\label{d4}
\end{align}

For points far from the Misner string, we set $\theta \sim \pi$, while $\theta \sim 0$ near the Misner string. Notice that in the limit, $B\rightarrow0$, the Ohmic conductivities (\ref{d1})-(\ref{d4}) reduce to the results reported earlier  \cite{Khan:2025fne}. Furthermore, these conductivities \eqref{d1}-\eqref{d4} decrease in the presence of the magnetic field. This stems from the fact that in the presence of the magnetic field ($B$), a fraction of the charge carriers are drifted in a direction orthogonal to the electric field ($E$) causing a lower Ohmic conductivity. In the literature, this phenomenon is called the magneto-resistance \cite{Kim:2010zq}, in which the Ohmic resistance 
%defined as the inverse of the Ohmic conductivity
 increases with the magnetic field. 

%\textcolor{red}{The above  expressions of Ohmic conductivities \eqref{d1}-\eqref{d4} suggest a reduction in Ohmic conductivity attributable to the presence of a magnetic field. This phenomenon can be interpreted as a conductivity analogue of magneto-resistance \cite{Kim:2010zq}, wherein Ohmic resistance increases due to the magnetic field. As demonstrated by the author in \cite{Ziman}, magneto-resistance is proportional to $B^2$ in the weak field limit. However, in this context, we are addressing conductivity rather than resistivity, thus the reduction in Ohmic conductivity due to the magnetic field can be understood as analogous to the magneto-resistance effect.
%Magneto-resistance in literature is given by $\frac{\Delta\rho}{\rho(B=0)}=\frac{\rho(B)-\rho(B=0)}{\rho(B=0)}$  \cite{Kim:2010zq}, and we found Magneto-resistance $\propto\frac{B^2}{T_{min}^4}$ both  near and far from Misner's string as given below, \begin{align}  \frac{\Delta\rho_{\ph\ph,U(1)}}{\rho_{\ph\ph,U(1)}(B=0)}\Bigg|^{fM}_{T\sim T_{min}}=\frac{81B^2 \csc^2\th }{16 \pi ^4 T^4_{min}},\hspace{2mm} \frac{\Delta\rho_{\ph\ph,U(1)}}{\rho_{\ph\ph,U(1)}(B=0)}\Bigg|^{nM}_{T\sim T_{min}}=\frac{81B^2 \csc^2\th }{8 \pi ^4 T^4_{min}} \end{align}\begin{align}  \frac{\Delta\rho_{\ph\ph,\text{thermal}}}{\rho_{\ph\ph,\text{thermal}}(B=0)}\Bigg|^{fM}_{T\sim T_{min}}=\frac{81B^2 \csc^2\th }{32 \pi ^4 T^4_{min}},\hspace{2mm} \frac{\Delta\rho_{\ph\ph,\text{thermal}}}{\rho_{\ph\ph,\text{thermal}}(B=0)}\Bigg|^{nM}_{T\sim T_{min}}=\frac{243B^2 \csc^2\th }{32 \pi ^4 T^4_{min}} \end{align}

To clarify further, we plot the holographic Ohmic conductivities ($\s_{\ph\ph,U(1)}$ and $\s_{\ph\ph,thermal}$) against the temperature ($T$) near the threshold value ($T\sim T_{min}$). Here, we set the NUT parameter $n=0.2$. For points far from the Misner string, we set $\th=0.9\pi$, and for points closer to the Misner string, we set $\th=0.1\pi$.
\begin{figure}[H]
     \centering
     \begin{subfigure}[b]{0.495\textwidth}
         \centering
         \includegraphics[width=\textwidth]{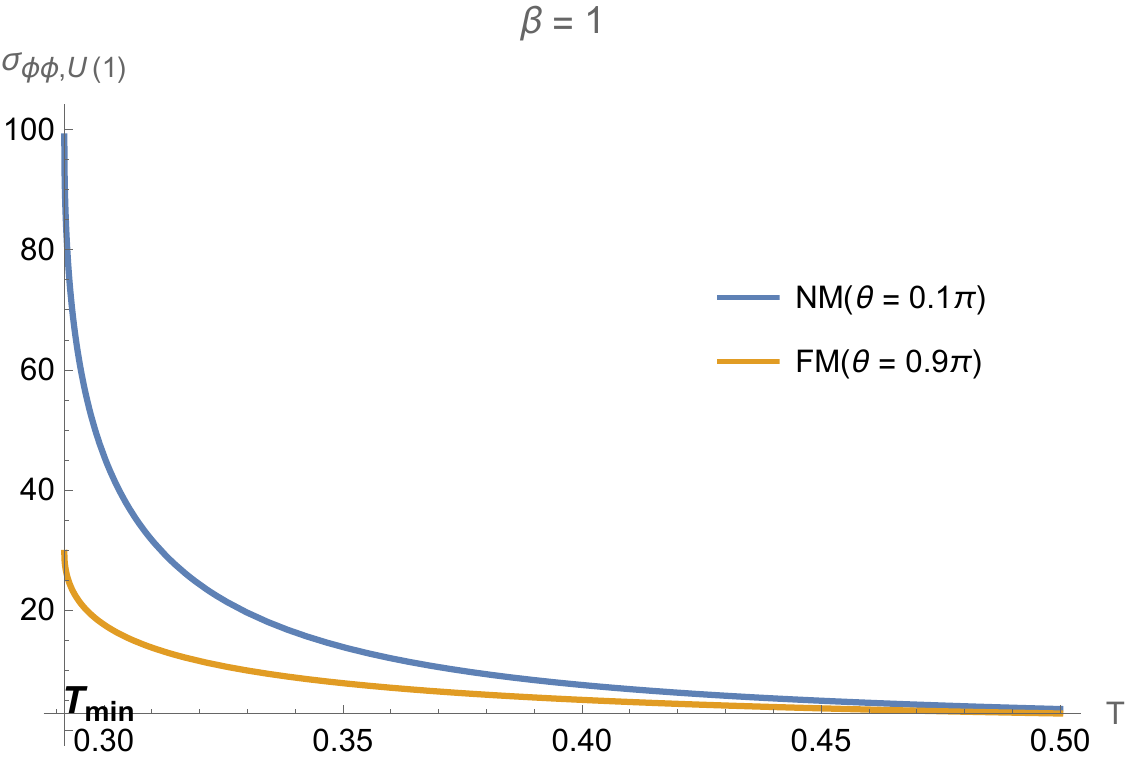}
         \caption{$\s_{\ph \ph,U(1)}$ vs $T$ for $B=0.01 $}\label{fig3a}
              \end{subfigure}
              \hfill
     \begin{subfigure}[b]{0.495\textwidth}
         \centering
         \includegraphics[width=\textwidth]{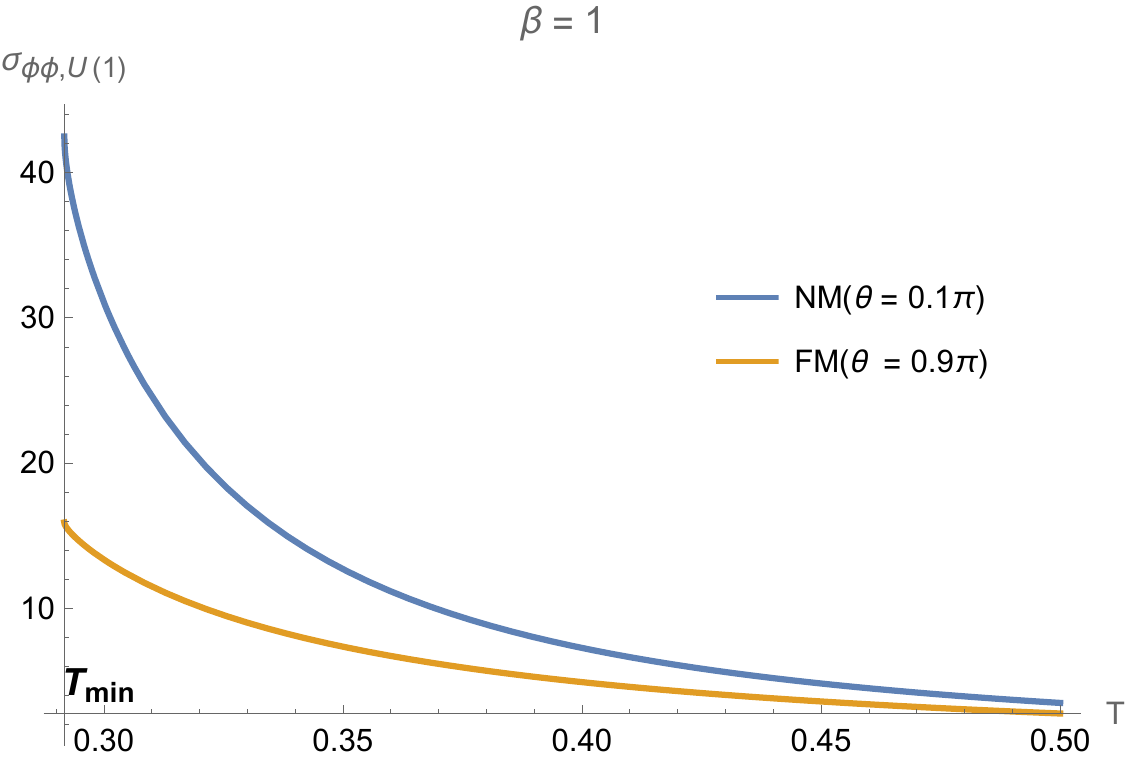}
         \caption{$\s_{\ph \ph ,U(1)}$ vs $T$ for $B=0.1$}\label{fig3b}
             \end{subfigure}
             \caption{$\s_{\phi\ph,U(1)}$ vs temperature plot. Here, we set for $E=0.1$ and $J_t=10$.}
        \label{figure3}
\end{figure}

Like Hall conductivity, the $U(1)$ Ohmic conductivity $\s_{\ph\ph,U(1)}$ increases as we approach the minimum temperature ($T_{min}$) as shown in Figure \ref{figure3}. However, the increase in conductivity near the Misner string is more prominent due to the novel frame-dragging effects as explained before (\ref{am}) \cite{Khan:2025fne}. Moreover, $\s_{\ph\ph,U(1)}$ decreases with an increase in the magnetic field, as illustrated in (\ref{d1})-(\ref{d2}).  

\begin{figure}[H]
     \centering
     \begin{subfigure}[b]{0.495\textwidth}
         \centering
         \includegraphics[width=\textwidth]{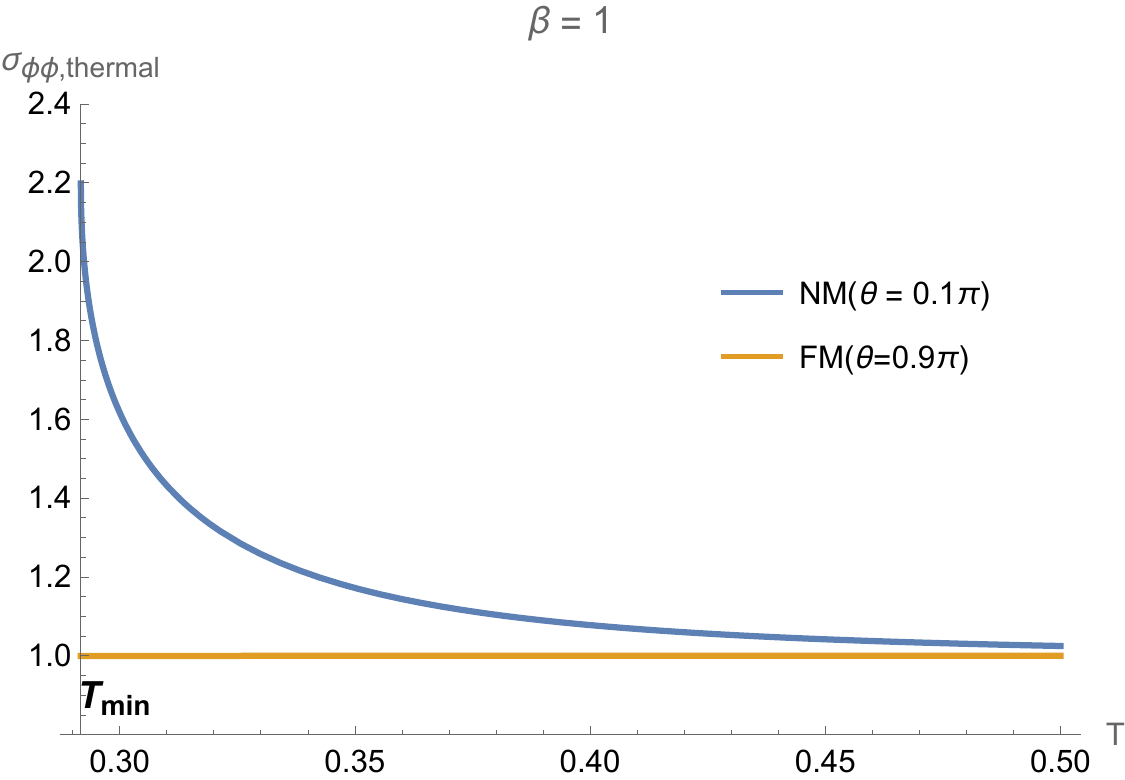}
         \caption{$\s_{\ph\ph,\text{thermal}}$ vs $T$ for $B=0.01$ }\label{fig4a}
              \end{subfigure}
              \hfill
     \begin{subfigure}[b]{0.495\textwidth}
         \centering
         \includegraphics[width=\textwidth]{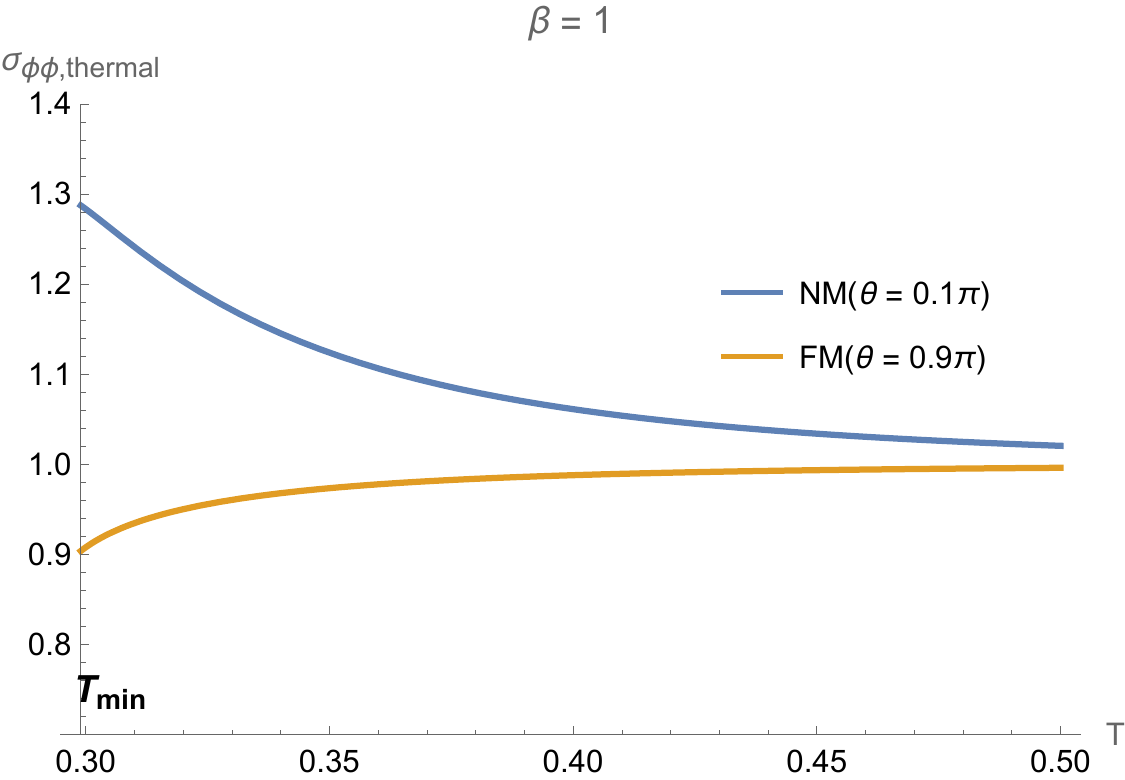}
         \caption{$\s_{\ph\ph,\text{thermal}}$ vs $T$ for $B=0.1$}\label{fig4b}
             \end{subfigure}
        
             \caption{$\s_{\phi\ph,\text{thermal}}$ vs temperature plot. Here, we set $E=0.1$ and $J_t=10$.}
        \label{figure4}
\end{figure}

On the other hand, the thermal Ohmic conductivity, $\s_{\ph\ph,\text{thermal}}$ also increases near the threshold temperature ($T\sim T_{min}$) for points closer to the Misner string. However, for points located further away, the conductivity remains constant  $(\s_{\ph\ph,\text{thermal}}\sim1)$ when the magnetic field is small ($B=0.01$). Interestingly, as we increase the magnetic field (say, $B=0.1$), the conductivity decreases quadratically (\ref{d3})-(\ref{d4}) in $B$ and saturates to unity, regardless of the location of the Misner string, as shown in Figure \ref{figure4}.   

As before, both the externally added $U(1)$ charge carriers and the thermally produced  charge pairs are further drifted toward the Misner string due to frame dragging (\ref{am}). This leads to the larger conductivities near Misner string as shown below %\footnote{\textcolor{red}{here $\th$ is the angular position for near the Misner string  e.i. $\th\sim0$  }}
 \begin{align}
     \frac{\sigma_{\ph\ph,U(1)}\bigg|_{T \sim T_{min}}^{nM}}{\sigma_{\ph\ph,U(1)}\bigg|_{T \sim T_{min}}^{fM}}  =\frac{18 \sqrt{3} E n\cot\left(\frac{\th}{2}\right)}{\pi ^2\sin\theta T^2_{min }}>>1\hspace{1mm},\hspace{2mm}\frac{\sigma_{\ph\ph,thermal}\bigg|_{T \sim T_{min}}^{nM}}{\sigma_{\ph\ph,thermal}\bigg|_{T \sim T_{min}}^{fM}}=\frac{18nE\cot\left(\frac{\th}{2}\right)}{\pi^2\sin\theta T_{min}^2}>>1.\label{ratio3}
\end{align}

In addition, at low temperatures, the number of $U(1)$ charge carriers is more significant than the thermally produced charge carriers. This leads to a greater contribution from $U(1)$, irrespective of the position of the Misner string. This can be illustrated by examining the following ratios
\begin{align}
  \frac{\s_{\ph\ph,U(1)}}{\s_{\ph\ph,\text{thermal}}}\Bigg|^{nM}_{T\sim T_{min}} = \frac{9\sqrt{3}J_t}{4\pi^2T_{min}^2 }>>1\hspace{1mm},\hspace{2mm}\frac{\s_{\ph\ph,U(1)}}{\s_{\ph\ph,\text{thermal}}}\Bigg|_{T\sim T_{min}}^{fM}=\frac{9J_t}{4\pi^2T_{min}^2 }>>1.\label{ratio4}
\end{align}

\subsubsection{Relationship between Hall and Ohmic conductivity}\label{rel}
It is worth noting that the authors in \cite{A.~O'Bannon} demonstrated that the change in Ohmic conductivity due to the magnetic field ($B$) is related to the Hall conductivity as $\s_{Hall}^2=-\s_{Ohmic}(B)\Delta\s_{Ohmic}$, where $\Delta\s_{Ohmic}=\s_{Ohmic}(B)-\s_{Ohmic}(B=0)$. Here $\s_{Ohmic}$ and $\s_{Hall}$ denote the Ohmic and Hall conductivity, respectively. Interestingly, a similar relation (up to an overall factor) exists between the conductivities (\ref{HallUafmb1})-\eqref{hallthnMb1}, (\ref{d1})-(\ref{d4}) associated with the TN-AdS black holes. Below, we show these relations for both far and near the Misner string.

The relationship between the $U(1)$ conductivites far away from the Misner string can be expressed as
\begin{align}
    &\s_{\th\ph,U(1)}^2\Bigg|^{fM}_{T\sim T_{min}}=\left(\frac{81BJ_t}{16\pi^4T_{min}^4}\right)^2 ,\hspace{2mm}\left(\s_{\ph\ph,U(1)}\Delta\s_{\ph\ph,U(1)}\right)\Bigg|^{fM}_{T\sim T_{min}}=-\left(\frac{81BJ_t}{16\pi^4T_{min}^4}\right)^2\csc^2\th,\\
    &\s_{\th\ph,U(1)}^2\Bigg|^{fM}_{T\sim T_{min}}=-\sin^2\th\left(\s_{\ph\ph,U(1)}\Delta\s_{\ph\ph,U(1)}\right)\Bigg|^{fM}_{T\sim T_{min}}.\label{r1}
\end{align}

On the other hand, we have a trivial relation for the thermal conductivity far from the Misner string as shown below %\footnote{\textcolor{red}{here we used $B<<T_{min}^2$}}
\begin{align}
 \s_{\th\ph,\text{thermal}}^2\Bigg|^{fM}_{T\sim T_{min}}\approx0, (\s_{\ph\ph,\text{thermal}}\Delta\s_{\ph\ph,\text{thermal}})\Bigg|^{fM}_{T\sim T_{min}}=-\frac{81B^2\csc^2\th }{32  \pi ^4  T_{min}^4}\approx0.\label{R10}
\end{align}

%It is essential to investigate the variations in Hall and Ohmic conductivity as influenced by the magnetic field. According to the expressions for Hall conductivity and Ohmic conductivity presented in Section 2 of \cite{A.~O'Bannon}, a relationship exists between the changes in the magnetic field, Ohmic conductivity, and Hall conductivity, expressed as \\ $\s_{xy}^2=-\s_{xx}(B)\Delta\s_{xx}$, where $\Delta\s_{xx}=\s_{xx}(B)-\s_{xx}(B=0)$. Below, we establish the relationship between the change in Hall conductivity and Ohmic conductivity with the change in the magnetic field from 0 to B in near and far from Misner's string in a weak magnetic field, based on the aforementioned expressions (\ref{d1})-(\ref{d4}).\footnote{Here we rescaled Magnetic filed $B$ in $\s_{\ph\ph,U(1)/\text{thermal}}$ by $\sin\th$}
%\begin{align}&
 %   \s_{\th\ph,U(1)}^2\Bigg|^{fM}_{T\sim T_{min}}=\left(\frac{81BJ_t}{16\pi^4T_{min}^4}\right)^2 ,\hspace{2mm}
   %\left(\s_{\ph\ph,U(1)}\Delta\s_{\ph\ph,U(1)}\right)\Bigg|^{fM}_{T\sim T_{min}}=-\left(\frac{81BJ_t}{16\pi^4T_{min}^4}\right)^2\csc^2\th\\& 
 %  \s_{\th\ph,U(1)}^2\Bigg|^{fM}_{T\sim T_{min}}=-\sin^2\th\left(\s_{\ph\ph,U(1)}\Delta\s_{\ph\ph,U(1)}\right)\Bigg|^{fM}_{T\sim T_{min}}\\&
%\s_{\th\ph,\text{thermal}}^2\Bigg|^{fM}_{T\sim T_{min}}=\left(\frac{81 B E n }{16 \pi ^4 T_{\min }^4}\right)^2\approx0,(\s_{\ph\ph,\text{thermal}}\Delta\s_{\ph\ph,\text{thermal}})\Bigg|^{fM}_{T\sim T_{min}}=-\frac{81B^2\csc^2\th }{32  \pi ^4  T_{min}^4}\approx0\end{align}

A similar relationship can be derived for the points near the Misner region. The conductivities arises from the $U(1)$ charge carriers can be expressed as 
\begin{align}
     \s_{\th\ph,U(1)}^2\Bigg|^{nM}_{T\sim T_{min}}=&\left(\frac{729 \sqrt{3} B E n J_t\cot\left(\frac{\th}{2}\right)}{32 \pi ^6  \sin\th T_{min}^6}\right)^2,\\(\s_{\ph\ph,U(1)}\Delta\s_{\ph\ph,U(1)})\Bigg|^{nM}_{T\sim T_{min}}=&-2\csc^2\th\left(\frac{729 \sqrt{3} B E n J_t\cot\left(\frac{\th}{2}\right)}{32 \pi ^6  \sin\th T_{min}^6}\right)^2,\\
     \s_{\th\ph,U(1)}^2\Bigg|^{nM}_{T\sim T_{min}}=&-\frac{\sin^2\th}{2}(\s_{\ph\ph,U(1)}\Delta\s_{\ph\ph,U(1)})\Bigg|^{nM}_{T\sim T_{min}}, \label{r2}
\end{align}
while the expressions for thermal conductivities are given below
\begin{align}
    \s_{\th\ph,\text{thermal}}^2\Bigg|^{nM}_{T\sim T_{min}}=&\left(\frac{81 B E n\cot\left(\frac{\th}{2}\right) }{8 \pi ^4\sin\th T_{\min }^4}\right)^2,\\
(\s_{\ph\ph,\text{thermal}}\Delta\s_{\ph\ph,\text{thermal}})\Bigg|^{nM}_{T\sim T_{min}}=&-\frac{3\csc^2\th}{2}\left(\frac{81 B E n\cot\left(\frac{\th}{2}\right) }{8 \pi ^4\sin\th T_{\min }^4}\right)^2\\
\s_{\th\ph,\text{thermal}}^2\Bigg|^{nM}_{T\sim T_{min}}=&-\frac{2\sin^2\th}{3}(\s_{\ph\ph,\text{thermal}}\Delta\s_{\ph\ph,\text{thermal}})\Bigg|^{nM}_{T\sim T_{min}}. \label{r3}
\end{align}

To summarize, the relations (\ref{r1}), (\ref{R10}), (\ref{r2}), and (\ref{r3}) demonstrate that the change in magnetic field ($B$) within the Ohmic conductivity induces the Hall conductivity up to an overall factor depending upon the position of the Misner string \cite{A.~O'Bannon}.  In other words, this factor encodes the location of the Misner string.

%Comments about this for fM
 %    \begin{align}
  %       &
   %    \s_{\th\ph,U(1)}^2\Bigg|^{nM}_{T\sim T_{min}}=\left(\frac{729 \sqrt{3} B E n J_t}{16 \pi ^6  \th^2T_{min}^6}\right)^2 \\&(\s_{\ph\ph,U(1)}\Delta\s_{\ph\ph,U(1)})\Bigg|^{nM}_{T\sim T_{min}}=-2\csc^2\th\left(\frac{729 \sqrt{3} B E n J_t}{16 \pi ^6  \th^2T_{min}^6}\right)^2\\&
    %   \s_{\th\ph,U(1)}^2\Bigg|^{nM}_{T\sim T_{min}}=-\frac{\sin^2\th}{2}(\s_{\ph\ph,U(1)}\Delta\s_{\ph\ph,U(1)})\Bigg|^{nM}_{T\sim T_{min}}\\&\s_{\th\ph,\text{thermal}}^2\Bigg|^{nM}_{T\sim T_{min}}=\left(\frac{81 B E n }{4 \pi ^4 \th^2T_{\min }^4}\right)^2,\hspace{2mm}
%(\s_{\ph\ph,\text{thermal}}\Delta\s_{\ph\ph,\text{thermal}})\Bigg|^{nM}_{T\sim T_{min}}=-\frac{3\csc^2\th}{2}\left(\frac{81 B E n }{4 \pi ^4 \th^2T_{\min }^4}\right)^2\\&
%\s_{\th\ph,U(1)}^2\Bigg|^{nM}_{T\sim T_{min}}=-\frac{2\sin^2\th}{3}(\s_{\ph\ph,U(1)}\Delta\s_{\ph\ph,U(1)})\Bigg|^{nM}_{T\sim T_{min}}
%\end{align}

%comment for nM\\
%Where $\Delta\s_{\ph\ph,U(1)/\text{thermal}}=\s_{\ph\ph,U(1)/\text{thermal}}(B)-\s_{\ph\ph,U(1)/\text{thermal}}(B=0)$

\subsubsection{Comparison between Ohmic and Hall conductivity}
In this section, we compare the Hall (\ref{HallUafmb1})-\eqref{hallthnMb1} and Ohmic conductivity (\ref{d1})-(\ref{d4}) associated with TN-AdS black holes in the small $B$ limit, both near and far from the Misner string. Notice that the holographic Ohmic conductivity is greater than its Hall conductivity for both near and far from the Misner string (see Figures \ref{figure1}-\ref{figure4}). We justify this statement by taking the following ratios
\begin{align}
  \frac{\s_{\ph\ph,U(1)}}{\s_{\th\ph,U(1)}}\Bigg|^{fM}_{T\sim T_{min}} = \frac{4\pi^2T_{min}^2}{9B}>>1\hspace{1mm},\hspace{2mm}  \frac{\s_{\ph\ph,U(1)}}{\s_{\th\ph,U(1)}}\Bigg|^{nM}_{T\sim T_{min}} =\frac{4\pi^2T_{min}^2}{9B}>>1,\label{ratio6}
\end{align}
\begin{align}
  \frac{\s_{\ph\ph,thermal}}{\s_{\th\ph,thermal}}\Bigg|^{fM}_{T\sim T_{min}} >>1\hspace{1mm},\hspace{2mm} \frac{\s_{\ph\ph,thermal}}{\s_{\th\ph,thermal}}\Bigg|^{nM}_{T\sim T_{min}} =\frac{4\pi^2T_{min}^2}{9B}>>1.\label{ratio5}
\end{align}

This stems from the fact that far from the Misner string, only dominant contribution to Hall conductivity appears through the magnetic field $(B)$. In the limit $B<E$, most of the $U(1)$ carriers are drifted through electric field, producing more contribution to the Ohmic transport. In case of thermally produced charge pairs, most of them are effectively drifted by the external electric field, while the Lorentz force contribution due to $B$ field can be ignored as compared with the electric field contributions. On top of it, as explained before, it produces opposite currents for thermally produced pairs resulting in a low Hall conductivity.

In the literature, the ratio of Ohmic conductivity to the Hall conductivity is referred to as the inverse Hall angle and it is denoted by $\cot\th_H$ \cite{Kim:2010zq}. When we are sufficiently far away from the Misner string, the Hall angle in the low temperature regime scales as the inverse of the Ohmic conductivity (\ref{d1}), as shown below\footnote{In low temperature regions, $U(1)$ conductivity dominates over thermal conductivity both near and far from the Misner string. Therefore, we can approximate $\s_{\phi\ph}\approx\s_{\phi\phi,U(1)}$ and $\s_{\th\ph}\approx\s_{\th\phi,U(1)}$.}
\begin{align}&
\cot\th_H\Bigg|^{fM}_{T\sim T_{min}}= \frac{\s_{\ph\ph}}{\s_{\th\ph}}\Bigg|^{fM}_{T\sim T_{min}} = \frac{4\pi^2T_{min}^2}{9B},\hspace{2mm}\s_{\ph\ph}|_{T\sim T_{min}}^{fM}\sim \frac{1}{T_{min}^{2}} \rightarrow  \cot\th_H\Bigg|^{fM}_{T\sim T_{min}}\sim \left(\s_{\ph\ph}|_{T\sim T_{min}}^{fM}\right)^{-1}.\label{fmratio}\end{align}

 Such scaling $(\cot\th_H\sim(\s_{\ph\ph})^{-1})$ is known as the Drude result \cite{Lee:2010ii}-\cite{Hartnoll:2009ns}. On the other hand, the Hall angle near the Misner string scales as the square root of the Ohmic conductivity \eqref{d2}, as illustrated below
 \begin{align}
 & \cot\th_H\Bigg|^{nM}_{T\sim T_{min}} =\frac{4\pi^2T_{min}^2}{9B},\s_{\ph\ph}|_{T\sim T_{min}}^{nM}\sim  \frac{1}{T_{min}^{4}} \rightarrow  \cot\th_H\Bigg|^{nM}_{T\sim T_{min}}\sim \left(\s_{\ph\ph}|_{T\sim T_{min}}^{nM}\right)^{-1/2},\label{nmratio}
\end{align}
which is referred as the non-Drude result \cite{Anderson:1991ixg}, \cite{A.W. Tyler }.

Below, we summarize the Hall conductivity (\ref{HallUafmb1})-\eqref{hallthnMb1} and the Ohmic conductivity (\ref{d1})-(\ref{d4}) associated with the TN-AdS black hole for both near and far from the Misner string in the small magnetic field limit ($B<<E<1$) and low temperature limit.
\begin{table}[H]
    \begin{center}
   
\renewcommand{\arraystretch}{1.7}
\begin{tabular}{|c|c|c|c|c|}
 \hline
 $\s$ &    $fM $  & $nM $ 
 \\ 
 \hline
 $\s_{\ph\ph,U(1)}$ &$\frac{9J_t}{4\pi^2T_{min}^2}\left(1-\frac{81B^2 \csc^2\th }{16 \pi ^4 T^4_{min}}\right)$
   & $ \frac{81 \sqrt{3} E n J_t\cot\left(\frac{\th}{2}\right)}{8 \pi ^4 \sin\th T_{\min }^4}\left(1-\frac{81B^2 \csc^2\th }{8 \pi ^4 T^4_{min}}\right)$\\
  \hline
 $\s_{\ph\ph,\text{thermal}}$ & $ 1-\frac{81B^2\csc^2\th }{32  \pi ^4  T_{min}^4} $&  $  \frac{9 En\cot\left(\frac{\th}{2}\right)}{2\pi ^2 \sin\th T_{min}^2}\left(1-\frac{243B^2\csc^2\th }{32  \pi ^4  T_{min}^4}\right)$ \\
 \hline
  $\s_{\th\ph,U(1)}$ &$\frac{81BJ_t}{16\pi^4T_{min}^4}$ &  $\frac{729 \sqrt{3} B E n J_t\cot\left(\frac{\th}{2}\right)}{32 \pi ^6  \sin\th T_{min}^6}$\\ \hline
   $\s_{\th\ph,\text{thermal}}$ &$0$ &  $\frac{81 B E n\cot\left(\frac{\th}{2}\right) }{8 \pi ^4\sin\th T_{min }^4}$\\ \hline
\end{tabular}
 \caption{Hall and Ohmic conductivity in the small magnetic field limit ($B<<1$) at low temperatures $(T\sim T_{min})$.}
    \label{tableltn1}
         
    \end{center}
\end{table}

\subsection{High temperature regime\label{3.2}}
In this section, we examine the Ohmic conductivity $\s_{\ph\ph}$ (\ref{condu1dcT})-\eqref{condthT} and the Hall conductivity $\s_{\th\ph}$ (\ref{hallu1T})-\eqref{hallthermal T} in the high temperature regime. We define the high temperature limit as $T >> T_{min}$. In particular, we investigate these conductivities both near and far from the Misner string for $\b=1$, while considering the small magnetic field i.e., $B<<1$. 

To begin with, we express the angular velocity (\ref{angb1}) in the small magnetic field and high temperature limit, which yields
\begin{align}
    |\tilde\Omega|=\frac{9 n \cot \left(\frac{\theta }{2}\right) }{8 \left(3 n^2+1\right)^2 }\left(\frac{T_{min}}{T}\right)^4\left(1-\frac{9 B^2 \csc ^2(\theta ) }{16 \left(3 n^2+1\right)^2 }\left(\frac{T_{min}}{T}\right)^4\right)<<1.\label{oth}
\end{align}
Notice that in the high temperature limit ($T>>T_{min}$), the effects due to frame dragging are negligible ($\tilde\Omega\sim \frac{1}{T^4}$), as shown above (\ref{oth}).

\subsubsection{Ohmic conductivity}
The holographic Ohmic conductivity ($\s_{\phi\phi}$) (\ref{condu1dcT})-\eqref{condthT}, both near and far from the Misner string in the high temperature ($T>>T_{min}$) and low magnetic field ($B<<1$) regime, can be expressed as follows
\begin{align}
  &\sigma_{\ph\ph,U(1)}\bigg|^{fM}_{T>>T_{min}}\approx\frac{9  J_t}{4 (3+9n^2)}\left(\frac{T_{min}}{T}\right)^2\left(1-\frac{81B^2\csc^2\th}{16(3+9n^2)^2}\left(\frac{T_{min}}{T}\right)^4\right),\label{highdcu1fmb1}\\&\sigma_{\ph\ph,U(1)}\bigg|_{T>>T_{min}}^{nM}\approx\frac{9  J_t}{4  (3+9n^2)}\left(\frac{T_{min}}{T}\right)^2\left(1-\frac{81B^2\csc^2\th}{16(3+9n^2)^2}\left(\frac{T_{min}}{T}\right)^4\right),\label{highdcu1nmb1} \\
&\sigma_{\ph\ph,thermal}\Bigg|^{fM}_{T>>T_{min}}\approx1-\frac{81B^2 \csc^2\th}{32(3+9n^2)^2}\left(\frac{T_{min}}{T}\right)^4,\label{highdcthfmb1}\\&\sigma_{\ph\ph,thermal}\Bigg|^{nM}_{T>>T_{min}}\approx1-\frac{243B^2E^2 n^2 \cot^2\left(\frac{\th}{2}\right)}{128( 1+ 3 n^2)^4 \sin^4\th }\left(\frac{T_{min}}{T}\right)^8,\label{highdcthb1}
\end{align}
where $0<<\th<\pi$ label the points far away from the Misner string ($fM$) and $0<\th<<\pi$ represents region near the Misner string ($nM$).

\begin{figure}[H]
     \centering
     \begin{subfigure}[b]{0.495\textwidth}
         \centering
         \includegraphics[width=\textwidth]{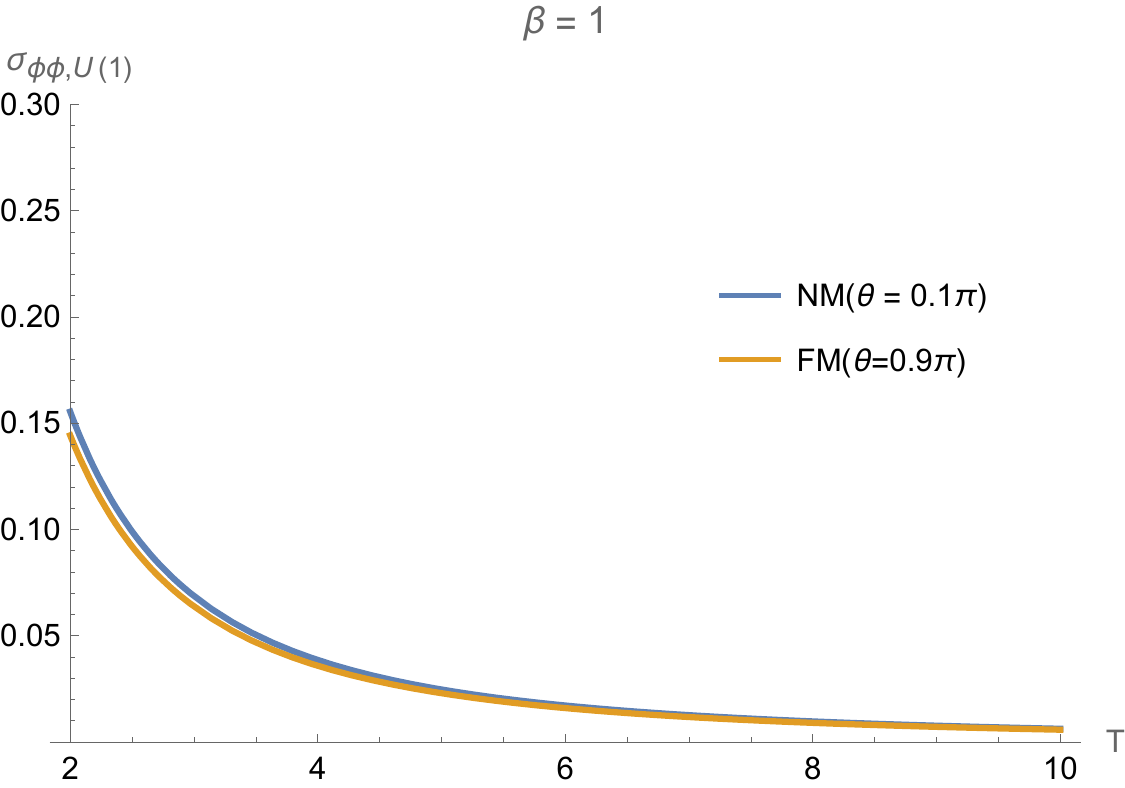}
         \caption{$\s_{\ph\ph,U(1)}$ vs temperature for $B=0.01$  }\label{fig6a}
              \end{subfigure}
              \hfill
     \begin{subfigure}[b]{0.495\textwidth}
         \centering
         \includegraphics[width=\textwidth]{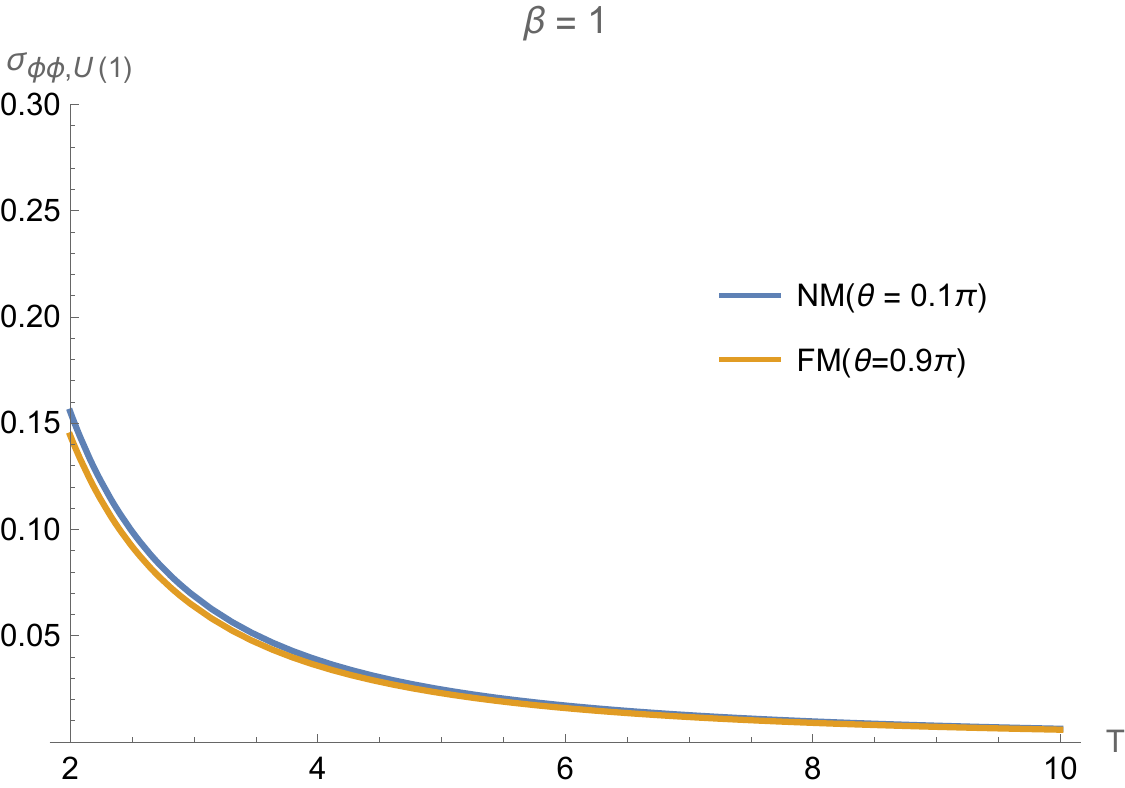}
         \caption{$\s_{\ph\ph,(U1)}$ vs temperature for $B=0.1$ }\label{fig6b}
             \end{subfigure}
        
             \caption{$\s_{\phi\ph}$ vs temperature plot. Here, we set $E=0.1$, $J_t=10$, and $n=0.2$. }
        \label{figure6}
\end{figure}

\begin{figure}[H]
     \centering
     \begin{subfigure}[b]{0.495\textwidth}
         \centering
         \includegraphics[width=\textwidth]{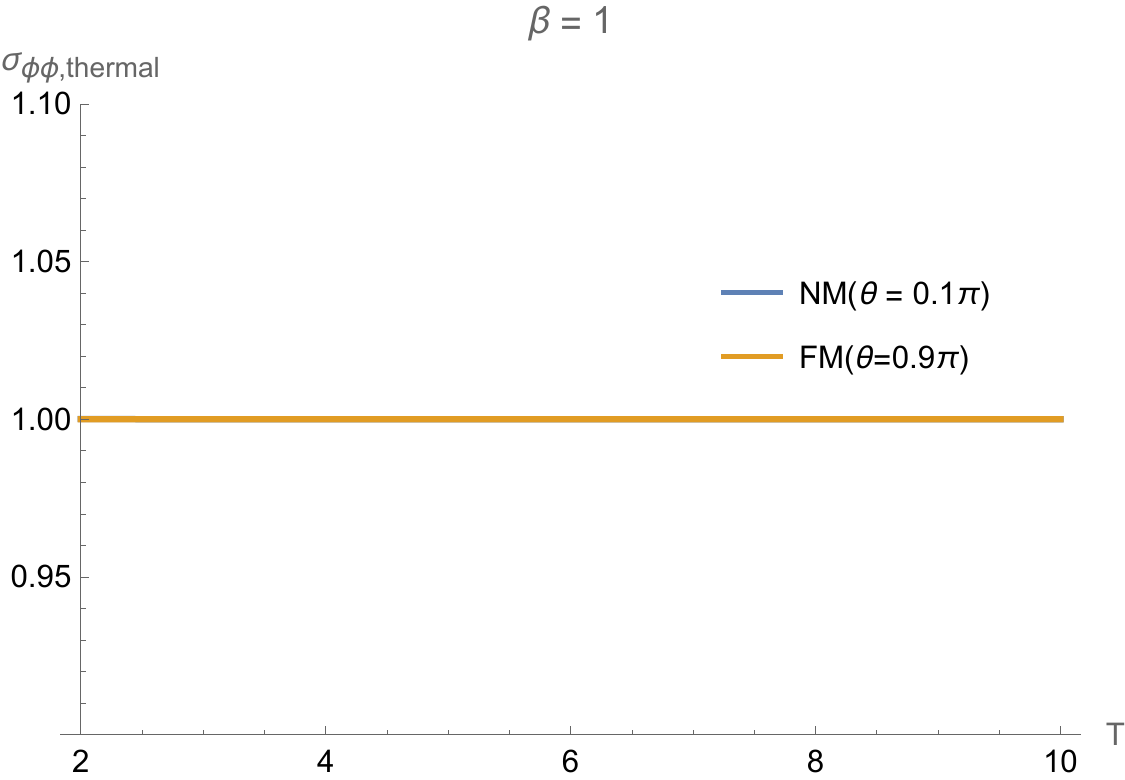}
         \caption{$\s_{\ph\ph,U(1)}$ vs temperature at $B=0.01$  }\label{fig6ath}
              \end{subfigure}
              \hfill
     \begin{subfigure}[b]{0.495\textwidth}
         \centering
         \includegraphics[width=\textwidth]{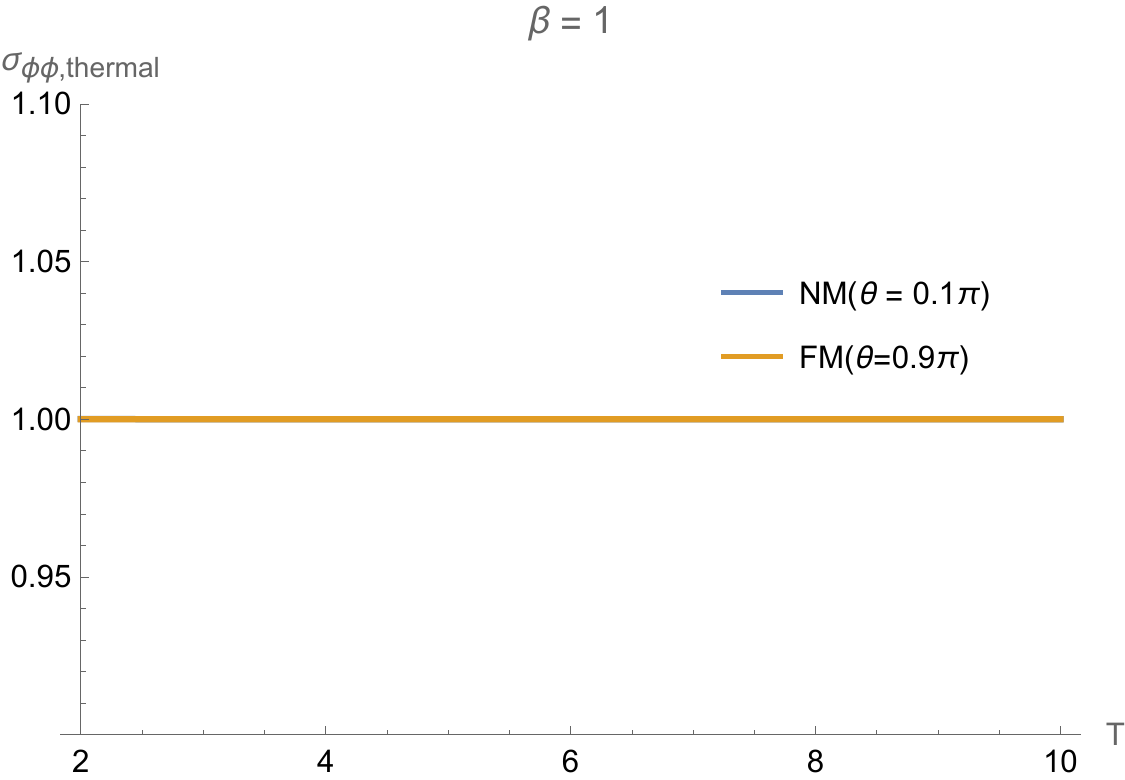}
         \caption{$\s_{\ph\ph,(U1)}$ vs temperature at $B=0.1$ }\label{fig6bth}
             \end{subfigure}
        
             \caption{$\s_{\phi\ph}$ vs temperature plot. Here, we set $E=0.1$, $J_t=10$, and $n=0.2$. }
        \label{figure6th}
\end{figure}

It is interesting to notice that the Ohmic conductivity from the externally added $U(1)$ charge carriers ($\sigma_{\ph\ph,U(1)}$) (\ref{highdcu1fmb1})-(\ref{highdcu1nmb1}) decreases with an increase in temperature and the effects of magnetic field on $\sigma_{\ph\ph,U(1)}$ is negligible, as shown in Figure \ref{figure6}. On the other hand, the contribution due to the thermally produced charge pair ($\sigma_{\ph\ph,thermal}$) (\ref{highdcthfmb1})-(\ref{highdcthb1}) saturates to unity, irrespective of the location of the Misner string and the magnetic field (see Figure \ref{figure6th}). The decrease in the Ohmic conductivity in the presence of the $B$ field is an artifact of the Lorentz force, which has been explained previously below \eqref{d4}. On top of it, the effects due to the frame dragging is diminished significantly at high temperatures. As a result, it cannot drift thermally produced pairs, causing a saturation in the corresponding Ohmic conductivities \cite{Khan:2025fne}.

In addition, at high temperatures, the number of thermally charged pairs exceeds that of $U(1)$ charge carriers, leading to higher thermal Ohmic conductivity irrespective of the position of the Misner string, as shown in the ratio below.
\begin{align}
    \frac{\sigma_{\ph\ph,\text{thermal}}}{\sigma_{\ph\ph,U(1)}}\Bigg|_{T >> T_{min}}^{fM}=\frac{4 \pi (3+9n^2)}{9  J_t}\left(\frac{T}{T_{min}}\right)^2>>1\hspace{1mm},
    \frac{\sigma_{\ph\ph,\text{thermal}}}{\sigma_{\ph\ph,U(1)}}\Bigg|_{T >> T_{min}}^{nM}=\frac{4 \pi (3+9n^2)}{9  J_t}\left(\frac{T}{T_{min}}\right)^2>>1.\label{Highr1}
\end{align}

\subsubsection{Hall conductivity}
In this Section, we examine the holographic Hall conductivity $(\s_{\th\ph})$ (see Appendix B \eqref{hallu1T}-\eqref{hallthermal T}) in the high temperature regime, while considering a small magnetic field, $B<<1$. Under this approximation, $\s_{\th\ph}$ both near and far from the Misner string can be expressed as follows\footnote{When we are far from the Misner string, $\s_{\th\phi,\text{thermal}} $ (\ref{hallthfmTh}) nearly vanishes, provided $BEnT^4_{min}<T^4$.}
\begin{align}&
    \s_{\th\phi,U(1)}\bigg|^{fM}_{T >>T_{min}}=\frac{9 B J_t }{16 \left(3 n^2+1\right)^2 }\left(\frac{T_{\min }}{T}\right)^4, \label{hallu1fmTh}\\&\s_{\th\ph,U(1)}\bigg|^{nM}_{T >> T_{min}}=\frac{9 B J_t }{16 \left(3 n^2+1\right)^2 }\left(\frac{T_{\min }}{T}\right)^4,\label{hallu1nmth}\\
&\s_{\th\phi,\text{thermal}}\bigg|^{fM}_{T >> T_{min}}\approx0, \label{hallthfmTh}\\  & \s_{\th\phi,\text{thermal}}\bigg|^{nM}_{T >> T_{min}}=\frac{9 B E n\cot\left(\frac{\th}{2}\right) }{8 \sin\th \left(3 n^2+1\right)^2}\left(\frac{T_{min}}{T}\right)^4.\label{hallthnmTh}
\end{align}
%, irrespective of the angular distance from the Misner string. 

\begin{figure}[H]
     \centering
     \begin{subfigure}[b]{0.495\textwidth}
         \centering
         \includegraphics[width=\textwidth]{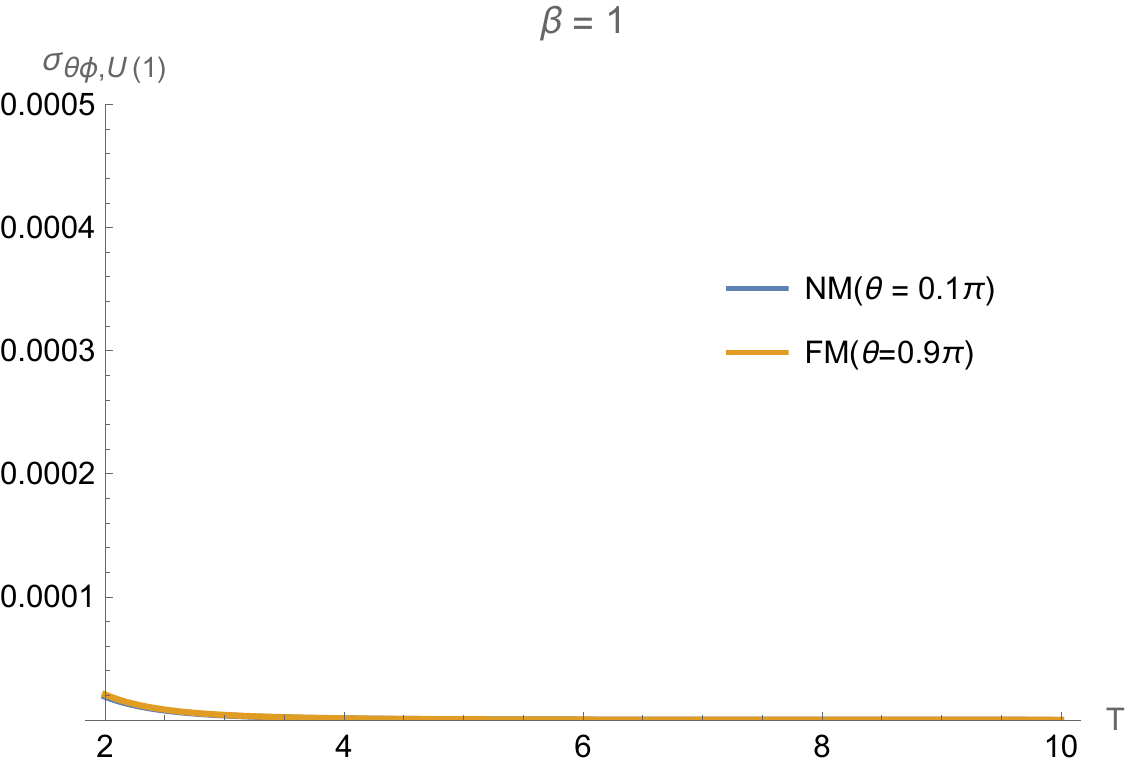}
         \caption{$\s_{\th\ph,U(1)}$ vs temperature for $B=0.01$ }\label{fig7a}
              \end{subfigure}
              \hfill
     \begin{subfigure}[b]{0.495\textwidth}
         \centering
         \includegraphics[width=\textwidth]{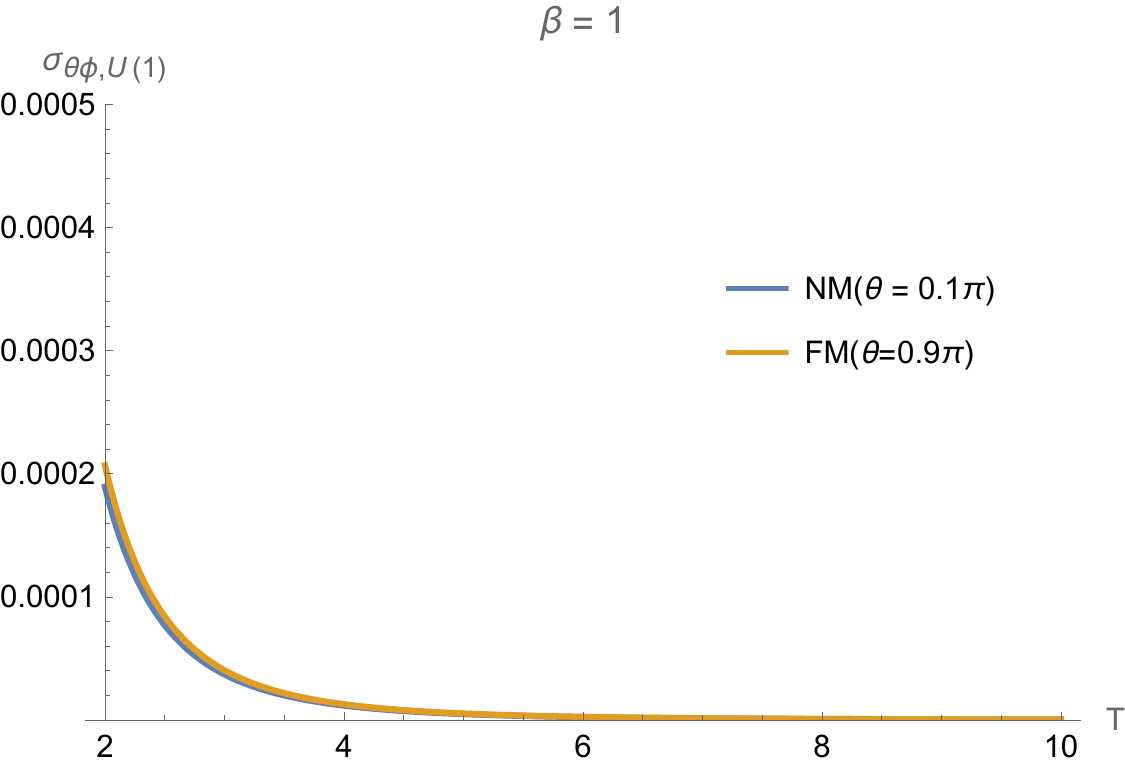}
         \caption{$\s_{\th \ph ,U(1)}$ vs temperature for $B=0.1$ }\label{fig7b}
             \end{subfigure}
        
             \caption{$\s_{\th\ph,U(1)}$ vs temperature plot. Here, we set $E=0.1$, $J_t=10$, and $n=0.2$.}
        \label{figure7}
\end{figure}

\begin{figure}[H]
     \centering
     \begin{subfigure}[b]{0.495\textwidth}
         \centering
         \includegraphics[width=\textwidth]{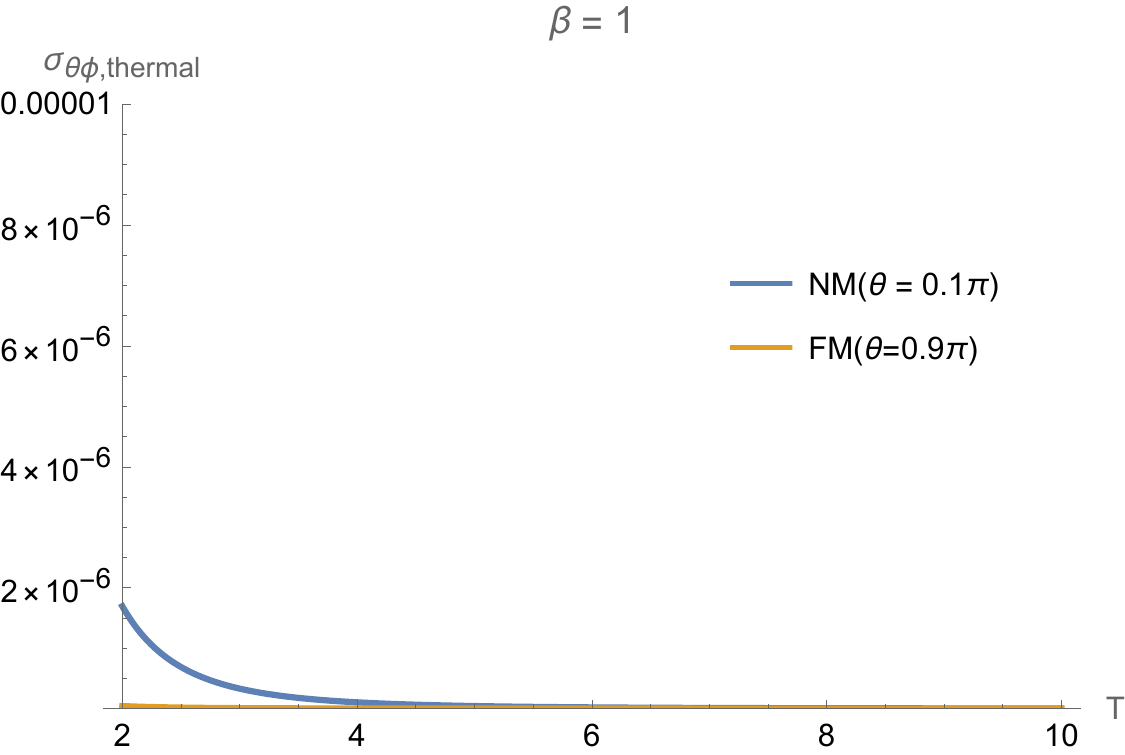}
         \caption{$\s_{\th\ph,\text{thermal}}$ vs temperature for $B=0.01$ }\label{fig8b}
              \end{subfigure}
              \hfill
     \begin{subfigure}[b]{0.495\textwidth}
         \centering
         \includegraphics[width=\textwidth]{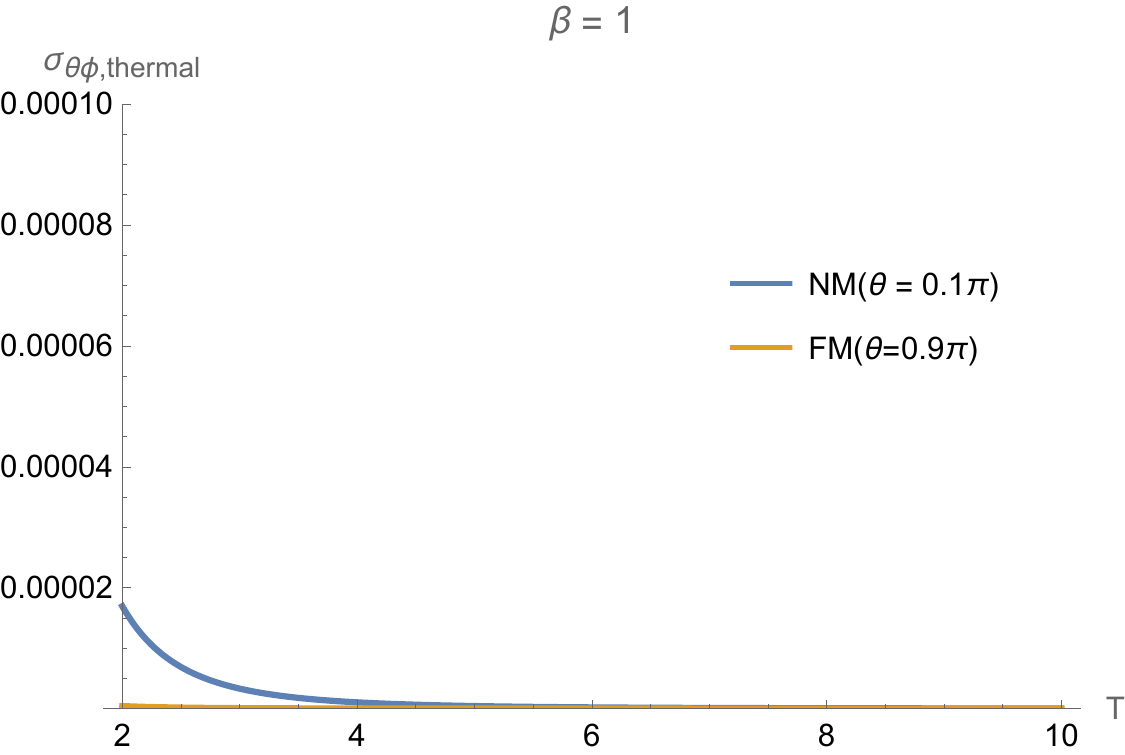}
         \caption{$\s_{\th \ph ,\text{thermal}}$ vs temperature for $B=0.1$ }\label{fig8a}
             \end{subfigure}
        
             \caption{$\s_{\th\ph,\text{thermal}}$ vs temperature plot. Here, we set $E=0.1$, $J_t=10$, and $n=0.2$.}
        \label{figure8}
\end{figure}

The above expressions \eqref{hallu1fmTh}-\eqref{hallu1nmth} show that the Hall conductivity due to the $U(1)$ charge carriers ($\s_{\th\ph,U(1)}$) decreases with increasing temperature and varies linearly with the magnetic field ($B$), as illustrated in Figure \ref{figure7}. This result is independent of the position of the Misner string because the frame-dragging effects in the high temperature regime are suppressed, as demonstrated in (\ref{oth}). In addition, the thermal contribution \eqref{hallthfmTh}-\eqref{hallthnmTh} to the holographic Hall conductivity is also negligible, regardless of the angular distance from the Misner string, as shown in Figure \ref{figure8}. 

The results \eqref{hallu1fmTh}-\eqref{hallu1nmth} is an artifact of low frame-dragging effects at high temperatures. In other words, Hall conductivity due to $U(1)$ charge carriers are only affected due to external magnetic field. Equation \eqref{hallthfmTh} is understandable as there are no frame-dragging and The Lorentz force induces opposing currents in particle-antiparticle pairs, resulting in a net zero Hall current within the thermal plasma.. However, due to frame dragging (as explained below \eqref{hall conductivity thermal}), there will be a non-zero Hall transport \eqref{hallthnmTh} near the Misner string.

It is worth noting that the contribution due to the $U(1)$ charge carriers to the Hall conductivity dominates over the thermal contribution, even in the high temperature regime, as demonstrated below %\footnote{\textcolor{red}{here $\th$ is the angular position for near the Misner string  e.i. $\th\sim0$  }}
\begin{align}
   \frac{\s_{\th\ph,U(1)}^{fM}}{\s_{\th\ph,\text{thermal}}^{fM}} >>1\hspace{1mm},\hspace{2mm}
   \frac{\s_{\th\ph,U(1)}^{nM}}{\s_{\th\ph,\text{thermal}}^{nM}} =\frac{  J_t\tan\left(\frac{\th}{2}\right)}{2En \csc\th}>>1.\label{dr01}
\end{align}
This observation is in contrast with the usual Ohmic conductivity behaviour as discussed above \eqref{Highr1}, where the thermally produced charge carrier dominates in the high temperature regime \cite{Khan:2025fne}.

This phenomenon can be understood as follows. The Hall current produced by thermally generated charge carriers arises solely from the frame-dragging effect. On the other hand, the Hall current associated with the $U(1)$ charge carriers is influenced by both frame-dragging and the  Lorentz force (i.e., $\vec {F} = q (\vec{v}\times \vec{B})$) (as discussed in Section \ref{hall conductivity}) \cite{Rath:2021ryd}, \cite{Blackman:1993pbp}. At high temperatures, frame-dragging effects become negligible, regardless of the angular position of the Misner string. As a result, the leading contribution comes from the Lorentz force. This leads to a higher $U(1)$ Hall conductivity both near and far from the Misner string as demonstrated in (\ref{dr01}).

%\textcolor{red}{The Hall current of thermally produced charge carriers is generated due to the frame-dragging effect, which becomes negligible at high temperatures regardless of angular position. In contrast, the Hall current of $U(1)$ charge carriers also experiences additional drift from frame-dragging; however, it is not solely generated by this effect. Consequently, at elevated temperatures, while the Hall current of thermally produced charge carriers vanishes, the Hall current of $U(1)$ charge carriers remains finite due to the non-zero Lorentz force $q(\vec{v}\times \vec{B})$. }

\subsubsection{Relationship between Hall and Ohmic conductivity}\label{rel2}
As discussed above in Section \ref{rel}, we also observe identical relationships between the Ohmic (\ref{highdcu1fmb1})-(\ref{highdcthb1}) and  Hall \eqref{hallu1fmTh}-\eqref{hallthnmTh} conductivities at higher temperatures ($T>>T_{min}$) and lower magnetic fields ($B<<1$). Below, we write down these relations for both near and away from the Misner string. 

When we are sufficiently far from the Misner string, the relationships between the conductivities are provided below
\begin{align}
    \s_{\th\ph,U(1)}^2\Bigg|^{fM}_{T>> T_{min}}=&-\sin^2\th\left(\s_{\ph\ph,U(1)}\Delta\s_{\ph\ph,U(1)}\right)\Bigg|^{fM}_{T>> T_{min}},\label{e11}
   \end{align}  
   \begin{align}
  \s^2_{\th\phi,\text{thermal}}\bigg|^{fM}_{T >> T_{min}}=0,\hspace{3mm}&(\s_{\ph\ph,\text{thermal}}\Delta\s_{\ph\ph,\text{thermal}})\bigg|^{fM}_{T>>T_{min}}
\approx0.\label{e11a}
\end{align}
On the other hand, when we approach the Misner string, the relationships between the conductivities can be expressed as
\begin{align}
    \s_{\th\ph,U(1)}^2\Bigg|^{nM}_{T>> T_{min}}=&-\sin^2\th\left(\s_{\ph\ph,U(1)}\Delta\s_{\ph\ph,U(1)}\right)\Bigg|^{nM}_{T>> T_{min}},\label{r1ht}\\
    \s_{\th\ph,\text{thermal}}^2\Bigg|^{nM}_{T>> T_{min}}=&-\frac{2\sin^2\th}{3}(\s_{\ph\ph,\text{thermal}}\Delta\s_{\ph\ph,\text{thermal}})\Bigg|^{nM}_{T>> T_{min}}. \label{e14}
\end{align}

%Notice that the pre-factor in the relationship for the $U(1)$ charge carriers at higher temperature is the same for both near and far from the Misner string because the frame dragging effects are negligible. Th

Notice that the pre-factor (\ref{r1ht}) differs by a factor of half compared to lower temperatures (\ref{r2}). This arises because of the lower frame-dragging effects at higher temperatures. However, the other expressions (\ref{e11}), (\ref{e11a}), and (\ref{e14}) match exactly with the corresponding relations at lower temperatures (\ref{r1}), (\ref{R10}), and (\ref{r3}) respectively. Notice that one has to consider sub-leading corrections (in frame-dragging) in the Ohmic transport $\s_{\ph\ph,\text{thermal}}$ while deriving the relation \eqref{e14}.

\subsubsection{Comparison between Ohmic and Hall conductivity}
In this section, we compare the Ohmic (\ref{highdcu1fmb1})-(\ref{highdcthb1}) and Hall conductivity \eqref{hallu1fmTh}-\eqref{hallthnmTh} for both near and far from the Misner string at high temperature regime and small magnetic field limit. In this limit, the holographic Hall conductivity is found to be lower than the Ohmic conductivity, both near and far from the Misner string (see Figures \ref{figure6}-\ref{figure8}). This is demonstrated below by taking the following ratios 
\begin{align}
    \frac{\s_{\ph\ph,U(1)}}{\s_{\th\ph,U(1)}}\Bigg|_{T>>T_{min}}^{fM}=\frac{4 \left(3 +9n^2\right)}{9  B  }\left(\frac{T}{T_{min}}\right)^2>>1\hspace{2mm}, \frac{\s_{\ph\ph,U(1)}}{\s_{\th\ph,U(1)}}\Bigg|_{T>>T_{min}}^{nM}=\frac{4 \left(3 +9n^2\right)}{9  B  }\left(\frac{T}{T_{min}}\right)^2>>1.
\end{align}
\begin{align}
  \frac{\s_{\ph\ph,thermal}}{\s_{\th\ph,thermal}}\Bigg|^{fM}_{T>> T_{min}} >>1\hspace{1mm},\hspace{2mm} \frac{\s_{\ph\ph,thermal}}{\s_{\th\ph,thermal}}\Bigg|^{nM}_{T>> T_{min}} =\frac{8 \sin\th \left(3 n^2+1\right)^2}{9 B E n\cot\left(\frac{\th}{2}\right) }\left(\frac{T}{T_{\min }}\right)^4>>1.\label{ratio5h}
\end{align}

The Ohmic counterpart is always higher than the Hall component due to low $B$ field and negligible frame-dragging effect at high temperature. Below, we summaries the Ohmic conductivity and Hall conductivity associated with the TN-AdS black hole at high temperature for both near and far from the Misner string, while keeping the magnetic field small. 
\begin{table}[H]
    \begin{center}
   
\renewcommand{\arraystretch}{1.7}
\begin{tabular}{|c|c|c|c|c|}
 \hline
 $\s$ &    $fM $  & $nM $ 
 \\ 
 \hline
 $\s_{\ph\ph,U(1)}$ &$\frac{9  J_t}{4  (3+9n^2)}\left(\frac{T_{min}}{T}\right)^2\left(1-\frac{81B^2\csc^2\th}{16(3+9n^2)^2}\left(\frac{T_{min}}{T}\right)^4\right)$
   & $ \frac{9  J_t}{4  (3+9n^2)}\left(\frac{T_{min}}{T}\right)^2\left(1-\frac{81B^2\csc^2\th}{16(3+9n^2)^2}\left(\frac{T_{min}}{T}\right)^4\right)$\\
  \hline
 $\s_{\ph\ph,\text{thermal}}$ & $1-\frac{81B^2 \csc^2\th}{32(3+9n^2)^2}\left(\frac{T_{min}}{T}\right)^4$&  $1-\frac{243B^2E^2 n^2 \cot^2\left(\frac{\th}{2}\right)}{128( 1+ 3 n^2)^4 \sin^4\th }\left(\frac{T_{min}}{T}\right)^8$ \\
 \hline
  $\s_{\th\ph,U(1)}$ &$\frac{9 B J_t }{16 \left(3 n^2+1\right)^2 }\left(\frac{T_{min }}{T}\right)^4$ &  $\frac{9 B J_t }{16 \left(3 n^2+1\right)^2 }\left(\frac{T_{min }}{T}\right)^4$\\ \hline
   $\s_{\th\ph,\text{thermal}}$ &$0$ &  $\frac{9 B E n\cot\left(\frac{\th}{2}\right) }{8 \sin\th \left(3 n^2+1\right)^2}\left(\frac{T_{min}}{T}\right)^4.$\\ \hline
\end{tabular}
 \caption{Hall and Ohmic conductivity in the small magnetic field limit ($B<<1$) at high temperatures $(T>> T_{min})$. }
    \label{tableltn2}
         
    \end{center}
\end{table}
%Expressions \eqref{highdcu1fmb1}-\eqref{highdcthb1}  are similar as we found earlier which reveals that $\sigma_{\ph\ph,U(1)}$ falls with temperature,while $\s_{\ph\ph,thermal}$  remains constant. Similarly, figure \ref{figure6} indicates that $\s_{\ph\ph,U(1)}$ and $\s_{\ph\ph,thermal}$ are not affected by the position of the Misner 
%The ratio of conductivities for the near Misner ($nM$) string and far away ($fM$) from it can be expressed as
%\begin{align}
 %   \frac{\sigma_{\ph\ph,U(1)}\Bigg|_{T >> T_{min}}^{nM} }{\sigma_{\ph\ph,U(1)}\Bigg|_{T >> T_{min}}^{fM} }=1+O\left[\left(\frac{T_{min}}{T}\right)^4\right]\hspace{1mm},\hspace{2mm}\frac{\sigma_{\ph\ph,thermal}\Bigg|_{T >> T_{min}}^{nM} }{\sigma_{\ph\ph,thermal}\Bigg|_{T >> T_{min}}^{fM} }=1+O\left[\left(\frac{T_{min}}{T}\right)^4\right].\label{highratiob1}
%\end{align}

\section{Hall transports at finite magnetic field}
In this Section, we explore the Hall and Ohmic conductivities (see Appendix B (\ref{condu1dcT})-\eqref{hallthermal T}) at ``finite" magnetic fields $(B>1)$. In particular, we investigate the effects due to frame-dragging (or equivalently the NUT parameter ($n$)) in both low ($T\sim T_{min}$) and high temperature ($T>>T_{min}$) regimes, while taking into account the influence of the finite magnetic field ($B$).

In the previous Section \ref{low b 3}, we studied the Hall transport in the low magnetic field ($B<<1$). To be specific, we first expanded the conductivities (\ref{condu1dcT})-\eqref{hallthermal T} in the small $B$ limit and then considered the low and high temperature expansions. However, in the present Section, we keep the conductivities (\ref{condu1dcT})-\eqref{hallthermal T} exact in the magnetic field $(B)$ and investigate their behaviour at low and high temperature regions. In other words, we expand (\ref{condu1dcT})-\eqref{hallthermal T} only in the temperature $(T)$, while keeping magnetic field $(B)$ fixed.

%\textcolor{red}{The main difference between approaches used in low magnetic field $B<<1$ transport results in section \ref{low b 3} and the following result in finite Magnetic field $B>1$ in this section is that in section \ref{low b 3} first we expand \eqref{condu1dcT}-\eqref{hallthermal T} in small $B$ upto $B^2$ then in low and high temperature to get Ohmic and Hall conductivities in low magnetic field at low and high temperature.but in this section we directly expand \eqref{condu1dcT}-\eqref{hallthermal T} in low and high temperature without expanding in $B$ to get following result og Hall and Ohmic conductivify at Low and high temprature.}

\subsection{Low temperature regime}
In this section, we study the Ohmic ($\s_{\ph\ph}$) \eqref{condu1dcT}-\eqref{condthT} and Hall conductivity ($\s_{\th\ph}$) \eqref{hallu1T} -\eqref{hallthermal T} at low temperature ($T\sim T_{min}$) and finite magnetic field ($B$). It is worth noting that under these conditions, the angular velocity \eqref{angb1} can be expressed as
\begin{align} |\tilde\Omega|=\frac{2 n  \cot \left(\frac{\theta }{2}\right)}{B^2\csc ^2(\theta )}+\sqrt{O(T-T_{min})}.\label{ha}\end{align}

Notice that the angular velocity (\ref{ha}) is inversely proportional to the square of the magnetic field. This means that, at finite magnetic fields, the effects due to frame dragging are relatively smaller. However, these frame-dragging effects could become significant as we approach the Misner string.

\subsubsection{Hall conductivity}
In this Section, we examine the Hall conductivity ($\s_{\th\ph}$) \eqref{hallu1T}-\eqref{hallthermal T} under finite magnetic field $(B>1)$. At low temperatures ($T\sim T_{min}$), the $\s_{\th\ph}$ can be expressed both near and far from the Misner string as follows\footnote{The $\s_{\th\ph,\text{thermal}}$  vanishes for points that are far from the Misner string, provided $EnT^2_{min}<B^2$.}

\begin{align}
 &   \s_{\th\phi,U(1)}\bigg|^{fM}_{T \sim T_{min}}=\frac{  J_t\sin^2\th}{ B}+O\left(\sqrt{T-T_{min}}\right)\label{b1f},\\&\s_{\th\ph,U(1)}\bigg|^{nM}_{T \sim T_{min}}=\frac{  J_t\sin^2\th}{ B}\left(1+\frac{96 \pi ^4 E^2 n^2 \cot ^2\left(\frac{\theta }{2}\right)  T_{\min }^4}{81 B^4 \csc^2\th}\right)+O\left(\sqrt{T-T_{min}}\right)\label{b2f},\\&\s_{\th\phi,\text{thermal}}\bigg|^{fM}_{T \sim T_{min}}=O\left(\sqrt{T-T_{min}}\right)\label{b3f}, \\&\s_{\th\phi,\text{thermal}}\bigg|^{nM}_{T \sim T_{min}}=\frac{8 \pi ^2  E n \cot \left(\frac{\theta }{2}\right)  T_{\min }^2}{9 B^2 \csc ^2(\theta )}+O\left(\sqrt{T-T_{min}}\right).\label{b4f}
\end{align}
Here, the superscripts $fM$ and $nM$ denote points far and close to the Misner string, respectively.  For points that are far from the Misner string, the angle $\th\sim\pi$, while for points closer to the Misner string, $\th\sim0$.

\begin{figure}[H]
     \centering
     \begin{subfigure}[b]{0.495\textwidth}
         \centering
         \includegraphics[width=\textwidth]{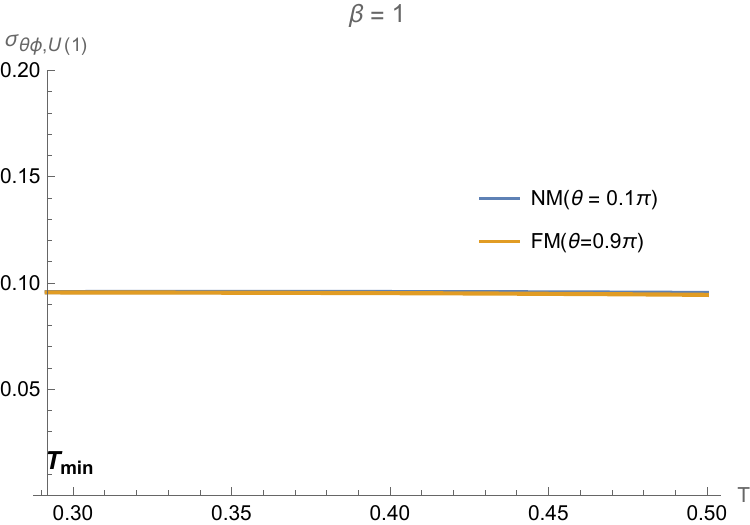}
         \caption{$\s_{\th\ph,U(1)}$ vs temperature }\label{fig1Aa}
              \end{subfigure}
              \hfill
     \begin{subfigure}[b]{0.495\textwidth}
         \centering
         \includegraphics[width=\textwidth]{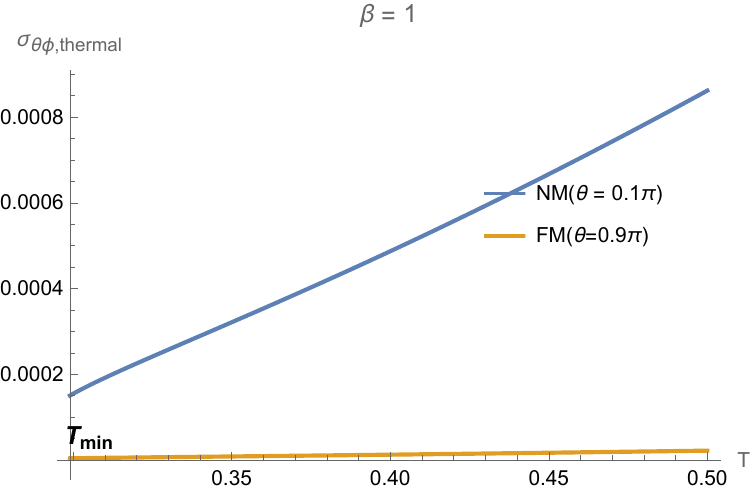}
         \caption{$\s_{\th\ph ,\text{thermal}}$ vs temperature}\label{fig1Ab}
             \end{subfigure}
        
             \caption{$\s_{\th\ph}$ vs temperature plot. Here we set $E=0.1$, $B=10$, $J_t=10$ and $n=0.2$.}
        \label{figure1a}
\end{figure}

It is important to notice that in the presence of finite magnetic field, the Hall conductivity due to externally added $U(1)$ charge carriers (\ref{b1f})-(\ref{b2f}) remains nearly constant in the low temperature regime, regardless of the position of the Misner string (see Figure \ref{fig1Aa}). On the other hand, the thermal contribution to the Hall conductivity (\ref{b3f})-(\ref{b4f}), is nearly zero at points that are far from the Misner string. However, as we get closer to the Misner string, the conductivity increases with a little rise in the temperature due to the frame-dragging, as illustrated in Figure \ref{fig1Ab}. This behavior contrasts with the Hall conductivity at lower magnetic fields \eqref{HallUafmb1}-\eqref{hallthnMb1}, where the conductivity decreases as the temperature increases.

%\eqref{b1f}-\eqref{b4f} indicates that frame - dragging effect in finite Magnetic field $B$ at low temperature are negligible for hall conductivity.
%We can also match with conductivity at low temprature from \cite{A.~O'Bannon} which comes out to be $\s_{xy}=\frac{J_t}{B}$, we get $\s_{\th\ph}\approx\frac{J_t\sin^2\th}{B}$ since we rescaled  Hall  conductivity by $\sin^2\th$ in section \ref{hall conductivity}.

Figure \ref{figure1a} illustrates that the $U(1)$ charge carriers are unaffected due to frame-dragging at finite magnetic field. However, the thermal charge carriers are influenced by the location of the Misner string. This is further reflected in the ratios of these conductivities, both near and far from the Misner string, as shown below
\begin{align}
     \frac{\sigma_{\th\ph,U(1)}\bigg|_{T \sim T_{min}}^{nM}}{\sigma_{\th\ph,U(1)}\bigg|_{T \sim T_{min}}^{fM}}  =1+\frac{96 \pi ^4 E^2 n^2 \cot ^2\left(\frac{\theta }{2}\right)  T_{\min }^4}{81 B^4 \csc^2\th}\approx1\hspace{1mm},\hspace{2mm}\frac{\sigma_{\th\ph,thermal}\bigg|_{T \sim T_{min}}^{nM}}{\sigma_{\th\ph,thermal}\bigg|_{T \sim T_{min}}^{fM}}>>1.\label{ratio1f}
\end{align}

Clearly, in the presence of finite magnetic fields $(B>E)$ and at low temperatures, the number of $U(1)$ charge carriers dominate over the thermally produced charge carriers. These $U(1)$ carriers are mostly drifted by the Lorentz force due to magnetic field $B$ producing a uniform current. The uniform current both near and away from the Misner string is an artifact of a smaller frame-dragging effect at finite magnetic  field $(B)$. This is further demonstrated below
\begin{align}
  \frac{\s_{\th\ph,U(1)}}{\s_{\th\ph,thermal}}\Bigg|^{nM}_{T\sim T_{min}} = \frac{9J_tB\tan\left(\frac{\th}{2}\right)}{8\pi^2En\ T_{min}^2}>>1\hspace{1mm},\hspace{2mm} \frac{\s_{\th\ph,U(1)}}{\s_{\th\ph,thermal}}\Bigg|^{fM}_{T\sim T_{min}}>>1.\label{ratio2f}
\end{align}

\subsubsection{Ohmic conductivity }
In this Section, we study the Ohmic conductivity ($\s_{\th\ph}$) \eqref{condu1dcT}-\eqref{condthT} at finite magnetic fields ($B>1$). At low temperatures ($T\sim T_{min}$), $\s_{\ph\ph}$ can be approximated both near and far from the Misner string as follows
%\begin{align}  & \s_{\phi\ph,U(1)}\bigg|^{fM}_{T \sim T_{min}}=\frac{36 \pi ^2 J_t T_{\min }^2}{81 B^2 \csc ^2(\theta )+16 \pi ^4 T_{\min }^4},\label{d1f}\\&\s_{\phi\ph,U(1)}\bigg|^{nM}_{T \sim T_{min}}=\frac{36 \pi ^2 J_t T_{\min }^2}{81 B^2 \csc ^2(\theta )+16 \pi ^4 T_{\min }^4}\left(1+\frac{7776 \pi ^4 E^2 n^2 \cot ^2\left(\frac{\theta }{2}\right) \csc ^2(\theta ) T_{\min }^4}{2 \left(81 B^2 \csc ^2(\theta )+16 \pi ^4 T_{\min }^4\right){}^2}\right), \label{d2f}\\& \s_{\phi\ph,\text{thermal}}\bigg|^{fM}_{T \sim T_{min}}=\frac{4 \pi ^2 T_{min}^2}{\sqrt{81 B^2 \csc ^2(\theta )+16 \pi ^4 T_{min}^4}},\label{d3f}\\&\s_{\phi\ph,\text{thermal}}\bigg|^{nM}_{T \sim T_{min}}=\frac{4 \pi ^2 T_{min}^2}{\sqrt{81 B^2 \csc ^2(\theta )+16 \pi ^4 T_{min}^4}}\left(1+\frac{2592 \pi ^4 E^2 n^2 \cot ^2\left(\frac{\theta }{2}\right) \csc ^2(\theta ) T_{\min }^4}{\left(81 B^2 \csc ^2(\theta )+16 \pi ^4 T_{\min }^4\right){}^2}\right).\label{d4f}\end{align}
\begin{align}
   & \s_{\phi\ph,U(1)}\bigg|^{fM}_{T \sim T_{min}}=\frac{36 \pi ^2 J_t T_{min }^2}{81 B^2 \csc ^2(\theta )},\label{d1f}\\&\s_{\phi\ph,U(1)}\bigg|^{nM}_{T \sim T_{min}}=\frac{36 \pi ^2 J_t T_{min }^2}{81 B^2 \csc ^2(\theta )}\left(1+\frac{96 \pi ^4 E^2 n^2 \cot ^2\left(\frac{\theta }{2}\right)  T_{min }^4}{ 81 B^4 \csc ^2(\theta )}\right), \label{d2f}\\& \s_{\phi\ph,thermal}\bigg|^{fM}_{T \sim T_{min}}=\frac{4 \pi ^2 T_{min}^2}{9 B \csc(\theta )},\label{d3f}\\&\s_{\phi\ph,thermal}\bigg|^{nM}_{T \sim T_{min}}=\frac{4 \pi ^2 T_{min}^2}{9 B \csc(\theta )}\left(1+\frac{32 \pi ^4 E^2 n^2 \cot ^2\left(\frac{\theta }{2}\right)  T_{min }^4}{81 B^4 \csc ^2(\theta )}\right).\label{d4f}
\end{align}
For points near the Misner string, $\th\sim0$, and for points far away from the Misner string, $\th\sim\pi$.

\begin{figure}[H]
     \centering
   
     \begin{subfigure}[b]{0.495\textwidth}
         \centering
         \includegraphics[width=\textwidth]{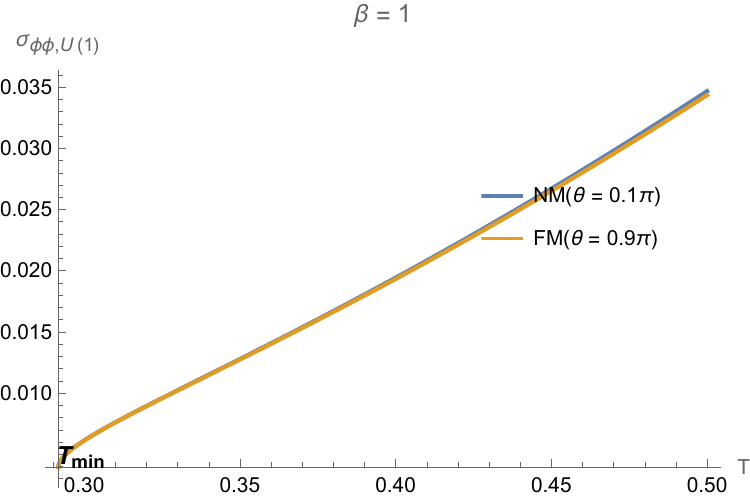}
         \caption{$\s_{\ph \ph ,U(1)}$ vs temperature plot}\label{fig3Ab}
             \end{subfigure}
             \hfill
               \begin{subfigure}[b]{0.495\textwidth}
         \centering
         \includegraphics[width=\textwidth]{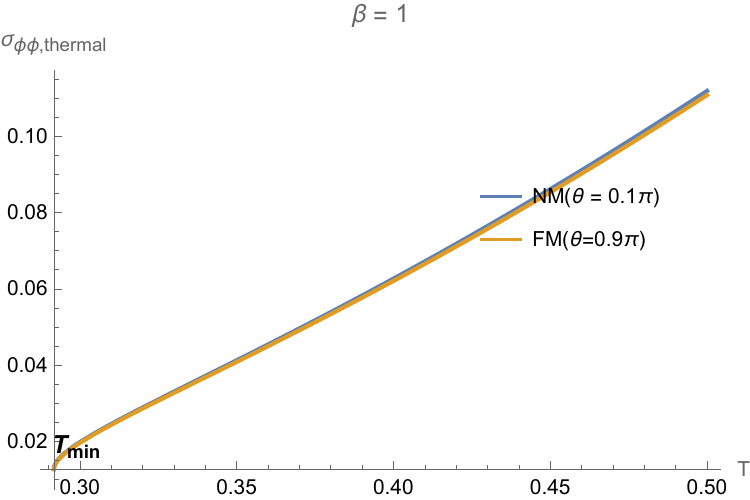}
         \caption{$\s_{\ph \ph,\text{thermal}}$ vs temperature plot}\label{fig3Aa}
              \end{subfigure}
              \hfill\caption{$\s_{\phi\ph}$ vs temperature plot. Here, we set for $E=0.1$, $B=10$, and $J_t=10$.}
        \label{figure3A}
\end{figure}

It is important to notice that the expressions \eqref{d1f}-\eqref{d4f} show that the frame dragging effects are suppressed at finite magnetic fields ($B>1$) and at low temperatures ($T\sim T_{min}$). To further clarify this point, we plot both the conductivities ($\s_{\phi\ph,U(1)}$ and $\s_{\phi\ph,thermal}$) against the temperature as shown in Figure \ref{figure3A}. In this analysis, we set the Nut parameter $n=0.2$. Figure \ref{figure3A} shows that both $\s_{\phi\ph,U(1)}$ and $\s_{\phi\ph,thermal}$ decrease as we lower the temperature, regardless of the position of the Misner string. This result is in contrast with the Ohmic conductivities \eqref{d1}-\eqref{d4} studied before for lower magnetic field $(B<<1)$ at lower temperature ($T\sim T_{min}$), where the conductivities decrease (or remain constant) with increase in temperature. This stems from the fact that at finite  $B(>E)$, the charge carriers are mostly driven by the magnetic field $(B)$, causing a larger contribution to the Hall coefficient, thereby lowering the corresponding contribution in the Ohmic counterpart.

%Similarly,\eqref{d1f}-\eqref{d4f} indicates that frame-dragging effects are also negligible for Ohmic conductivity at low temperature in a finite magnetic field B.
%We can also get the Low temprature Ohmic  conductivity from the expression of Ohmic conductivity from \cite{A.~O'Bannon}, which comes out to be $\s_{xx,U(1)}\sim \frac{J_tT^2}{B^2}$ which match with our result for Ohmic conductivity and $\s_{xx,\text{thermal}}\sim\frac{T^3}{B}\approx0$ but we got $\s_{\ph\ph,\text{thermal}}\sim\frac{T^2_{min}}{B}$

The fact that frame-dragging effects are negligible at higher magnetic fields (and at lower temperatures) can be better understood by examining the ratio of Ohmic conductivities both near and far from the Misner string as follows
 \begin{align}
     \frac{\sigma_{\ph\ph,U(1)}\bigg|_{T \sim T_{min}}^{nM}}{\sigma_{\ph\ph,U(1)}\bigg|_{T \sim T_{min}}^{fM}}  =1+\frac{96 \pi ^4 E^2 n^2 \cot ^2\left(\frac{\theta }{2}\right)  T_{min }^4}{ 81 B^4 \csc ^2(\theta )}\approx1,\\\frac{\sigma_{\ph\ph,thermal}\bigg|_{T \sim T_{min}}^{nM}}{\sigma_{\ph\ph,thermal}\bigg|_{T \sim T_{min}}^{fM}}=1+\frac{32 \pi ^4 E^2 n^2 \cot ^2\left(\frac{\theta }{2}\right)  T_{min }^4}{81 B^4 \csc ^2(\theta )}\approx1.\label{ratio3f}
\end{align}

In addition, under the influence of a finite magnetic field, the ratio of $U(1)$ charge to the thermal contribution to Ohmic conductivity can be expressed as
\begin{align}
  \frac{\s_{\ph\ph,U(1)}}{\s_{\ph\ph,\text{thermal}}}\Bigg|^{nM}_{T\sim T_{min}} \approx\frac{J_t}{B\csc\th }\hspace{1mm},\hspace{2mm}\frac{\s_{\ph\ph,U(1)}}{\s_{\ph\ph,\text{thermal}}}\Bigg|_{T\sim T_{min}}^{fM}=\frac{J_t}{B\csc\th }.\label{ratio4f}
\end{align}

Notice that the Ohmic conductivity is dominated by $U(1)$ charge carriers if charge density $J_t$ is higher than $B$ (for fixed $\th$). On the other hand, for $B>J_t$ thermal pairs dominate over the $U(1)$ carriers.

\subsubsection{Relationship between Hall and Ohmic conductivity}
In Section \ref{rel}, we discussed the relationships between the Hall and Ohmic conductivity at lower magnetic fields (see expressions (\ref{r1}), (\ref{R10}), (\ref{r2}), and (\ref{r3})). Interestingly, we also observed similar relationships between the conductivities (\ref{b1f})-(\ref{b2f}), \eqref{d1f}-\eqref{d4f} at finite magnetic fields. Below, we outline these relationships for both near and far from the Misner string.

The relationship between the conductivities due to the $U(1)$ charge carriers for both near and far away from the Misner string can be expressed as\footnote{Here, we use equation \eqref{d1}-\eqref{d4} to calculate $\s_{\ph\ph,U(1)}$ and $\s_{\ph\ph,thermal}$ in the absence of a magnetic field ($B$).}
\begin{align}
    %&\s_{\th\ph,U(1)}^2\Bigg|^{fM}_{T\sim T_{min}}=\left(\frac{  J_t\sin^2\th}{ B}\right)^2 ,\hspace{2mm}\left(\s_{\ph\ph,U(1)}\Delta\s_{\ph\ph,U(1)}\right)\Bigg|^{fM}_{T\sim T_{min}}=-\left(\frac{  J_t\sin\th}{ B}\right)^2,\\
    &\s_{\th\ph,U(1)}^2\Bigg|^{fM}_{T\sim T_{min}}=-\sin^2\th\left(\s_{\ph\ph,U(1)}\Delta\s_{\ph\ph,U(1)}\right)\Bigg|^{fM}_{T\sim T_{min}},\label{r1f}\\
    &\s_{\th\ph,U(1)}^2\Bigg|^{nM}_{T\sim T_{min}}=-\sin^2\th\left(\s_{\ph\ph,U(1)}\Delta\s_{\ph\ph,U(1)}\right)\Bigg|^{nM}_{T\sim T_{min}}.
\end{align}

A similar calculation yields the relationship between the conductivities due to the thermally produced charge carriers, as presented below
\begin{align}
 &\s_{\th\ph,thermal}^2\Bigg|^{fM}_{T\sim T_{min}}=0,\hspace{1mm} (\s_{\ph\ph,thermal}\Delta\s_{\ph\ph,\text{thermal}})\Bigg|^{fM}_{T\sim T_{min}}\approx0,\\
 &\s_{\th\ph,thermal}^2\Bigg|^{nM}_{T\sim T_{min}}= 0,\hspace{1mm}(\s_{\ph\ph,thermal}\Delta\s_{\ph\ph,\text{thermal}})\Bigg|^{nM}_{T\sim T_{min}}\approx0.
\end{align}

\subsubsection{Comparison between Ohmic and Hall conductivity}\label{sc1.4}
In this section, we compare the Hall (\ref{b1f})-(\ref{b2f}) and Ohmic \eqref{d1f}-\eqref{d4f} conductivities both near and far from the Misner string. Our analysis reveals that, in the presence of a finite magnetic field, the Hall conductivity due to the $U(1)$ charge carriers dominates over the Ohmic conductivity from the same charge carriers (see Figures \ref{figure1a}-\ref{figure3A}). On the other hand, the Ohmic conductivity from thermal charge carriers is greater than the thermal Hall conductivity, regardless of the location of Misner string. Below, we illustrate this observation by examining the following ratios
\begin{align}
  \frac{\s_{\ph\ph,U(1)}}{\s_{\th\ph,U(1)}}\Bigg|^{fM}_{T\sim T_{min}} = \frac{4\pi^2T_{min}^2}{9B}<1\hspace{1mm},\hspace{2mm}  \frac{\s_{\ph\ph,U(1)}}{\s_{\th\ph,U(1)}}\Bigg|^{nM}_{T\sim T_{min}} =\frac{4\pi^2T_{min}^2}{9B}<1,\label{ratio6fb}
\end{align}
\begin{align}
  \frac{\s_{\ph\ph,thermal}}{\s_{\th\ph,thermal}}\Bigg|^{fM}_{T\sim T_{min}} >>1\hspace{1mm},\hspace{2mm} \frac{\s_{\ph\ph,thermal}}{\s_{\th\ph,thermal}}\Bigg|^{nM}_{T\sim T_{min}} =\frac{B\tan\left(\frac{\th}{2}\right)}{2En\sin\th}>>1.\label{ratio5fb}
\end{align}

This phenomenon can be understood as follows. At a finite magnetic field, the Lorentz force obstructs the effective acceleration of $U(1)$ charge carriers along the direction of the electric field, leading to a decrease in longitudinal conductivity or the Ohmic conductivity due to $U(1)$ charge carriers \cite{ashcroft2011solid}. Consequently, the transverse component or the Hall conductivity, which is oriented perpendicular to the electric field, becomes the dominant factor from the same charge carriers.

On the other hand, thermal charge carriers are also influenced by the Lorentz force in a finite magnetic field. However, the difference in speed between particles and antiparticles in the transverse direction $(\hat \th)$ is negligible due to minimal frame-dragging effects. As a result, the Ohmic current generated by thermally produced charge carriers exceeds the Hall current produced by the same carriers.

%\textcolor{red}{In the presence of a finite magnetic field, the Hall current surpasses the Ohmic current. As outlined in Ashcroft and Mermin (Ch. 12), when a strong magnetic field is applied, the Lorentz force hampers the effective acceleration of electrons in the direction of the electric field, leading to a decrease in longitudinal conductivity. As a result, the main component of the current arises from the transverse drift velocity, which is oriented perpendicular to the electric field, signifying a regime where the Hall effect prevails..\cite{ashcroft2011solid}}
%Furthermore, notice that in the strong magnetic field limit, i.e., $B>1$, both the Hall (\ref{b1f})-(\ref{b2f}) and Ohmic \eqref{d1f}-\eqref{d4f} conductivities are inversely proportional to the magnetic field ($B$). Consequently, it is not possible to define the Ohmic conductivity in the limit $B\rightarrow0$. Therefore, the quantity $\Delta\s_{Ohmic}=\s_{Ohmic}(B)-\s_{Ohmic}(B=0)$ is not well defined.  As a result, we cannot establish a similar relationship between the Hall and Ohmic conductivities as discussed for lower magnetic fields in Section \ref{rel}.

In the following, we provide a summary of the Hall conductivity (\ref{b1f})-(\ref{b2f}) and the Ohmic conductivity \eqref{d1f}-\eqref{d4f} related to the TN-AdS black hole for both near and far from the Misner string in the strong magnetic field limit ($B>1$). 
\begin{table}[H]
    \begin{center}
   
\renewcommand{\arraystretch}{1.7}
\begin{tabular}{|c|c|c|c|c|}
 \hline
 $\s$ &    $fM $  & $nM $ 
 \\ 
 \hline
 $\s_{\ph\ph,U(1)}$ &$\frac{36 \pi ^2 J_t T_{min }^2}{81 B^2 \csc ^2(\theta )}$
   & $ \frac{36 \pi ^2 J_t T_{min }^2}{81 B^2 \csc ^2(\theta )}\left(1+\frac{96 \pi ^4 E^2 n^2 \cot ^2\left(\frac{\theta }{2}\right)  T_{min }^4}{ 81 B^4 \csc ^2(\theta )}\right)$\\
  \hline
 $\s_{\ph\ph,thermal}$ & $ \frac{4 \pi ^2 T_{min}^2}{9 B \csc(\theta )} $&  $ \frac{4 \pi ^2 T_{min}^2}{9 B \csc(\theta )}\left(1+\frac{32 \pi ^4 E^2 n^2 \cot ^2\left(\frac{\theta }{2}\right)  T_{min }^4}{81 B^4 \csc ^2(\theta )}\right)$ \\
 \hline
  $\s_{\th\ph,U(1)}$ &$\frac{  J_t\sin^2\th}{ B}$ &  $\frac{  J_t\sin^2\th}{ B}\left(1+\frac{96 \pi ^4 E^2 n^2 \cot ^2\left(\frac{\theta }{2}\right)  T_{min }^4}{81 B^4 \csc^2\th}\right)$\\ \hline
   $\s_{\th\ph,thermal}$ &$0$ &  $\frac{8 \pi ^2  E n \cot \left(\frac{\theta }{2}\right)  T_{min }^2}{9 B^2 \csc ^2(\theta )}$\\ \hline
\end{tabular}
 \caption{Hall and Ohmic conductivity in the strong magnetic field  ($B>1$) and at low temperatures $(T\sim T_{min})$.}
    \label{tableltn3}
         
    \end{center}
\end{table}

\subsection{High temperature regime}
In this Section, we examine the Ohmic ($\s_{\ph\ph}$) \eqref{condu1dcT}-\eqref{condthT} and the Hall conductivity ($\s_{\th\ph}$) \eqref{hallu1T}-\eqref{hallthermal T} at high temperatures, i.e., $T>> T_{min}$, and at finite magnetic field ($B>1$). It is important to highlight that, under these circumstances, the angular velocity \eqref{angb1} can be represented as
%\textcolor{red}{\begin{align}  |\tilde\Omega|=\frac{9 n \cot \left(\frac{\theta }{2}\right) }{8 \left(3 n^2+1\right)^2 }\left(\frac{T_{min}}{T}\right)^4\approx0.\label{htbo}\end{align}}
\begin{align}
   |\tilde\Omega|=  \frac{162 n T_{min}^4 \cot \left(\frac{\theta }{2}\right)}{81 B^2T_{\min }^4 \csc ^2(\theta )+{144 \left(3 n^2+1\right)^2 T^4}{}}<<1,\label{htbo}
\end{align}
which shows that the frame-dragging effect is negligible similar to the low temperature phase.

\subsubsection{Ohmic conductivity}
In this section, we analyze the Ohmic conductivity ($\s_{\ph\ph}$) as described in \eqref{condu1dcT}-\eqref{condthT} when subjected to finite magnetic fields ($B>1$). At high temperatures ($T>> T_{min}$), the expression for $\s_{\ph\ph}$ can be approximated both near and far away from the Misner string as follows
\begin{align}
   & \s_{\phi\ph,U(1)}\bigg|^{fM}_{T >> T_{min}}=\frac{16 \pi ^2 {J_t} T^2 T_{\min }^4}{9 B^2 \csc ^2(\theta ) T_{\min }^4+16 \left(3 n^2+1\right)^2 T^4},\label{d1fh}
   %\frac{3 J_t}{4 \left(3 n^2+1\right) }\left(\frac{T_{min}}{T}\right)^2,
   \\&\s_{\phi\ph,U(1)}\bigg|^{nM}_{T >> T_{min}}=\frac{16 \pi ^2 {J_t} T^2 T_{\min }^4}{9 B^2 \csc ^2(\theta ) T_{\min }^4+16 \left(3 n^2+1\right)^2 T^4}\left(1+\frac{1536 \pi ^4 E^2 n^2 T^4 \cot ^2\left(\frac{\theta }{2}\right) \csc ^2(\theta ) T_{\min }^8}{\left(9 B^2 \csc ^2(\theta ) T_{\min }^4+16 \left(3 n^2+1\right)^2 T^4\right){}^2}\right)
   %\frac{3 J_t}{4 \left(3 n^2+1\right) }\left(\frac{T_{min}}{T}\right)^2\left(1+\frac{27E^2 n^2 \cot ^2\left(\frac{\theta }{2}\right)}{8 \left(3 n^2+1\right)^2 \sin^2\th}\left(\frac{T_{min}}{T}\right)^4\right)
   , \label{d2fh}\\& \s_{\phi\ph,thermal}\bigg|^{fM}_{T >> T_{min}}=1-\frac{9 B^2  \csc ^2(\theta )}{32 \left(3 n^2+1\right)^2 }\left(\frac{T_{min}}{T}\right)^4,\label{d3fh}\\&\s_{\phi\ph,thermal}\bigg|^{nM}_{T >> T_{min}}=1-\frac{9 B^2  \csc ^2(\theta )}{32 \left(3 n^2+1\right)^2 }\left(\frac{T_{min}}{T}\right)^4+\frac{9 E^2 n^2 \cot ^2\left(\frac{\theta }{2}\right) \csc (\theta )^2 }{8\left(3 n^2+1\right)^2 }\left(\frac{T_{min}}{T}\right)^4.\label{d4fh}
\end{align}

\begin{figure}[H]
     \centering
     \begin{subfigure}[b]{0.495\textwidth}
         \centering
         \includegraphics[width=\textwidth]{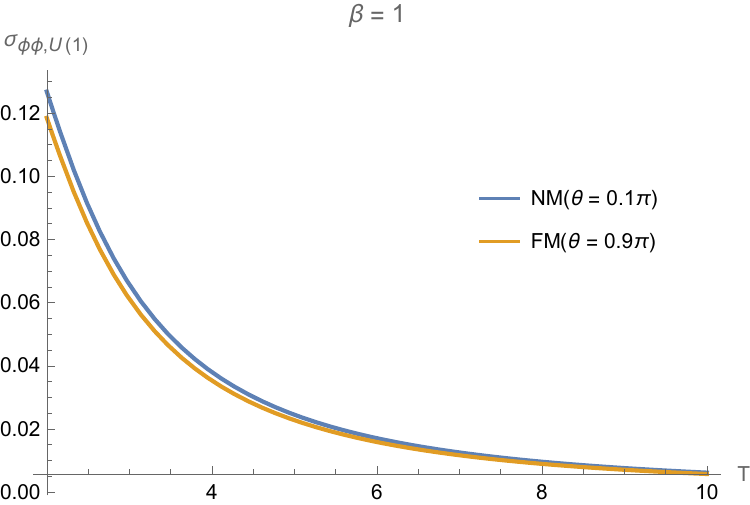}
         \caption{$\s_{\ph \ph,U(1)}$ vs temperature }\label{fig3Ba}
              \end{subfigure}
              \hfill
     \begin{subfigure}[b]{0.495\textwidth}
         \centering
         \includegraphics[width=\textwidth]{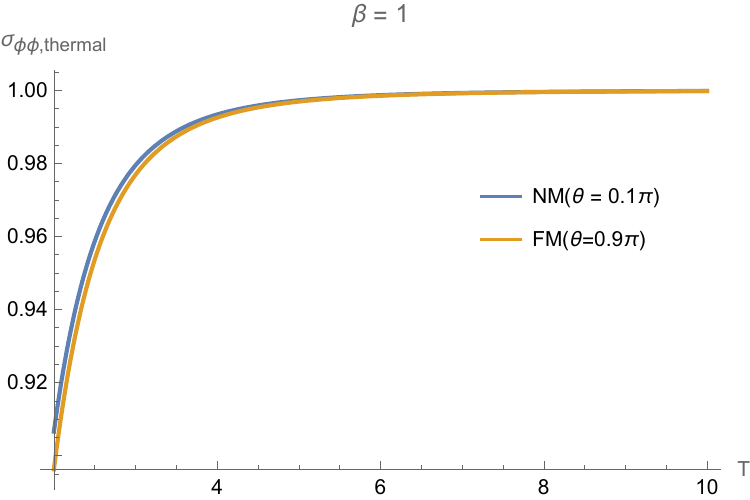}
         \caption{$\s_{\ph \ph ,\text{thermal}}$ vs temperature }\label{fig3Bb}
             \end{subfigure}
             \caption{$\s_{\phi\ph}$ vs temperature plot. Here, we set for $E=0.1$, $B=10$,  $J_t=10$ and $n=0.2$.}
        \label{figure3B}
\end{figure}

Notice that the Ohmic conductivity due to $U(1)$ carriers (\ref{d1fh})-(\ref{d2fh}) decreases as the temperature increases, regardless of the location of the Misner string, as shown in Figure \ref{fig3Ba}. On the other hand, the conductivity associated with thermal charge carriers (\ref{d3fh})-(\ref{d4fh}) increases with temperature and then saturates at nearly unity (see Figure \ref{fig3Bb}). This observation is similar to the conductivity behavior observed at lower magnetic (\ref{highdcu1fmb1})-(\ref{highdcthb1}).

At high temperatures thermally produced pairs populate over $U(1)$ carriers, thereby producing a significant contribution to the Ohmic conductivity as compared to $U(1)$ counterpart. This is further illustrated in the ratio expressed below

%It can be explicitly demonstrated that frame-dragging effects are negligible at higher magnetic fields and high temperatures by taking the ratios of Ohmic conductivities, both near and far from the Misner string as follows (\textcolor{red}{no need of this ratio})
 %\begin{align}
   %  \frac{\sigma_{\ph\ph,U(1)}\bigg|_{T \sim T_{min}}^{nM}}{\sigma_{\ph\ph,U(1)}\bigg|_{T \sim T_{min}}^{fM}}  =&1+\frac{27E^2 n^2 \cot ^2\left(\frac{\theta }{2}\right)}{8 \left(3 n^2+1\right)^2 \sin^2\th}\left(\frac{T_{min}}{T}\right)^4\approx1,\\\frac{\sigma_{\ph\ph,thermal}\bigg|_{T \sim T_{min}}^{nM}}{\sigma_{\ph\ph,thermal}\bigg|_{T \sim T_{min}}^{fM}}=&1+\frac{9 E^2 n^2 \cot ^2\left(\frac{\theta }{2}\right) \csc (\theta )^2 }{8\left(3 n^2+1\right)^2 }\left(\frac{T_{min}}{T}\right)^4\approx1.\label{ratio31f}
%\end{align}

%Furthermore, the thermal charge carriers dominate over the $U(1)$ charge carriers at high temperatures (see Figure \ref{figure3B}). This results in a larger $\sigma_{\phi\phi, thermal}$ both near and far from the Misner string, as illustrated below
\begin{align}
  \frac{\s_{\ph\ph,U(1)}}{\s_{\ph\ph,\text{thermal}}}\Bigg|^{nM}_{T>> T_{min}} \approx\frac{3 J_t}{4 \left(3 n^2+1\right) }\left(\frac{T_{min}}{T}\right)^2<1\hspace{1mm},\hspace{2mm}\frac{\s_{\ph\ph,U(1)}}{\s_{\ph\ph,\text{thermal}}}\Bigg|_{T>> T_{min}}^{fM}=\frac{3 J_t}{4 \left(3 n^2+1\right) }\left(\frac{T_{min}}{T}\right)^2<1.\label{ratio40f}
\end{align}

\subsubsection{Hall conductivity}
In this section, we explore the Hall conductivity ($\s_{\th\ph}$) \eqref{hallu1T}-\eqref{hallthermal T} at finite magnetic fields ($B>1$). In the high temperature ($T>> T_{min}$) limit, the  $\s_{\th\ph}$ can be expressed both near and far away from  Misner string as follows%\footnote{\textcolor{red}{under what condition eq (106) holds}}
\begin{align}
 &   \s_{\th\phi,U(1)}\bigg|^{fM}_{T >> T_{min}}=\frac{9 B {J_t} T_{\min }^4}{9 B^2 \csc ^2(\theta ) T_{\min }^4+16 \left(3 n^2+1\right)^2 T^4}\label{b1fh},\\&\s_{\th\ph,U(1)}\bigg|^{nM}_{T >> T_{min}}=\frac{9 B {J_t} T_{\min }^4}{9 B^2 \csc ^2(\theta ) T_{\min }^4+16 \left(3 n^2+1\right)^2 T^4}\left(1+\frac{124416 \pi ^4 B^2 E^2 {J_t}^2 n^2 T^4 \cot ^2\left(\frac{\theta }{2}\right)\csc^2\th T_{\min }^{16}}{\left(9 B^2 \csc ^2(\theta ) T_{\min }^4+16 \left(3 n^2+1\right)^2 T^4\right){}^4}\right)\label{b2fh},\\&\s_{\th\phi,\text{thermal}}\bigg|^{fM}_{T >> T_{min}}=0\label{b3fh}, \\&\s_{\th\phi,\text{thermal}}\bigg|^{nM}_{T >>T_{min}}=\frac{96 \pi ^2 B E n T^2 \cot \left(\frac{\theta }{2}\right)\csc\th T_{\min }^6}{\left(9 B^2 \csc ^2(\theta ) T_{\min }^4+16 \left(3 n^2+1\right)^2 T^4\right){}^{3/2}}.\label{b4fh}
\end{align}

\begin{figure}[H]
     \centering
     \begin{subfigure}[b]{0.495\textwidth}
         \centering
         \includegraphics[width=\textwidth]{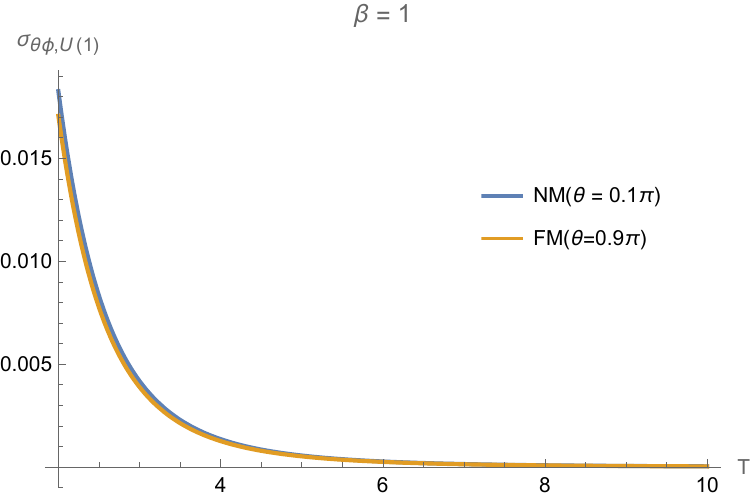}
         \caption{$\s_{\th\ph,U(1)}$ vs $T$ for $B=10$ }\label{fig1Ca}
              \end{subfigure}
              \hfill
     \begin{subfigure}[b]{0.495\textwidth}
         \centering
         \includegraphics[width=\textwidth]{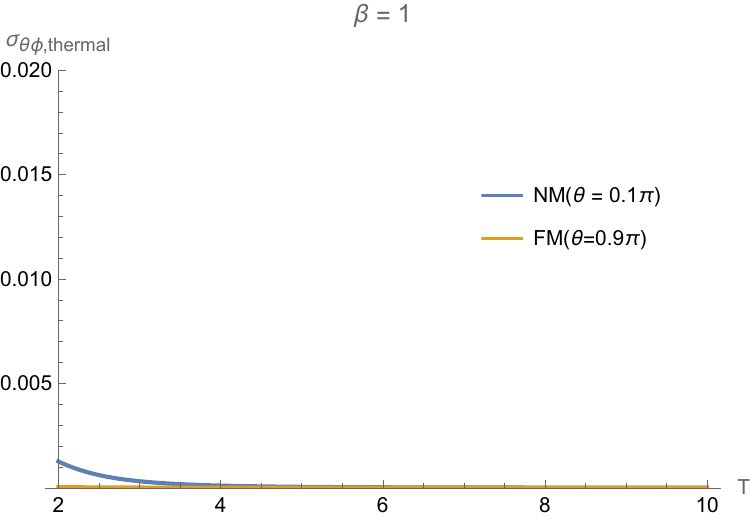}
         \caption{$\s_{\th\ph ,\text{thermal}}$ vs $T$ at $B=10$}\label{fig1Cb}
             \end{subfigure}
        
             \caption{$\s_{\th\ph}$ vs temperature plot. Here we set $E=0.1$, $J_t=10$ and $n=0.2$.}
        \label{figure1C}
\end{figure}

Notice that Hall conductivity (\ref{b1fh})-(\ref{b2fh}) that arises due to $U(1)$ charge carriers decreases with increase in temperature, regardless of the position of the Misner string, as shown in Figure \ref{fig1Ca}. In contrast, the thermal contribution is nearly negligible at points far from the Misner string (\ref{b3fh}). However, as we approach the Misner string, the thermal conductivity varies inversely with the fourth power of temperature (\ref{b4fh}), as shown in Figure \ref{fig1Cb}. This behaviour is identical with the conductivities observed earlier at lower magnetic field \eqref{hallu1fmTh}-\eqref{hallthnmTh}.

Furthermore, the number of $U(1)$ carriers remains unaffected at high temperature. This together with a smaller frame-dragging effect keeps the $U(1)$ Hall current finite and uniform with reference to the position of the Misner string. On the other hand, due to the over population of thermally produced charged pairs, that are drifted due to frame dragging and the magnetic field, one finds a finite Hall current \eqref{b4fh} near Misner string. Away from Misner string frame dragging becomes zero, and the particle-antiparticle pairs move in the same direction with same speed  due to $B$ field causing a net zero Hall current \eqref{b3fh}.
%Figure \ref{figure1C} demonstrates that the frame-dragging does not impact the $U(1)$ charge carriers in a strong magnetic field. In contrast, the thermal charge carriers are affected by the position of the Misner string. This finding is verified by comparing the ratios of these carriers. Conductivities both near and far from the Misner string, as shown below (\textcolor{red}{no need of this ratio})
%\begin{align}
 %    \frac{\sigma_{\th\ph,U(1)}\bigg|_{T >> T_{min}}^{nM}}{\sigma_{\th\ph,U(1)}\bigg|_{T >> T_{min}}^{fM}}  =1+\frac{27 E^2 n^2 \cot ^2\left(\frac{\theta }{2}\right) \csc ^2(\theta ) }{8 \left(3 n^2+1\right)^2 }\left(\frac{T_{min}}{T}\right)^4\approx1\hspace{1mm},\hspace{2mm}\frac{\sigma_{\th\ph,thermal}\bigg|_{T >> T_{min}}^{nM}}{\sigma_{\th\ph,thermal}\bigg|_{T >>T_{min}}^{fM}}>>1.\label{ratio11f}
%\end{align}

Finally, one can show that the Hall conductivity due to $U(1)$ dominates over the thermal contribution, irrespective of the angular position of the Misner string (see Figure \ref{figure1C}). In the following, we demonstrate this observation by examining the following ratios
\begin{align}
  \frac{\s_{\th\ph,U(1)}}{\s_{\th\ph,thermal}}\Bigg|^{nM}_{T >> T_{min}} = \frac{J_t\tan^2\left(\frac{\th}{2}\right)}{2En\csc\th}>>1\hspace{1mm},\hspace{2mm} \frac{\s_{\th\ph,U(1)}}{\s_{\th\ph,thermal}}\Bigg|^{fM}_{T >> T_{min}}>>1.\label{ratio20f}
\end{align}

This stems from the fact that in the limit of zero frame-dragging, thermally produced charge pairs gives rise to a net zero Hall current as explained above. Therefore, a major contribution to the Hall current appears due to $U(1)$ carriers, those are accelerated under Lorentz force in a nonzero magnetic field.

\subsubsection{Relationship between Hall and Ohmic conductivity}
In Section \ref{rel2}, we examined the relationships between Hall and Ohmic conductivity at lower magnetic fields (\ref{e11})-(\ref{e14}). Notably, we also found similar relationships between the conductivities described in expressions (\ref{d1fh})-(\ref{d4fh}), (\ref{b1fh})-(\ref{b4fh}) at finite magnetic fields. Below, we outline these relationships in both regions near and far from the Misner string.

The relationship between the conductivities from the $U(1)$ charge carriers, both near and far from the Misner string, can be expressed as\footnote{Here, we use equation (\ref{highdcu1fmb1})-(\ref{highdcthb1}) to calculate $\s_{\ph\ph,U(1)}$ and $\s_{\ph\ph,thermal}$ in the absence of the magnetic field ($B$).}
\begin{align}
    %&\s_{\th\ph,U(1)}^2\Bigg|^{fM}_{T\sim T_{min}}=\left(\frac{  J_t\sin^2\th}{ B}\right)^2 ,\hspace{2mm}\left(\s_{\ph\ph,U(1)}\Delta\s_{\ph\ph,U(1)}\right)\Bigg|^{fM}_{T\sim T_{min}}=-\left(\frac{  J_t\sin\th}{ B}\right)^2,\\
    &\s_{\th\ph,U(1)}^2\Bigg|^{fM}_{T >> T_{min}}=-\sin^2\th\left(\s_{\ph\ph,U(1)}\Delta\s_{\ph\ph,U(1)}\right)\Bigg|^{fM}_{T\sim T_{min}},\label{r1f1}\\
    &\s_{\th\ph,U(1)}^2\Bigg|^{nM}_{T >> T_{min}}=-\sin^2\th\left(\s_{\ph\ph,U(1)}\Delta\s_{\ph\ph,U(1)}\right)\Bigg|^{nM}_{T\sim T_{min}}.
\end{align}

A similar calculation reveals the relationship between the conductivities due to thermally produced charge carriers, as shown below
\begin{align}
 &\s_{\th\ph,thermal}^2\Bigg|^{fM}_{T >> T_{min}}=0,\hspace{2mm} (\s_{\ph\ph,thermal}\Delta\s_{\ph\ph,\text{thermal}})\Bigg|^{fM}_{T >> T_{min}}\approx0,\\
 &\s_{\th\ph,thermal}^2\Bigg|^{nM}_{T >> T_{min}}= 0,\hspace{2mm}(\s_{\ph\ph,thermal}\Delta\s_{\ph\ph,\text{thermal}})\Bigg|^{nM}_{T >> T_{min}}\approx0.
\end{align}

\subsubsection{Comparison between Ohmic and Hall conductivity}
In this section, we compare the Ohmic (\ref{d1fh})-(\ref{d4fh}) and Hall (\ref{b1fh})-(\ref{b4fh}) conductivities both near and far from the Misner string. Our analysis indicates that, at a finite magnetic field and high temperature, the Hall conductivity resulting from the externally added $U(1)$ charge carriers dominates over the Ohmic conductivity produced by the $U(1)$ charge carriers (see Figures \ref{figure3B}-\ref{figure1C}), irrespective of the location of the Misner string as discussed in Section \ref{sc1.4}.

On the other hand, the Ohmic conductivity from thermal charge carriers exceeds the thermal Hall conductivity both near and far from the Misner string. Below, we show this by taking the ratios of the above entities
\begin{align}
  \frac{\s_{\ph\ph,U(1)}}{\s_{\th\ph,U(1)}}\Bigg|^{fM}_{T >> T_{min}} = \frac{4(1+3n^2)}{3B}\left(\frac{T}{T_{min}}\right)^2<1\hspace{1mm},\hspace{2mm}  \frac{\s_{\ph\ph,U(1)}}{\s_{\th\ph,U(1)}}\Bigg|^{nM}_{T >> T_{min}} =\frac{4(1+3n^2)}{3B}\left(\frac{T}{T_{min}}\right)^2<1,\label{ratio60fb}
\end{align}
\begin{align}
  \frac{\s_{\ph\ph,thermal}}{\s_{\th\ph,thermal}}\Bigg|^{fM}_{T >> T_{min}}>>1\hspace{1mm},\hspace{2mm} \frac{\s_{\ph\ph,thermal}}{\s_{\th\ph,thermal}}\Bigg|^{nM}_{T >> T_{min}} =\frac{8\sin\th \left(3 n^2+1\right)^2 }{9 B E n \cot \left(\frac{\theta }{2}\right) }\left(\frac{T}{T_{min}}\right)^4>>1.\label{ratio50fb}
\end{align}

Equation \eqref{ratio60fb} indicates that for $B>E$ majority of the $U(1)$ carriers are drifted in the direction orthogonal to electric field $(E)$, causing a higher value in the Hall conductivity. On the other hand, in the thermal counterpart \eqref{ratio50fb}, most of the charge pairs are drifted in the same directions with same speed  causing a lower Hall transport as compared to its Ohmic counterpart.

In the following, we summarise the Ohmic conductivity (\ref{d1fh})-(\ref{d4fh}) and Hall conductivity (\ref{b1fh})-(\ref{b4fh}) associated with the TN-AdS black hole at high temperature for both near and far from the Misner string, while considering a finite magnetic field. 
\begin{table}[H]
    \begin{center}
   
\renewcommand{\arraystretch}{1.7}
\begin{tabular}{|c|c|c|c|c|}
 \hline
 $\s$ &    $fM $  & $nM $ 
 \\ 
 \hline
 $\s_{\ph\ph,U(1)}$ &$\frac{16 \pi ^2 {J_t} T^2 T_{min }^4}{9 B^2 \csc ^2(\theta ) T_{min }^4+16 \left(3 n^2+1\right)^2 T^4}$
   & $ \frac{16 \pi ^2 {J_t} T^2 T_{min }^4}{9 B^2 \csc ^2(\theta ) T_{min }^4+16 \left(3 n^2+1\right)^2 T^4}\left(1+\frac{1536 \pi ^4 E^2 n^2 T^4 \cot ^2\left(\frac{\theta }{2}\right) \csc ^2(\theta ) T_{min }^8}{\left(9 B^2 \csc ^2(\theta ) T_{min }^4+16 \left(3 n^2+1\right)^2 T^4\right){}^2}\right)$\\
  \hline
 $\s_{\ph\ph,\text{thermal}}$ & $1-\frac{9 B^2  \csc ^2(\theta )}{32 \left(3 n^2+1\right)^2 }\left(\frac{T_{min}}{T}\right)^4$&  $1-\frac{9 B^2  \csc ^2(\theta )}{32 \left(3 n^2+1\right)^2 }\left(\frac{T_{min}}{T}\right)^4+\frac{9 E^2 n^2 \cot ^2\left(\frac{\theta }{2}\right) \csc (\theta )^2 }{8\left(3 n^2+1\right)^2 }\left(\frac{T_{min}}{T}\right)^4$ \\
 \hline
  $\s_{\th\ph,U(1)}$ &$\frac{9 B J_t }{16 \left(3 n^2+1\right)^2 }\left(\frac{T_{min }}{T}\right)^4$ &  $\frac{9 B J_t }{16 \left(3 n^2+1\right)^2 }\left(\frac{T_{min}}{T}\right)^4\left(1+\frac{27 E^2 n^2 \cot ^2\left(\frac{\theta }{2}\right) \csc ^2(\theta ) }{8 \left(3 n^2+1\right)^2 }\left(\frac{T_{min}}{T}\right)^4\right)$\\ \hline
   $\s_{\th\ph,\text{thermal}}$ &$0$ &  $\frac{9 B E n\cot\left(\frac{\th}{2}\right) }{8 \sin\th \left(3 n^2+1\right)^2}\left(\frac{T_{min}}{T}\right)^4.$\\ \hline
\end{tabular}
 \caption{Hall and Ohmic conductivity in the finite magnetic field limit ($B>1$) at high temperatures $(T>> T_{min})$. }
    \label{tableltn4}
         
    \end{center}
\end{table}
\section{Conclusion and future directions}
To summarize, in the present work, we investigate the Hall transport associated with the four-dimensional Taub-NUT AdS black holes \cite{Liu:2023uqf}. To be specific, we obtained the holographic Ohmic and Hall conductivity for these black holes using a D-brane probe \cite{A.~O'Bannon}. Our analysis has been performed in the limit of low electric field $(E<<1)$, while the magnetic field $(B)$ could be both small or finite. We explore the effects of the Nut parameter ($n$), which leads to frame dragging for both small ($B<<1$) and finite magnetic fields ($B>1$).

Our calculations indicate that both the Ohmic ($\s_{\ph\ph}$) and Hall ($\s_{\th\ph}$) conductivities receive contributions from externally added $U(1)$ charge carriers and thermally generated charge pairs. The major finding of our paper is that, unlike the previous holographic studies \cite{A.~O'Bannon}-\cite{Lee:2010uy}, which report zero Hall conductivity in thermal plasma, our calculations demonstrate a non-vanishing Hall conductivity. This is attributed to the phenomenon of frame dragging, which induces an ``effective motion'' of (charges carriers within the thermal plasma) and hence a non-zero Hall current in a direction transverse to the applied electric field.  We first study these conductivities considering a low magnetic field, i.e., $B<<1$, in both low ($T\sim T_{min}$) and high ($T>> T_{min}$) temperature regimes.

In the low temperature and low magnetic field regime, both the conductivities $\s_{\th\ph}$ and $\s_{\ph\ph}$ due to the externally added $U(1)$ charge carriers increase as we approach the $T_{min}$, with a faster rate close to the Misner string. This behavior is due to the  significantly drift of $U(1)$ charge carriers  near the Misner string as a result of the frame-dragging. Furthermore, as we increase the magnetic field, the $U(1)$ charge carriers experience additional drift in the direction transverse to the electric field $(E)$ (due to Lorentz force $\vec{F}=q(\vec{v}\times \vec{B})$), leading to higher Hall conductivity. However, the Ohmic conductivity decreases notably with an increase in the magnetic field because a fraction of $U(1)$ charge carriers are accelerated in the direction perpendicular to the external electric field ($E$).

On the other hand, both the Hall and the Ohmic conductivities due to the thermally produced charge carriers increase as we approach the minimum temperature ($T_{min}$) for points closer to the Misner string. However, as we move away from the Misner string, the thermal Hall conductivity becomes negligible, and the thermal Ohmic conductivity remains nearly constant. Additionally, as we increase the magnetic field, the thermal Hall conductivity rises, while the thermal Ohmic conductivity decreases.

Our analysis reveals that the externally added $U(1)$ charge carriers are significantly greater than the thermal charge carriers at low temperatures. This leads to higher Ohmic and Hall conductivities due to the $U(1)$ charge carriers than their respective thermal counterpart. 

Furthermore, it is interesting to notice that when we are far from the Misner string, the Ohmic conductivity grows as $\s_{\ph\ph}\approx\s_{\ph\ph,U(1)}\sim T^{-2}$ and the Hall conductivity as $\s_{\th\ph}\approx\s_{\th\ph,U(1)}\sim T^{-4}$ \cite{Stojkovic}. These observations show that at low temperatures, the conductivity acts like a Fermi liquid \cite{Blake:2014yla}-\cite{Ding}, also known as the Drude result, where $\frac{\s_{\ph\ph}}{\s_{\th\ph}}\sim \s_{\ph\ph}^{-1}$ \cite{Lee:2010ii}-\cite{Hartnoll:2009ns}. The authors in \cite{Lee:2010ii} also reported similar observations, where they compute the Ohmic and Hall conductivity associated with the black hole in the asymptotic Lifshitz space-time in small electric and magnetic fields. On the other hand, as we approach closer to the Misner string, frame dragging begins to
dominate, leading to a conductivity behavior, $\s_{\ph\ph,U(1)}\sim T^{-4}$ and $\s_{\th\ph,U(1)}\sim T^{-6}$. This can be interpreted
as a new phase of a quantum liquid \cite{Khan:2025fne}, \cite{Ge:2016lyn}, where $\frac{\s_{\ph\ph}}{\s_{\th\ph}}\sim \s_{\ph\ph}^{-1/2}$, which is referred as the non-Drude result \cite{Anderson:1991ixg}, \cite{A.W. Tyler }.

In the high temperature ($T>>T_{min}$) and low magnetic field ($B<<1$) limit,  the effects due to the frame dragging are negligible. As a result, the behavior of conductivities, both near and far from the Misner string, are nearly identical. We observe that the Ohmic conductivity arises due to $U(1)$ charge carriers, decreases with increasing temperature. In contrast, the Ohmic conductivity, sourced due to thermally produced charge carriers, saturates at unity regardless of the magnetic field ($B$).  Similarly, the Hall conductivity due to $U(1)$ charge carriers also decreases with temperature, while the thermal contribution remains negligibly small \cite{Lee:2010ii}.% \textcolor{red}{These observations  for Ohmic conductivity ($U(1)$ and thermal)and Hall $U(1)$ are  consistent with the result of \cite{Lee:2010ii} for a four-dimensional space-time,and as usual Author in \cite{Lee:2010ii} also obtained zero Hall of thermally produced charge carriers}

Moreover, the number of thermally produced charged pairs is greater than that of $U(1)$ charge carriers in the high temperature limit, resulting in higher thermal Ohmic conductivity, both near and far from the Misner string. In contrast, the Hall conductivity due to $U(1)$ charge carriers remains greater than the thermal Hall conductivity, even at high temperatures. This is because the thermal Hall conductivity arises due to the frame dragging, which becomes negligible at high temperatures. %Consequently, unlike Ohmic conductivity, the Hall conductivity of $U(1)$ remains dominant even at high temperatures. 
%due to the influence of the Lorentz force.

In addition, we find that the change in the magnetic field induces a Hall conductivity (in both the low and high temperature limits) upto an overall factor that determines the position of the Misner string. Furthermore, we find that the Ohmic conductivity exceeds the Hall conductivity for both $U(1)$ and thermally produced charge carriers in both low and high temperature regimes. This is because the Hall conductivity primarily arises due to the magnetic field, and we are working in the small magnetic field limit $B<<1$ and in particular for $B<E$.

Next, we investigate the Ohmic and Hall conductivities at finite magnetic field ($B>1$) at both low ($T\sim T_{min}$) and high temperature ($T>>T_{min}$) regimes. Our analysis reveals that the frame-dragging effects are relatively smaller at finite magnetic field. At low temperatures ($T\sim T_{min}$), we find the Hall conductivity due to the $U(1)$ charge carriers is nearly constant, regardless of the location of the Misner string \cite{A.~O'Bannon}, \cite{Hartnoll:2007ai}. On the other hand, near the Misner string, the thermal contribution to the Hall conductivity varies quadratically with temperature, and it nearly vanishes at locations far from the Misner string. However, the $U(1)$ contribution to the Hall conductivity dominates over the thermal contribution at low temperatures and in a finite magnetic field.  

Moreover, the Ohmic conductivity resulting from the $U(1)$ charge carriers and those produced thermally, decreases as we approach the minimum temperature, regardless of the location of the Misner string. Our findings show that at low temperatures and finite magnetic field ($B>1>E$),  the contributions due to $U(1)$ charge scale as $\sim T_{\text{min}}^2$. This observation is consistent with the existing literature \cite{A.~O'Bannon}, where they found a zero Ohmic conductivity in the limit $T\rightarrow0$. Notably, the contribution due to thermal charge carriers to the Ohmic conductivity dominates over the $U(1)$ contribution because in the regime where $B > E$, the $U(1)$ charge carriers are primarily drifted in response to the magnetic field. Consequently, the Hall conductivity associated with $U(1)$ charge carriers is greater than the Ohmic conductivity stemming due to the same charge carriers. In contrast, the thermal Ohmic conductivity dominates over the thermal Hall conductivity, as frame-dragging effects are relatively smaller at finite magnetic fields.

At higher temperatures ($T>>T_{min}$) and in the presence of a finite magnetic field ($B>1$), the effects due to frame-dragging are negligible. This results in similar behavior for both Ohmic and Hall conductivity, both near and far from the Misner string. At higher temperatures, the thermal charge carriers become more significant than the $U(1)$ charge carriers, leading to higher thermal Ohmic conductivity, which further saturates at unity. On the other hand, the Hall conductivity associated with $U(1)$ charge carriers decreases with increasing temperature and becomes more dominant than the thermal contribution, due to the negligible frame dragging. Furthermore, the Hall conductivity due to $U(1)$ charge carriers dominates over the Ohmic conductivity associated with the same charge carriers. Conversely, the Ohmic conductivity due to thermal charge carriers exceeds the corresponding Hall conductivity.  

Finally, we observe that, a change in the magnetic field induces a Hall conductivity similar to what has been observed in the limit of a small magnetic field. This phenomenon occurs at both the low and high temperature limits and differs by an overall factor that depends on the location of the Misner string. 

Below, we present a list of interesting future projects that are worth pursuing.

$\bullet$ The authors in \cite{Karch:2008uy} study the stress-energy tensor of flavor fields for the planar $AdS_5$ black hole by using the probe $D$-brane approach. In particular, they obtained the Ohmic and Hall conductivities \cite{Karch:2007pd}, \cite{A.~O'Bannon} and identified infrared divergences in the stress-energy tensor, arising due to the constant rate of energy and momentum loss of the flavor fields. It would be an interesting direction to explore the stress-energy tensor of the flavor field associated with TN-$AdS_4$ black holes and examine the frame-dragging effects on it. 

$\bullet$ It would be interesting to explore the thermoelectric transport \cite{Donos:2014cya} of TN-Ad$S_4$ black holes. To be specific, to study momentum relaxation, anisotropy, and back-reaction in the context of these black holes, as well as examine the effects of frame-dragging on those entities.

%$\bullet$ Investigating the thermoelectric components within the boundary theory of TN-$AdS_4$ black holes presents an intriguing research opportunity. However, several challenges must be addressed. Notably, there is no meaningful separation between the contributions of the flavor fields and the adjoint fields to the heat currents. Additionally, when external work is applied to the system, the heat current is not stationary within the probe D-brane method, as discussed by the author in \cite{Karch:2008uy}. In summary, to effectively extract the thermoelectric components, additional elements such as momentum relaxation, anisotropy, or back-reaction are required. 

We hope to address some of these problems in the near future.

\section*{Acknowledgments}
AK and DR are indebted to the authorities of Indian Institute of Technology, Roorkee for their unconditional support towards researches in basic sciences. HR would like to thank the authorities of Saha Institute of Nuclear Physics, Kolkata, for their support. DR acknowledges the Mathematical Research Impact Centric Support (MATRICS) grant (MTR/2023/000005) received from ANRF, India.

\appendix
\section{\(A_t',H',K'\) from conserved charges\label{A,H,K expr}}

In this Appendix, we provide the expressions for $K'(z)$, $A'_t(z)$, and $H'(z)$ in terms of the conserved charge ($b,c$ and $d$) and the component of Taub-NUT space-time metric.
To begin with, we introduce the $DBI$ action as defined in \eqref{DBI indp of A,H}
\begin{align}
    L_{DBI}=-\Bigg(g_{\theta\theta}\bigg[|g_{tt}|g_{zz}g_{\phi \phi}+|g_{tt}|H^{'2}-g_{\phi \phi}A_t^{'2}-2|g_{t \phi}|A_t'H'+g_{zz}(|g_{t \phi}|^2-E^2)\bigg]\nonumber\\ -A_t'^2 B^2+2 A_t' B E K'+B^2 g_{zz}|g_{tt}|+K'^2 \left(-E^2+|g_{tt}| g_{\phi \phi}+|g_{t\phi}|^2\right)\Bigg)^{1/2}.\label{LDBI app}
\end{align}

Since $L_{DBI}$ is only dependent on $z$ derivatives of $A_t,H,K$ but not on  $A_t,H,K$, therefore we get three constants of motion  for \(L_{DBI}\) 
\begin{align}
b=&\hspace{1mm}\frac{\partial L_{DBI}}{\partial K'}=-\frac{A'_t B E+K' \left(|g_{tt} |g_{\phi \phi}+|g_{t\phi}|^2-E^2\right)}{L_{DBI}},\label{bL}\\
c=&\hspace{1mm}\frac{\partial L_{DBI}}{\partial A_t'}=\frac{A'_t \left(B^2+g_{\theta\theta}g_{\phi \phi}\right)-B. E. K'+g_{\theta\theta}|g_{t\phi}|H'}{L_{DBI}},\\
d=&\hspace{1mm}\frac{\partial L_{DBI}}{\partial H'}  =\frac{{g_{\theta \theta}}(A_t'|g_{t\phi}|-|g_{tt}|H')}{L_{DBI}}.\label{dL}
\end{align}

 We invert \eqref{bL}-\eqref{dL} to get following expressions for $A_t',H',K'$
\begin{align}
    A^{'2}_t=&\frac{g_{\theta \theta} g_{zz} \left[(c |g_{tt}|+d| g_{t\phi }|) \left(|g_{tt}| g_{\phi \phi }-E^2+g_{t\phi }^2\right)-B b E |g_{tt}|\right]^2}{G},\label{At}\\
    H^{'2}=&\frac{g_{zz} \left[g_{\theta \theta} (c| g_{t\phi }|-d g_{\phi \phi}) \left(|g_{tt}| g_{\phi \phi}-E^2+g_{t\phi }^2\right)-B^2 d \left(|g_{tt}| g_{\phi \phi }+g_{t\phi}^2\right)-B b E g_{\theta \theta} |g_{t\phi }|\right]^2}{g_{\theta\theta}G},\label{Hp}\\
    K^{'2}=&-\frac{g_{\theta \theta } g_{zz} \left[B^2 b |g_{tt}|+B E (c| g_{tt}|+d|g_{t\phi }|)+b g_{\theta \theta} \left(|g_{tt}| g_{\phi \phi }+g_{t\phi }^2\right)\right]^2}{G},\label{Kp}
\end{align}
where we define
\begin{align} &
G=\left(|g_{tt}| g_{\phi \phi}+g_{t\phi }^2\right) \bigg[B^2 \left(g_{\theta \theta} |g_{tt}| \left(b^2-\left(|g_{tt}| g_{\phi \phi }+g_{t\phi }^2\right)\right)+d^2 \left(|g_{tt}| g_{\phi \phi }+g_{t\phi}^2\right)\right) \nonumber\\& +2 B b E g_{\theta \theta} (c |g_{tt}|+d |g_{t\phi }|)+g_{\theta \theta} \bigg(b^2 g_{\theta \theta } \left(|g_{tt}| g_{\phi \phi }+g_{t\phi }^2\right)-c^2 |g_{tt}| \left(-E^2+|g_{tt}| g_{\phi \phi }+g_{t\phi}^2\right)\nonumber\\&-2 cd |g_{t\phi }| \left(-E^2+|g_{tt}| g_{\phi \phi}+g_{t\phi}^2\right)+\left(E^2-|g_{tt}| g_{\phi \ph}-g_{t\phi}^2\right) \left( g_{\theta \theta} \left(|g_{tt}| g_{\phi \phi}+g_{t\phi}^2\right)-d^2 g_{\phi \phi}\right)\bigg)\bigg]  .
\end{align}

Substituting \eqref{At}-\eqref{Kp} into $L_{DBI}$ \eqref{LDBI app}, we obtain the following on-shell Lagrangian 
\begin{equation}
    L_{DBI}^{on-shell}=-\frac{(L^4+n^{2}z^2)L^2\sin\theta}{z^4} {\frac{\xi}{\sqrt{\xi \chi-\alpha^2}}},
\end{equation}
where  $\xi,\chi,\alpha$ are defined in \eqref{c}-\eqref{a} in Section \ref{hall conductivity}. 

\section{Conductivities expression in terms of temperature\label{exect condT} }

In this Appendix, we provide the expressions of Ohmic and Hall conductivity \eqref{condu1dc}-\eqref{hall conductivity thermal} in terms of temperature. We start with the following expressions of conductivities \eqref{condu1dc}-\eqref{hall conductivity thermal}, given below %as introduced in Section \ref{hall conductivity} 
{\begin{align} &
    \sigma_{\phi\phi ,U(1)}^{2}=\frac{ J_t^2 \left(L^4+n^2 z_h^2\right)^2}{\left(B^2 z_h^4 \csc^2\th+\left(L^4+n^2 z_h^2\right)^2\right)^2}\left[z_h^4+\frac{2 ez_h  \tilde\Omega  \left(L^4+n^2 z_h^2\right)^2 }{J_t \sin\th}\sqrt{1+\frac{z_h^4 (J_t^2+B^2 \csc^2\th)}{\left(z_h^2 n^2+L^4\right)^2}}\right.\nonumber\\&\hspace{18mm}\left.+\frac{3 \tilde \Omega ^2 e^2z_h^2\left(L^4+n^2 z_h^2\right)^2}{\sin^2\th }-\frac{4 e^2 J_t L^6 z_h^{10}  \left(\left(L^4+n^2 z_h^2\right)^2-B^2 z_h^4 \csc^2\th\right)\csc^2\th}{\left(3 L^4+z_h^2 \left(L^2+3 n^2\right)\right) \left(B^2 z_h^4 \csc^2\th+\left(L^4+n^2 z_h^2\right)^2\right)^2}\right],\label{condu1dcA}
\end{align}

\begin{align} &
    \sigma_{\phi\phi ,\text{thermal}}^{2}=\frac{ \left(L^4+n^2 z_{h}^2\right)^2}{B^2 z_{h}^4 \csc^2\th+\left(L^4+n^2 z_{h}^2\right)^2}\left[\frac{4 B^2 E^2L^6 z_{h}^{10} \csc^4\th }{\left(3 L^4+z_h^2 \left(L^2+3 n^2\right)\right) \left(B^2 z_h^4 \csc^2\th+\left(L^4+n^2 z_h^2\right)^2\right)^2}\right.\nonumber\\&\left.\hspace{22mm}+1+\frac{\tilde\Omega ^2e^2  \left(L^4+n^2 z_h^2\right)^2}{z_h^2\sin^2\th}\right],\label{condthdcA}\end{align} }

   { \begin{align}
\sigma_{\th\ph,U(1)}^2=&\frac{J_t^2 z_h^8 B^2}{\left(B^2\csc^2\th+\left(z_h^2 n^2+L^4\right)^2\right)^2}\left[1
 +\frac{2  e z_h  \tilde\Omega   \left(L^4+n^2 z_h^2\right)^2 }{ J_tz_h^4\sin\th}\sqrt{1+\frac{z_h^4 \left(B^2 \csc^2\th+J_t^2\right)}{\left(L^4+n^2 z_h^2\right)^2}} \right. \nonumber\\&  \left.+\frac{3\tilde \Omega ^2 e^2\left(z_h^2 n^2+L^4\right)^2}{z_h^2\sin^2\th}-\frac{8 z_h^{6}  e^2 L^6 \csc^2\th \left(z_h^2 n^2+L^4\right)^2}{\left(z_h^2 \left(L^2+3 n^2\right)+3 L^4\right) \left(z_h^4 B^2 \csc^2\th+\left(z_h^2 n^2+L^4\right)^2\right)^2}\right],\label{hallu1A} \\
 \s^2_{\th\ph,\text{thermal}}=&\frac{B^2 \tilde\Omega ^2 e^2 z_h^2 \left(z_h^2 n^2+L^4\right)^2}{\sin^2\th\left(z_h^4 B^2 \csc^2\th+\left(z_h^2 n^2+L^4\right)^2\right)}.\label{hall conductivity thermalA}
  \end{align}}
  
  By inverting \eqref{htn} and using definition of $T_{min}$, we can express $z_h$ in terms of $T$ and $T_{min}$ as
\begin{align}
    z_h=\frac{3 }{2 \pi   T \pm 2 \pi \sqrt{ T^2-T_{min}^2}}.\hspace{1mm}\label{zhT}
\end{align}

Substituting $z_h$ (\ref{zhT}) in $\tilde\Omega$ (\ref{omega at z*}), we obtain the following expression of the (frame-dragging) angular velocity,
\begin{align}
   \big| \tilde\Omega\big|
   =&\hspace{1mm}\frac{162  n\csc\th(\cos\th+\beta)}{\left(\left(4 \pi ^2  \left(2 T \tilde{T}-T_{min }^2+2 T^2\right)+9 n^2\right)^2+81B^2\csc^2\th\right)} .\label{angb}
 \end{align}
 
 Finally by substituting $z_h$ \eqref{zhT}, and $\tilde\Omega$ \eqref{angb} in Ohmic and Hall conductivities \eqref{condu1dcA}-\eqref{hall conductivity thermalA}, we obtain the Hall and Ohmic conductivities in terms of temperature $(T)$, given below
{\begin{align}&
    \sigma_{\phi\phi,U(1)}=\frac{9 {J_t}  \left(4 \pi ^2(T+\tilde T)^2+9 n^2\right)}{\left(81 B^2 \csc^2\th+\left(4 \pi ^2  (T+\tilde T)^2+9 n^2\right)^2\right)}\left[\frac{972 E^2 n^2  (\beta +\cos (\theta ))^2 \left(4 \pi ^2  (T+\tilde T)^2+9 n^2\right)^2}{\sin ^4(\theta )\left(81 B^2 \csc^2\th+\left(4 \pi ^2  (T+\tilde T)^2+9 n^2\right)^2\right)^2}\right.\nonumber \\ &\hspace{16mm}+1-\frac{1296 \pi ^2 E^2 \csc^2\th (T+\tilde T)^2 \left(\left(9 n^2+4 \pi ^2 (T+\tilde T)^2\right)^2-81 B^2 \csc^2\th\right)}{\left(27 n^2+12 \pi ^2 (T+\tilde T)^2+9\right) \left(81 B^2 \csc^2\th+\left(9 n^2+4 \pi ^2 (T+\tilde T)^2\right)^2\right)^2}\nonumber \\&\hspace{16mm}+\left.\frac{4 E n (\beta +\cos (\theta )) \left(9 n^2+4 \pi ^2 (T+\tilde T)^2\right)^2 \sqrt{\frac{81 \left(B^2 \csc^2\th+J_t^2\right)}{\left(9 n^2+4 \pi ^2 (T+\tilde T)^2\right)^2}+1}}{\sin^2\th J_t \left(81 B^2 \csc^2\th+\left(9 n^2+4 \pi ^2 (T+\tilde T)^2\right)^2\right)}\right]^{1/2},\label{condu1dcT}
\end{align}}
{\begin{align}
    &\sigma_{\ph\phi,\text{thermal}}=\frac{  \left(4 \pi ^2 L^4 (T+\tilde T)^2+9 n^2\right)}{\left(81 B^2 \csc^2\th+\left(4 \pi ^2 L^4 (T+\tilde T)^2+9 n^2\right)^2\right)^{1/2}}\Bigg[1+\nonumber\\&\hspace{26mm}\left.\frac{324 E^2 n^2  (\beta +\cos (\theta ))^2 \left(4 \pi ^2 L^4 (T+\tilde T)^2+9 n^2\right)^2}{\sin^4\th\left(81 B^2 \csc^2\th+\left(4 \pi ^2 L^4 (T+\tilde T)^2+9 n^2\right)^2\right)^2}+\right.\nonumber \\ &\hspace{26mm}\left.\frac{104976 \pi ^2 B^2 E^2 L^6 \csc^4\th (T+\tilde T)^2}{\left(12 \pi ^2 L^4 (T+\tilde T)^2+9 L^2+27 n^2\right) \left(81 B^2 \csc^2\th+\left(4 \pi ^2 L^4 (T+\tilde T)^2+9 n^2\right)^2\right)^2}  \right]^{1/2},\label{condthT}
\end{align}}
{\begin{align}&
    \sigma_{\th\ph,U(1)}=\frac{81 B J_t}{\left(81 B^2 \csc^2\th+\left(9 n^2+4 \pi ^2 (T+\tilde T)^2\right)^2\right)}\left[\frac{972 E^2 n^2 (\beta +\cos (\theta ))^2 \left(9 n^2+4 \pi ^2 (T+\tilde T)^2\right)^2}{\sin ^4(\theta ) \left(81 B^2 \csc^2\th+\left(9 n^2+4 \pi ^2 (T+\tilde T)^2\right)^2\right)^2}\right.\nonumber \\ & \hspace{16mm}+1-\frac{2592 \pi ^2 E^2 \csc^2\th (T+\tilde T)^2 \left(9 n^2+4 \pi ^2 (T+\tilde T)^2\right)^2}{\left(27 n^2+12 \pi ^2 (T+\tilde T)^2+9\right) \left(81 B^2 \csc^2\th+\left(9 n^2+4 \pi ^2 (T+\tilde T)^2\right)^2\right)^2}\nonumber \\&\hspace{16mm}\left.+\frac{4 E n  (\beta +\cos (\theta )) \left(9 n^2+4 \pi ^2 (T+\tilde T)^2\right)^2 \sqrt{\frac{81 \left(B^2 \csc^2\th+J_t^2\right)}{\left(9 n^2+4 \pi ^2 (T+\tilde T)^2\right)^2}+1}}{J_t \sin^2\th\left(81 B^2 \csc^2\th+\left(9 n^2+4 \pi ^2 (T+\tilde T)^2\right)^2\right)}\right]^{1/2} ,\label{hallu1T}
\end{align}}\\
\begin{align}
\sigma_{\th\phi,\text{thermal}}=\frac{162 B E n \csc^2\th (\beta +\cos (\theta )) \left(9 n^2+4 \pi ^2 (T+\tilde T)^2\right)}{\left(81 B^2 \csc^2\th+\left(9 n^2+4 \pi ^2 (T+\tilde T)^2\right)^2\right)^{3/2}}.\label{hallthermal T}
\end{align}
\section{Conductivity analysis for \(\mathbf{\beta=0,-1}\) at low magnetic  field  \label{Appc}}

It is worth noting that one can also analyze the Hall conductivity $(\s_{\th\ph})$ and the Ohmic conductivity $(\s_{\ph\ph})$ for the other two cases, i.e., $\b=0$ and $\b=-1$, which will further verify our claims that frame dragging effects are dominated as we approach the Misner string.
In the case $\b=0$, the Misner string is located\footnote{In the present analysis, we study the case where  location of the Misner string for $\beta=0$ is at $\theta=\pi$.} at $\th=0$ and $\th=\pi$. However, for $\b=-1$, the Misner string is located only at $\th=\pi$.

To start our discussion for $\b=0,-1$, we first re-write the angular velocity (\ref{omega at z*}) for $\b=0,-1$ in terms of the Hawking temperature (\ref{htn}) and the minimum temperature ($T_{min}$)  similar to \eqref{angb1}. The angular velocity can be expressed for $\b=0,-1$ can be expressed  respectively as
\begin{align}
   \big| \tilde\Omega\big|_{\b=0}
   =&\hspace{1mm}\frac{162  n }{\left(\left(4 \pi ^2  \left(2 T \tilde{T}-T_{min }^2+2 T^2\right)+9 n^2\right)^2+81B^2\csc^2\th\right)}\cot \left({\theta }{}\right),\label{angb0}
 \end{align}
 \begin{align}
   \big| \tilde\Omega\big|_{\b=-1}
   =&\hspace{1mm}\frac{162  n }{\left(\left(4 \pi ^2  \left(2 T \tilde{T}-T_{min }^2+2 T^2\right)+9 n^2\right)^2+81B^2\csc^2\th\right)}\tan \left(\frac{\theta }{2}\right).\label{angbn1}
 \end{align}
\subsection{Ohmic and Hall conductivities at low temperature regime}
 The angular velocity (\ref{angb0}) in the low temperature $(T\sim T_{min})$ and in the small magnetic field limit ($B<<1$)  can be approximated as,
\begin{align}
    |\tilde\Omega|_{\b=0}=\frac{81 n \cot \left({\theta }{}\right)}{8 \pi ^4  T_{\min }^4}\left(1-\frac{81B^2\csc^2\th}{16\pi^4T_{min}^4}\right)+O(\sqrt{T-T_{min}}).\label{amb0}
\end{align}

To proceed further, we express the Hall and Ohmic conductivities  (\ref{condu1dcT})-(\ref{hallthermal T}) in the regions near and far from the Misner string for $\beta=0$. In the low temperature regime ($T\sim T_{min}$), we find the following expressions for the Hall conductivity 
\begin{align}
 &   \s_{\th\phi,U(1)}\bigg|^{fM}_{T \sim T_{min}}=\frac{81BJ_t}{16\pi^4T_{min}^4}+O\left(\sqrt{T-T_{min}}\right)\label{HallUafmb0},\\&\s_{\th\ph,U(1)}\bigg|^{nM}_{T \sim T_{min}}=\frac{729 \sqrt{3} B E n J_t\cot\th}{32 \pi ^6\sin\th   T_{min}^6}+O\left(\sqrt{T-T_{min}}\right)\label{HallU1nMb0},%\\&\s_{\th\phi,\text{thermal}}\bigg|^{fM}_{T \sim T_{min}}=\frac{81 B E n }{16 \pi ^4 T_{\min }^4}+O\left(\sqrt{T-T_{min}}\right),
 \\&\s_{\th\phi,\text{thermal}}\bigg|^{fM}_{T \sim T_{min}}=O\left(\sqrt{T-T_{min}}\right),\label{hallthfMb0}\\&\s_{\th\phi,\text{thermal}}\bigg|^{nM}_{T \sim T_{min}}=\frac{81 B E n\cot\th }{8 \pi ^4\sin\th T_{\min }^4}+O\left(\sqrt{T-T_{min}}\right).\label{hallthnMb0}
\end{align}
where $0<<\th<\pi$ measures the proximity to the Misner string. Below we plot the Hall conductivities against the temperature for both near and far from the Misner string. 
\begin{figure}[H]
     \centering
     \begin{subfigure}[b]{0.495\textwidth}
         \centering
         \includegraphics[width=\textwidth]{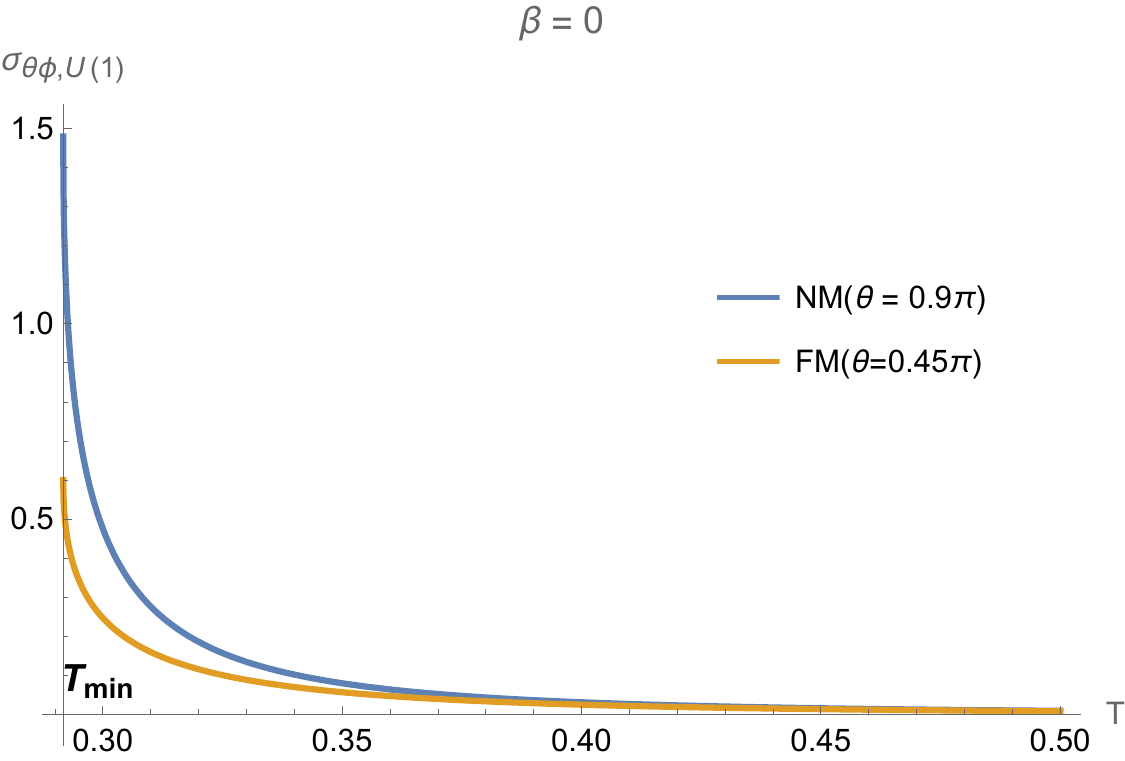}
         \caption{$\s_{\th\ph,U(1)}$ vs $T$ at $B=0.01$ }\label{fig13a}
              \end{subfigure}
              \hfill
     \begin{subfigure}[b]{0.495\textwidth}
         \centering
         \includegraphics[width=\textwidth]{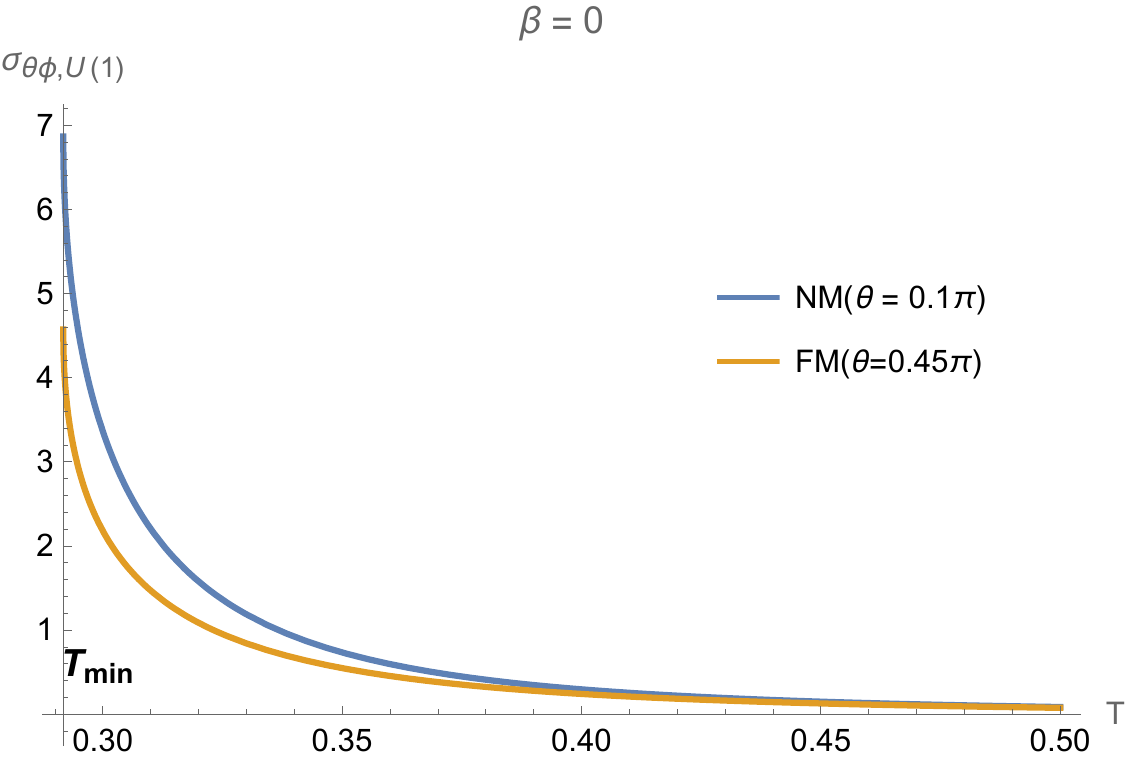}
         \caption{$\s_{\th\ph ,U(1)}$ vs $T$ at $B=0.1$}\label{fig13b}
             \end{subfigure}
        
             \caption{$\s_{\th\ph,U(1)}$ vs $T$ plot for $\b=0$, $E=0.1$ and $J_t=10$ at small $B.$}
        \label{figure13}
\end{figure}
\begin{figure}[H]
     \centering
     \begin{subfigure}[b]{0.495\textwidth}
         \centering
         \includegraphics[width=\textwidth]{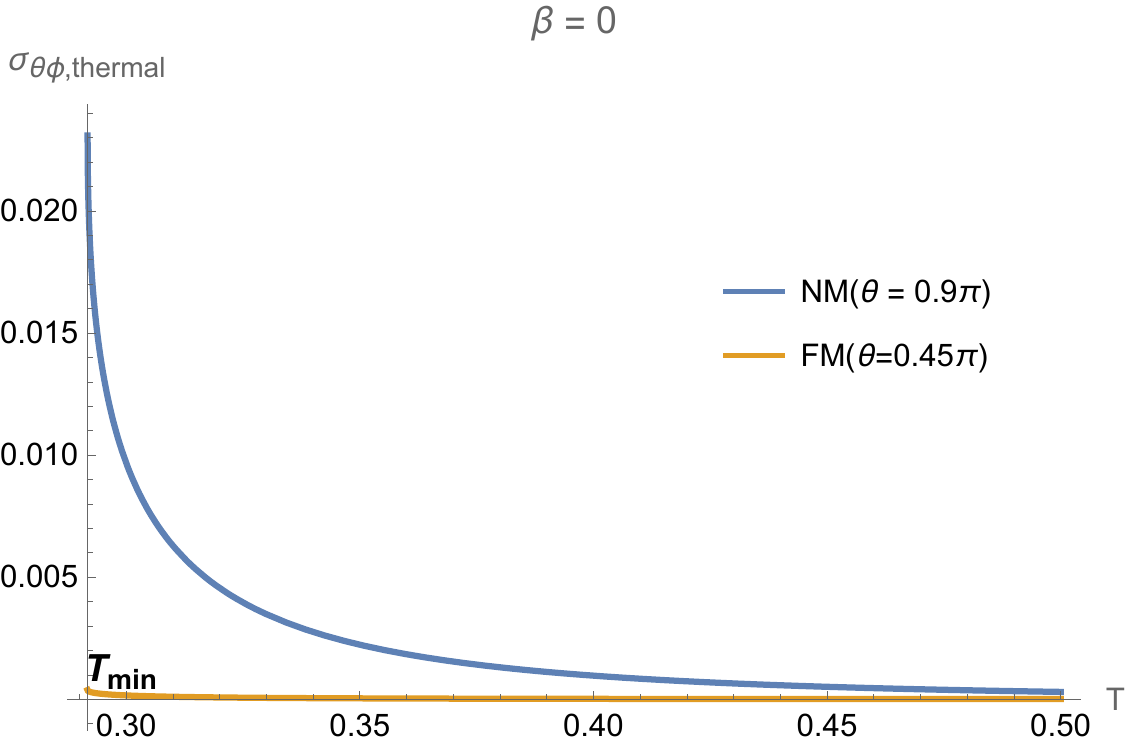}
         \caption{$\s_{\th\ph,\text{thermal}}$ vs $T$ at $B=0.01$ }\label{fig14a}
              \end{subfigure}
              \hfill
     \begin{subfigure}[b]{0.495\textwidth}
         \centering
         \includegraphics[width=\textwidth]{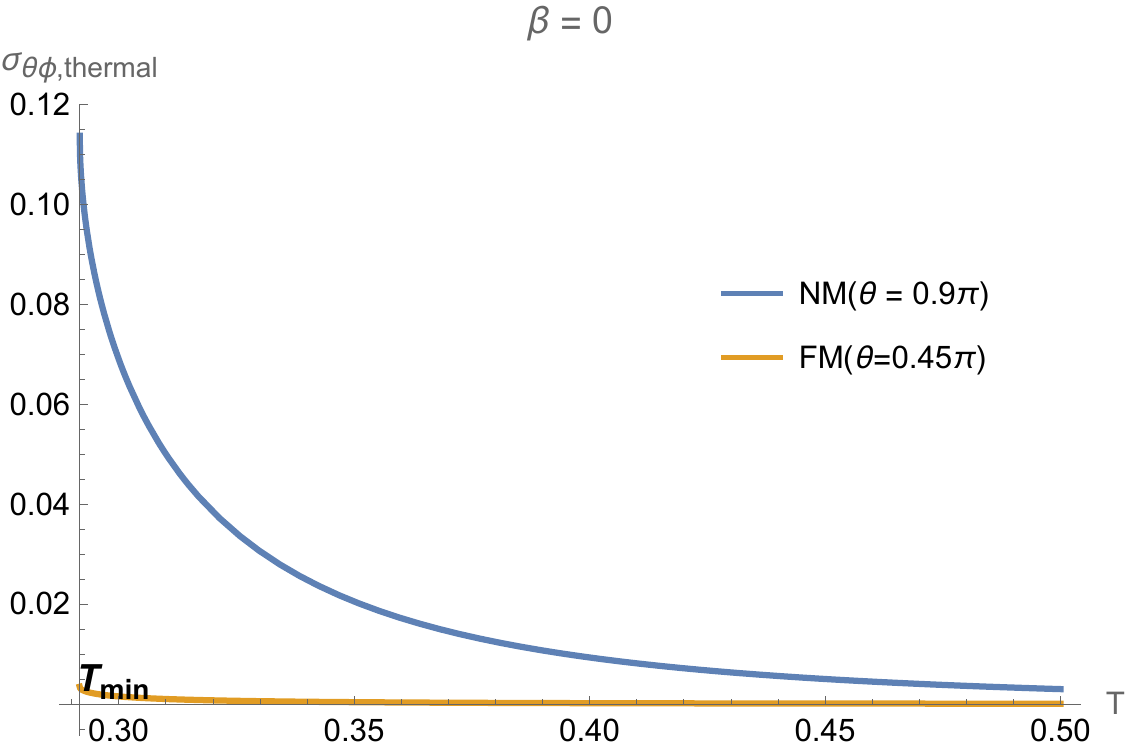}
         \caption{$\s_{\th\ph ,\text{thermal}}$ vs $T$ at $B=0.1$}\label{fig14b}
             \end{subfigure}
        
             \caption{$\s_{\th\ph,\text{Thermal}}$ vs $T$ plot for $\b=0$, $E=0.1$ and $J_t=10$ at small $B$.}
        \label{figure14}
\end{figure}

The Ohmic conductivity for \(\b=0\) can be expressed as
\begin{align}
   & \s_{\phi\ph,U(1)}\bigg|^{fM}_{T \sim T_{min}}=\frac{9J_t}{4\pi^2T_{min}^2}\left(1-\frac{81B^2 \csc^2\th }{16 \pi ^4 T^2_{min}}\right)+O\left(\sqrt{T-T_{min}}\right),\\&\s_{\phi\ph,U(1)}\bigg|^{nM}_{T \sim T_{min}}=\frac{81 \sqrt{3} E n J_t \cot(\th)}{4 \pi ^4 \sin\th T_{\min }^4}\left(1-\frac{81B^2 \csc^2\th }{8 \pi ^4 T^2_{min}}\right)+O\left(\sqrt{T-T_{min}}\right), \\& \s_{\phi\ph,\text{thermal}}\bigg|^{fM}_{T \sim T_{min}}=1-\frac{81B^2\csc^2\th }{32  \pi ^4  T_{min}^4},\\&\s_{\phi\ph,\text{thermal}}\bigg|^{nM}_{T \sim T_{min}}=\frac{9 En\cot(\th)}{2\pi ^2 \sin\th T_{min}^2}\left(1-\frac{243B^2\csc^2\th }{32  \pi ^4  T_{min}^4}\right)+O\left(\sqrt{T-T_{min}}\right).
\end{align}
To clarify further, we plot these Ohmic conductivities against the temperature for both near and far from the Misner string, as shown below.
\begin{figure}[H]
     \centering
     \begin{subfigure}[b]{0.495\textwidth}
         \centering
         \includegraphics[width=\textwidth]{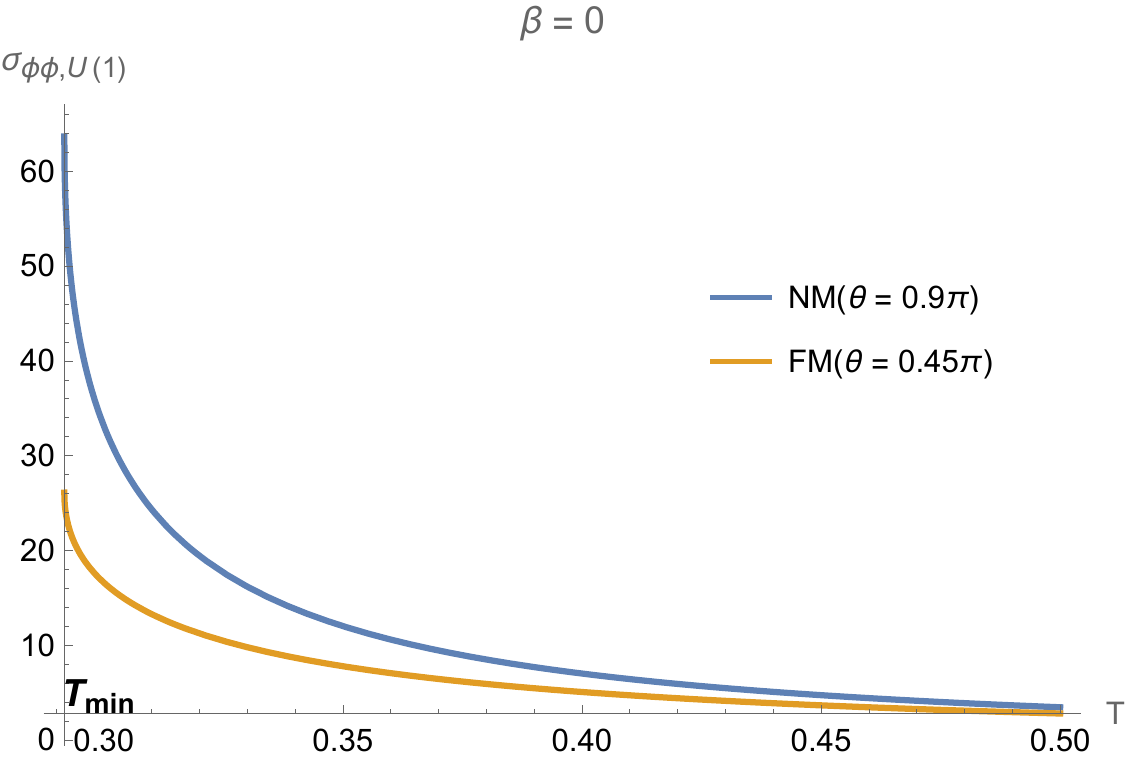}
         \caption{$\s_{\ph \ph,U(1)}$ vs $T$ at $B=0.01$ }\label{fig11a}
              \end{subfigure}
              \hfill
     \begin{subfigure}[b]{0.495\textwidth}
         \centering
         \includegraphics[width=\textwidth]{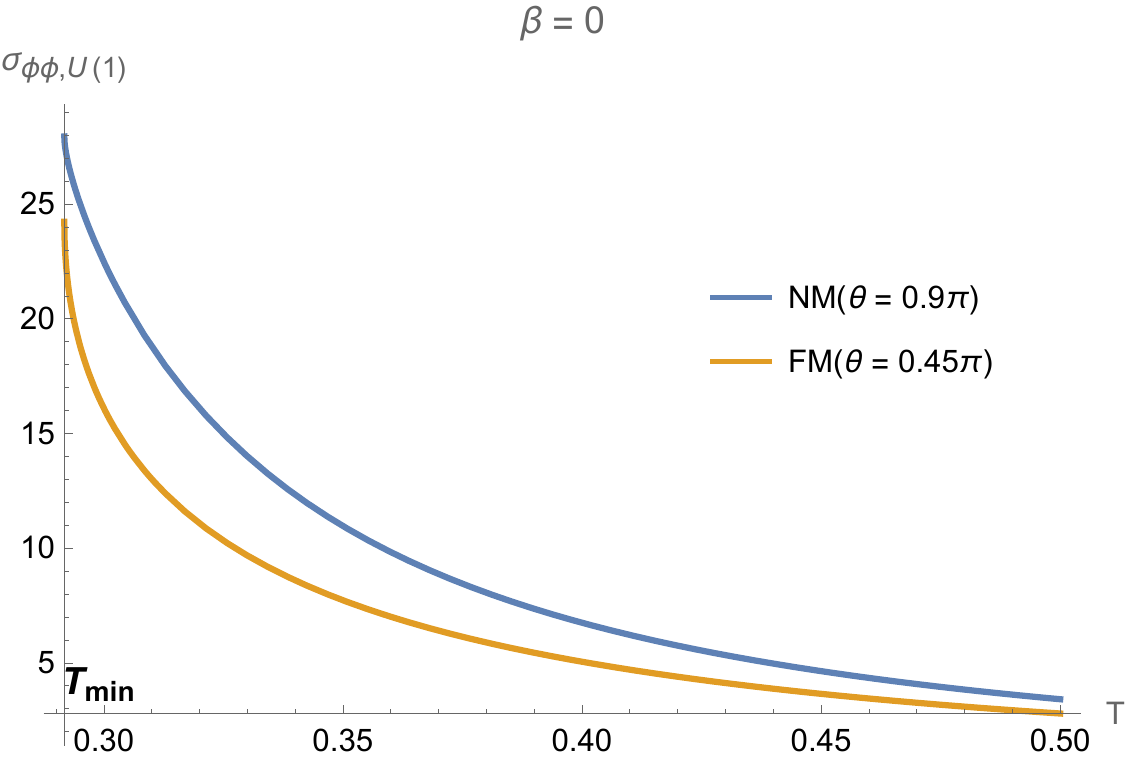}
         \caption{$\s_{\ph \ph ,U(1)}$ vs $T$ at $B=0.1$}\label{fig11b}
             \end{subfigure}
             \caption{$\s_{\phi\ph,U(1)}$ vs $T$ plot for $\b=0$, $E=0.1$ and $J_t=10$ at small $B$. }
        \label{figure11}
\end{figure}
\begin{figure}[H]
     \centering
     \begin{subfigure}[b]{0.495\textwidth}
         \centering
         \includegraphics[width=\textwidth]{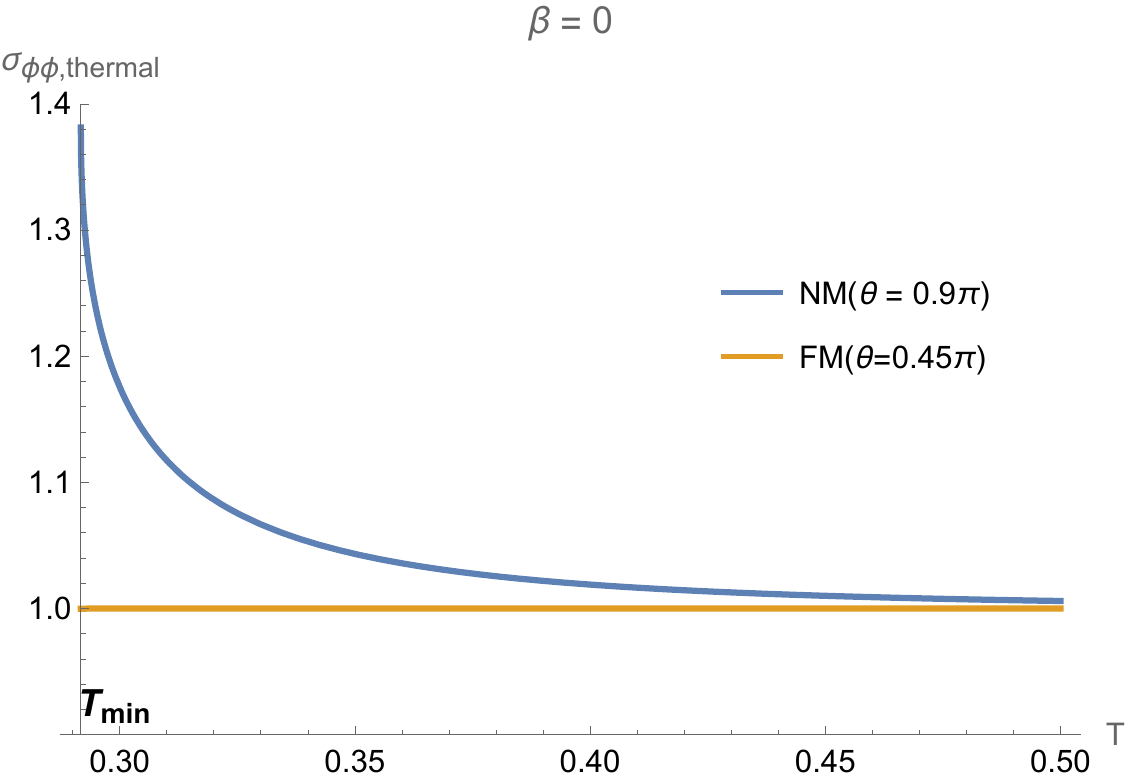}
         \caption{$\s_{\ph\ph,\text{Thermal}}$ vs $T$ at $B=0.01$ }\label{fig12a}
              \end{subfigure}
              \hfill
     \begin{subfigure}[b]{0.495\textwidth}
         \centering
         \includegraphics[width=\textwidth]{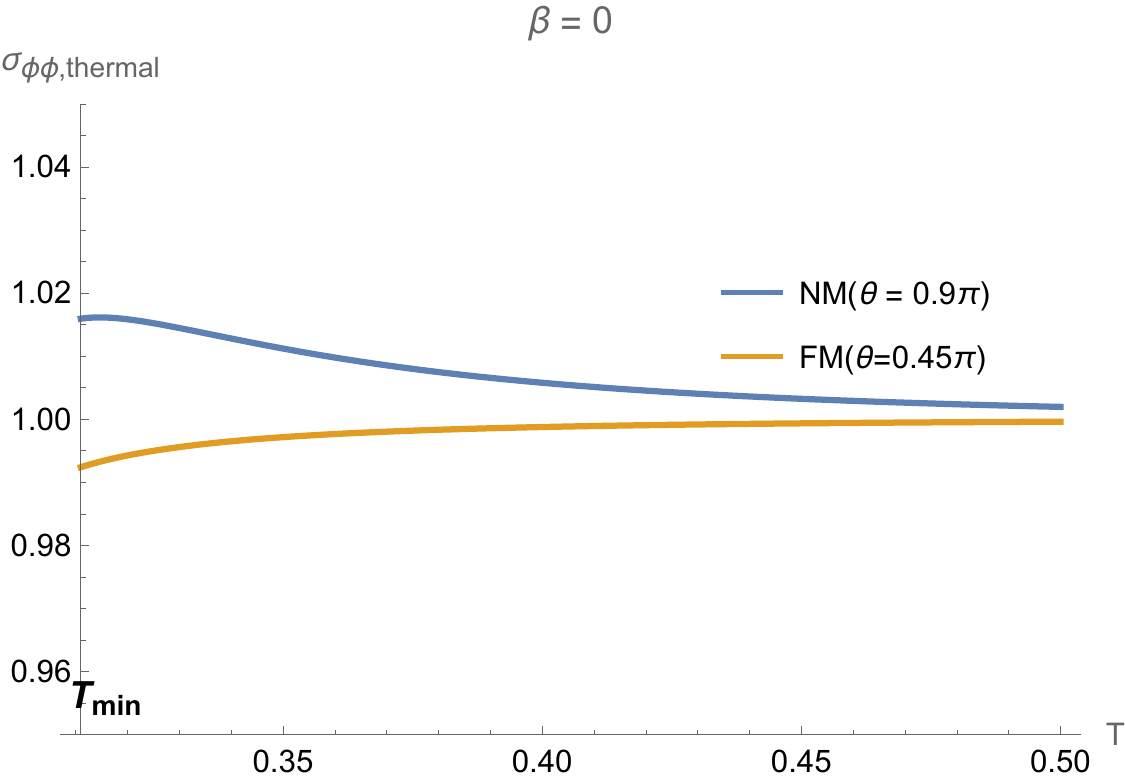}
         \caption{$\s_{\ph\ph,\text{Thermal}}$ vs $T$ at $B=0.1$}\label{fig12b}
             \end{subfigure}
        
             \caption{$\s_{\phi\ph,\text{Thermal}}$ vs $T$ plot for $\b=0$, $E=0.1$ and $J_t=10$ at small $B$.}
        \label{figure12}
\end{figure}

Like $\b=1$, both Hall and Ohmic conductivities due to $U(1)$ carriers and thermally produced charge pairs are higher near the Misner as compared to far from it, and can be shown in following ratios 
\begin{align}
     \frac{\sigma_{\th\ph,U(1)}\bigg|_{T \sim T_{min}}^{nM}}{\sigma_{\th\ph,U(1)}\bigg|_{T \sim T_{min}}^{fM}}  =\frac{18 \sqrt{3} E n\cot\left({\th}{}\right)}{\pi ^2\sin\theta T^2_{min }}>>1\hspace{1mm},\hspace{2mm}\frac{\sigma_{\th\ph,thermal}\bigg|_{T \sim T_{min}}^{nM}}{\sigma_{\th\ph,thermal}\bigg|_{T \sim T_{min}}^{fM}}>>1,\label{ratio1b0}
\end{align}
\begin{align}
     \frac{\sigma_{\ph\ph,U(1)}\bigg|_{T \sim T_{min}}^{nM}}{\sigma_{\ph\ph,U(1)}\bigg|_{T \sim T_{min}}^{fM}}  =\frac{18 \sqrt{3} E n\cot\left({\th}{}\right)}{\pi ^2\sin\theta T^2_{min }}>>1\hspace{1mm},\hspace{2mm}\frac{\sigma_{\ph\ph,thermal}\bigg|_{T \sim T_{min}}^{nM}}{\sigma_{\ph\ph,thermal}\bigg|_{T \sim T_{min}}^{fM}}=\frac{18nE\cot\left({\th}{}\right)}{\pi^2\sin\theta T_{min}^2}>>1.\label{ratio3b0}
\end{align}

Next, we compute the conductivities for the case \(\b=-1\). The angular velocity (\ref{angbn1}) in the low temperature $(T\sim T_{min})$ and small magnetic field limit ($B<<1$) can be approximated as
\begin{align}
    |\tilde\Omega|_{\b=-1}=\frac{81 n \tan\left(\frac{\theta }{2}\right)}{8 \pi ^4  T_{\min }^4}\left(1-\frac{81B^2\csc^2\th}{16\pi^4T_{min}^4}\right)+O(\sqrt{T-T_{min}}).\label{ambn1}
\end{align}

Using (\ref{ambn1}), we can express the Hall and Ohmic conductivity for \(\b=-1\) as follows
\begin{align}
 &   \s_{\th\phi,U(1)}\bigg|^{fM}_{T \sim T_{min}}=\frac{81BJ_t}{16\pi^4T_{min}^4}+O\left(\sqrt{T-T_{min}}\right)\label{HallUafmbn1},\\&\s_{\th\ph,U(1)}\bigg|^{nM}_{T \sim T_{min}}=\frac{729 \sqrt{3} B E n J_t\tan\left(\frac{\th}{2}\right)}{32 \pi ^6\sin\th   T_{min}^6}+O\left(\sqrt{T-T_{min}}\right)\label{HallU1nMbn1},%\\&\s_{\th\phi,\text{thermal}}\bigg|^{fM}_{T \sim T_{min}}=\frac{81 B E n }{16 \pi ^4 T_{\min }^4}+O\left(\sqrt{T-T_{min}}\right),
 \\&\s_{\th\phi,\text{thermal}}\bigg|^{fM}_{T \sim T_{min}}=O\left(\sqrt{T-T_{min}}\right),\label{hallthfMbn1}\\&\s_{\th\phi,\text{thermal}}\bigg|^{nM}_{T \sim T_{min}}=\frac{81 B E n\tan\left(\frac{\th}{2}\right) }{8 \pi ^4\sin\th T_{\min }^4}+O\left(\sqrt{T-T_{min}}\right),\\\label{hallthnMbn1}
   & \s_{\phi\ph,U(1)}\bigg|^{fM}_{T \sim T_{min}}=\frac{9J_t}{4\pi^2T_{min}^2}\left(1-\frac{81B^2 \csc^2\th }{16 \pi ^4 T^2_{min}}\right)+O\left(\sqrt{T-T_{min}}\right),\\&\s_{\phi\ph,U(1)}\bigg|^{nM}_{T \sim T_{min}}=\frac{81 \sqrt{3} E n J_t \tan\left(\frac{\th}{2}\right)}{4 \pi ^4 \sin\th T_{\min }^4}\left(1-\frac{81B^2 \csc^2\th }{8 \pi ^4 T^2_{min}}\right)+O\left(\sqrt{T-T_{min}}\right), \\& \s_{\phi\ph,\text{thermal}}\bigg|^{fM}_{T \sim T_{min}}=1-\frac{81B^2\csc^2\th }{32  \pi ^4  T_{min}^4}+O\left(\sqrt{T-T_{min}}\right),\\&\s_{\phi\ph,\text{thermal}}\bigg|^{nM}_{T \sim T_{min}}=\frac{9 En \tan\left(\frac{\th}{2}\right)}{2\pi ^2 \sin\th T_{min}^2}\left(1-\frac{243B^2\csc^2\th }{32  \pi ^4  T_{min}^4}\right)+O\left(\sqrt{T-T_{min}}\right).
\end{align}

For further clarity, we plot these (Hall and Ohmic) conductivities against the temperature for both points far and close to the Misner string.

\begin{figure}[H]
     \centering
     \begin{subfigure}[b]{0.495\textwidth}
         \centering
         \includegraphics[width=\textwidth]{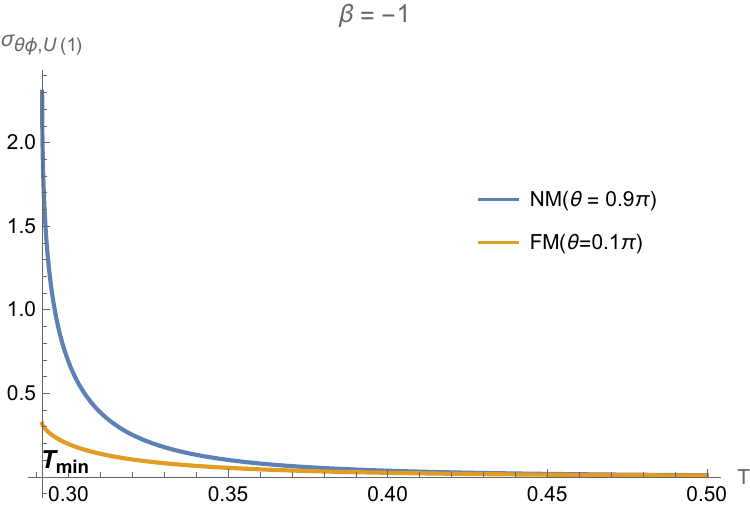}
         \caption{$\s_{\th\ph,U(1)}$ vs $T$ at $B=0.01$ }\label{fig9a}
              \end{subfigure}
              \hfill
     \begin{subfigure}[b]{0.495\textwidth}
         \centering
         \includegraphics[width=\textwidth]{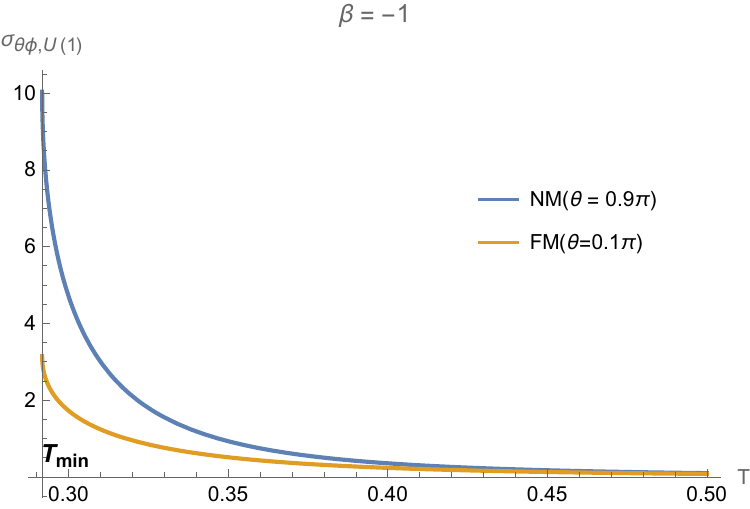}
         \caption{$\s_{\th\ph ,U(1)}$ vs $T$ at $B=0.1$}\label{fig9b}
             \end{subfigure}
        
             \caption{$\s_{\th\ph,U(1)}$ vs $T$ plot for $\b=-1$, $E=0.1$ and $J_t=10$ at small $B$.}
        \label{figure9}
\end{figure}
\begin{figure}[H]
     \centering
     \begin{subfigure}[b]{0.495\textwidth}
         \centering
         \includegraphics[width=\textwidth]{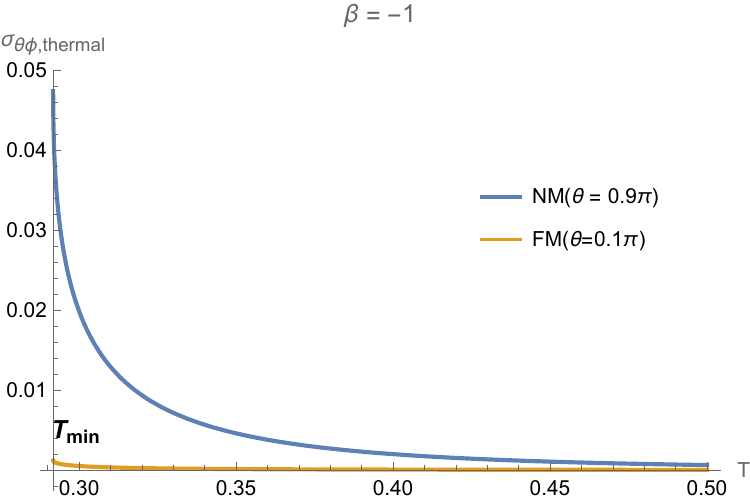}
         \caption{$\s_{\th\ph,\text{thermal}}$ vs $T$ at $B=0.01$ }\label{fig10a}
              \end{subfigure}
              \hfill
     \begin{subfigure}[b]{0.495\textwidth}
         \centering
         \includegraphics[width=\textwidth]{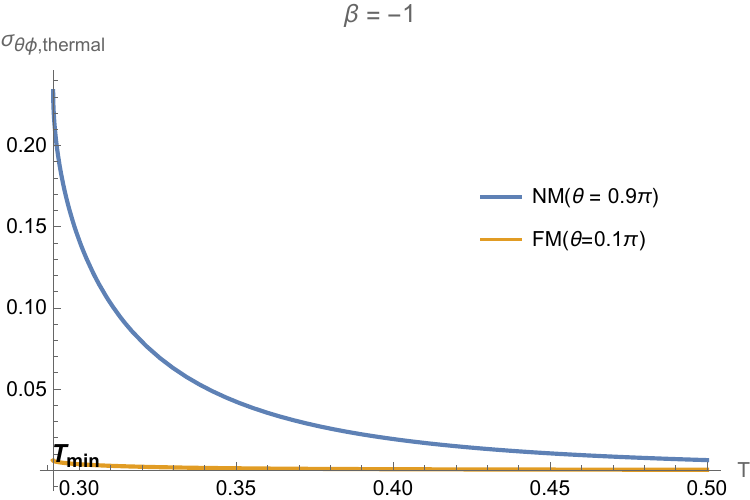}
         \caption{$\s_{\th\ph ,\text{thermal}}$ vs $T$ at $B=0.1$}\label{fig10b}
             \end{subfigure}
        
             \caption{$\s_{\th\ph,\text{Thermal}}$ vs $T$ plot for $\b=-1$, $E=0.1$ and $J_t=10$ at small $B$.}
        \label{figure10}
\end{figure}

\begin{figure}[H]
     \centering
     \begin{subfigure}[b]{0.495\textwidth}
         \centering
         \includegraphics[width=\textwidth]{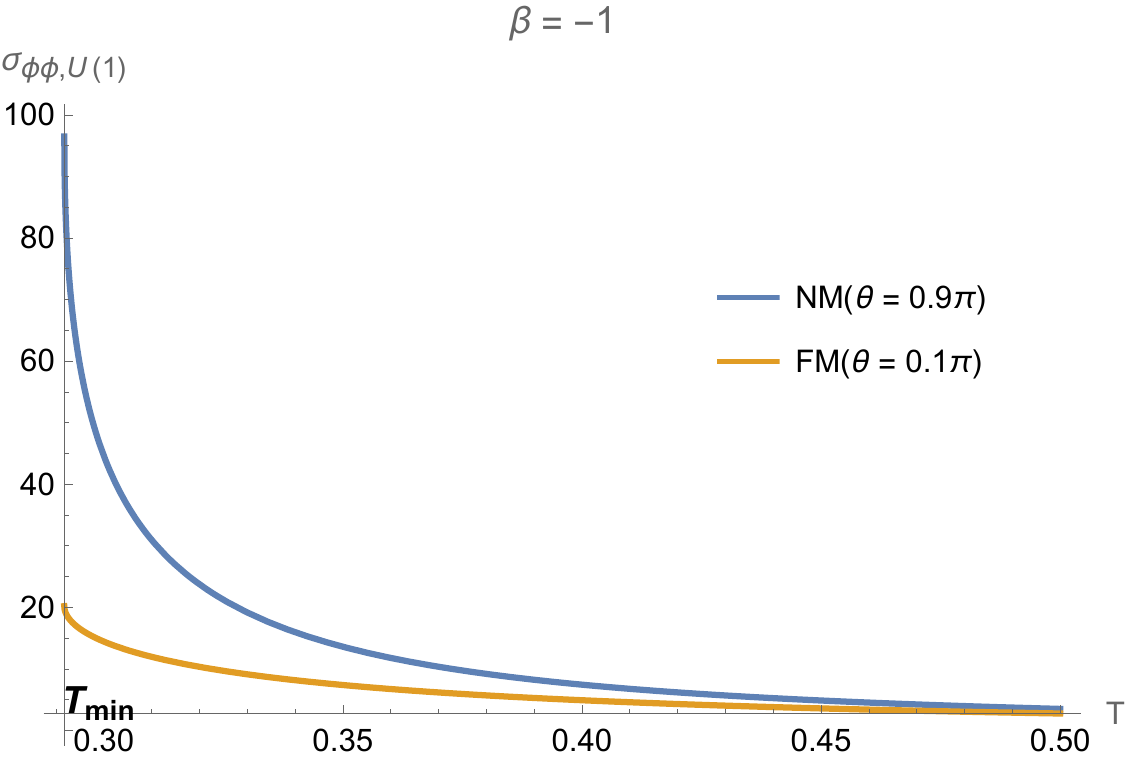}
         \caption{$\s_{\ph \ph,U(1)}$ vs $T $ at $ B=0.01$ }\label{fig19a}
              \end{subfigure}
              \hfill
     \begin{subfigure}[b]{0.495\textwidth}
         \centering
         \includegraphics[width=\textwidth]{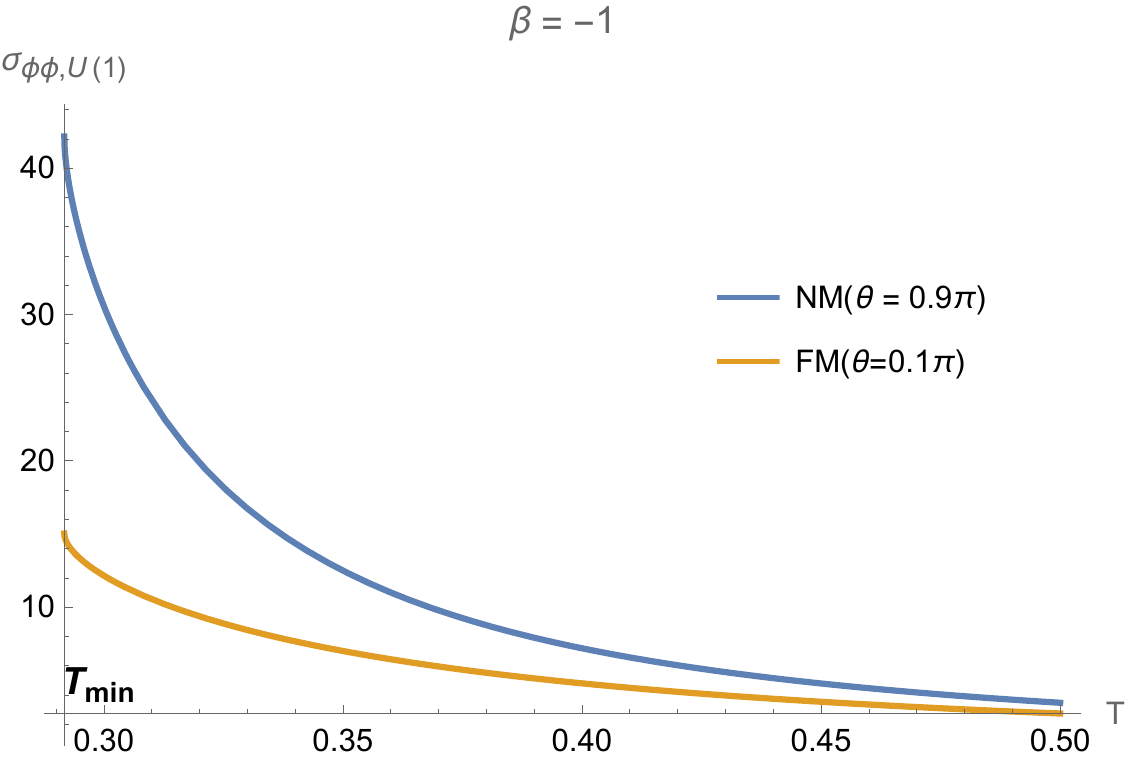}
         \caption{$\s_{\ph \ph ,U(1)}$ vs $T$ at $B=0.1$}\label{fig19b}
             \end{subfigure}
             \caption{$\s_{\phi\ph,U(1)}$ vs $T$ plot for $\b=-1$, $E=0.1$ and $J_t=10$ at small $B$.}
        \label{figure19}
\end{figure}
\begin{figure}[H]
     \centering
     \begin{subfigure}[b]{0.495\textwidth}
         \centering
         \includegraphics[width=\textwidth]{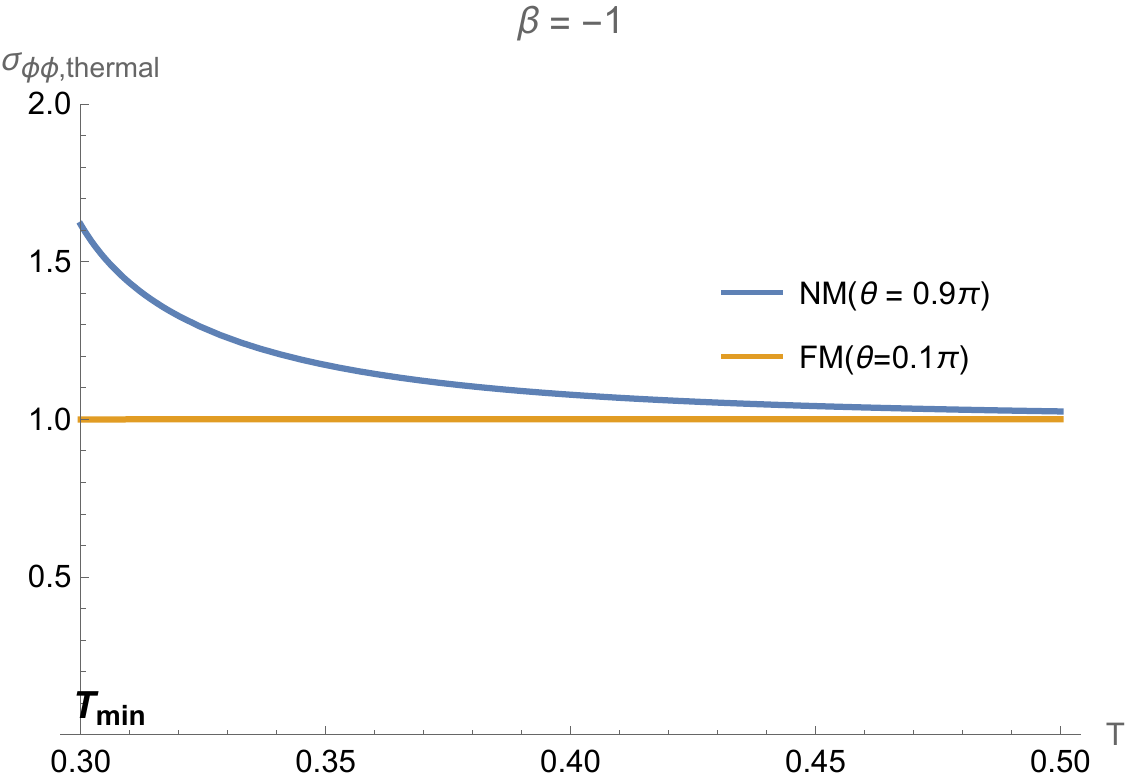}
         \caption{$\s_{\ph\ph,\text{Thermal}}$ vs $T$ at $B=0.01$ }\label{fig20a}
              \end{subfigure}
              \hfill
     \begin{subfigure}[b]{0.495\textwidth}
         \centering
         \includegraphics[width=\textwidth]{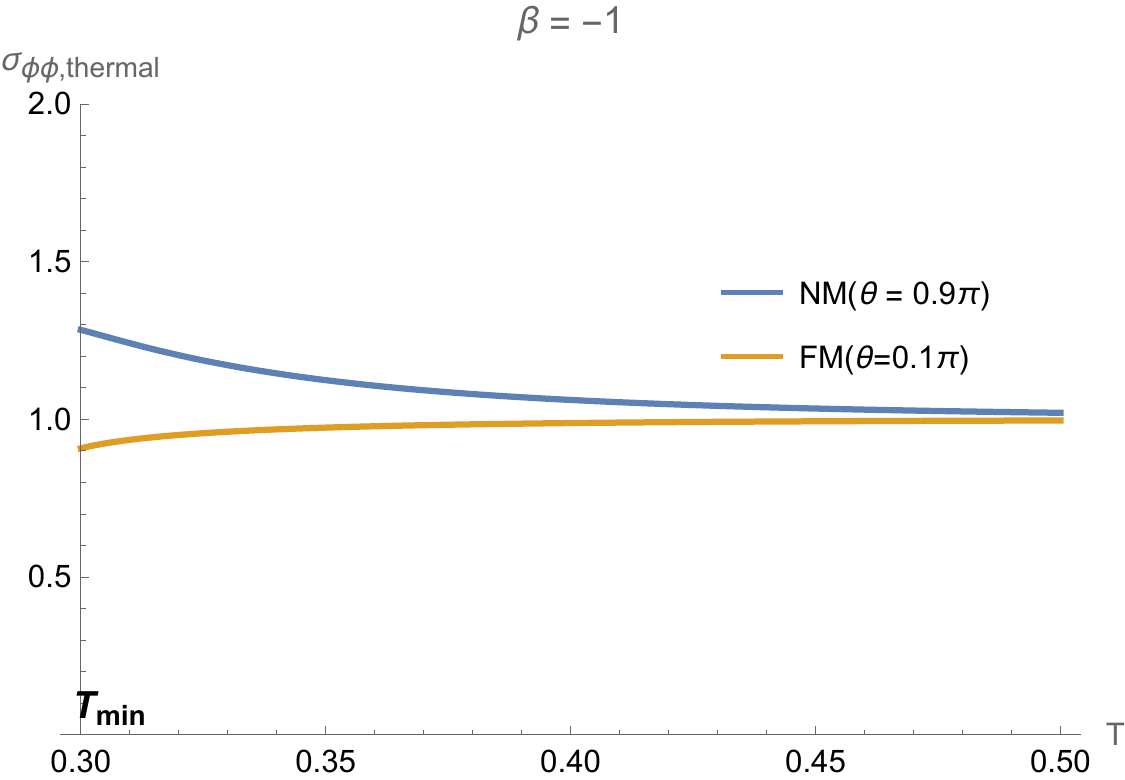}
         \caption{$\s_{\ph\ph,\text{Thermal}}$ vs $T$ at $B=0.1$}\label{fig20b}
             \end{subfigure}
        
             \caption{$\s_{\phi\ph,\text{Thermal}}$ vs $T$ plot for $\b=-1$, $E=0.1$ and $J_t=10$ at small $B$.}
        \label{figure20}
\end{figure}

Like previous examples with $\b=1$ and $\b=0$, both Hall and Ohmic conductivities due to $U(1)$ carriers and thermally produced charge pairs are higher near the Misner string as compared to far from it, which is shown below. 
\begin{align}
     \frac{\sigma_{\th\ph,U(1)}\bigg|_{T \sim T_{min}}^{nM}}{\sigma_{\th\ph,U(1)}\bigg|_{T \sim T_{min}}^{fM}}  =\frac{18 \sqrt{3} E n\tan\left(\frac{\th}{2}\right)}{\pi ^2\sin\theta T^2_{min }}>>1\hspace{1mm},\hspace{2mm}\frac{\sigma_{\th\ph,thermal}\bigg|_{T \sim T_{min}}^{nM}}{\sigma_{\th\ph,thermal}\bigg|_{T \sim T_{min}}^{fM}}>>1.\label{ratio1bn}
\end{align}
\begin{align}
     \frac{\sigma_{\ph\ph,U(1)}\bigg|_{T \sim T_{min}}^{nM}}{\sigma_{\ph\ph,U(1)}\bigg|_{T \sim T_{min}}^{fM}}  =\frac{18 \sqrt{3} E n\tan\left(\frac{\th}{2}\right)}{\pi ^2\sin\theta T^2_{min }}>>1\hspace{1mm},\hspace{2mm}\frac{\sigma_{\ph\ph,thermal}\bigg|_{T \sim T_{min}}^{nM}}{\sigma_{\ph\ph,thermal}\bigg|_{T \sim T_{min}}^{fM}}=\frac{18nE\tan\left(\frac{\th}{2}\right)}{\pi^2\sin\theta T_{min}^2}>>1.\label{ratio3bn}
\end{align}

Next, we analyze the ratio of $U(1)$ carriers to thermal pairs for both Hall and Ohmic conductivities\footnote{The ratios in expressions (\ref{ratio2b})-(\ref{ratio5bn}) are the same for both cases: $\b=0$ and $\b=-1$.}.
\begin{align}
  \frac{\s_{\th\ph,U(1)}}{\s_{\th\ph,thermal}}\Bigg|^{nM}_{T\sim T_{min}} = \frac{9\sqrt{3}J_t}{4\pi^2\ T_{min}^2}>>1\hspace{1mm},\hspace{2mm} \frac{\s_{\th\ph,U(1)}}{\s_{\th\ph,thermal}}\Bigg|^{fM}_{T\sim T_{min}}>>1.\label{ratio2b}
\end{align}
\begin{align}
  \frac{\s_{\ph\ph,U(1)}}{\s_{\ph\ph,\text{thermal}}}\Bigg|^{nM}_{T\sim T_{min}} = \frac{9\sqrt{3}J_t}{4\pi^2T_{min}^2 }>>1\hspace{1mm},\hspace{2mm}\frac{\s_{\ph\ph,U(1)}}{\s_{\ph\ph,\text{thermal}}}\Bigg|_{T\sim T_{min}}^{fM}=\frac{9J_t}{4\pi^2T_{min}^2 }>>1.\label{ratio4b}
\end{align}
Notice that, similar to $\b=1$, both Hall and Ohmic conductivities are dominated by $U(1)$ carriers both near and the far away from Misner string.

Finally, we conclude by comparing Ohmic and Hall conductivity due to both type of charge carriers
\begin{align}
  \frac{\s_{\ph\ph,U(1)}}{\s_{\th\ph,U(1)}}\Bigg|^{fM}_{T\sim T_{min}} = \frac{4\pi^2T_{min}^2}{9B}>>1\hspace{1mm},\hspace{2mm}  \frac{\s_{\ph\ph,U(1)}}{\s_{\th\ph,U(1)}}\Bigg|^{nM}_{T\sim T_{min}} =\frac{4\pi^2T_{min}^2}{9B}>>1,\label{ratio6bn}
\end{align}

\begin{align}
  \frac{\s_{\ph\ph,thermal}}{\s_{\th\ph,thermal}}\Bigg|^{fM}_{T\sim T_{min}} >>1\hspace{1mm},\hspace{2mm} \frac{\s_{\ph\ph,thermal}}{\s_{\th\ph,thermal}}\Bigg|^{nM}_{T\sim T_{min}} =\frac{4\pi^2T_{min}^2}{9B}>>1,\label{ratio5bn}
\end{align}
which reveals a higher Ohmic conductivity as compared to its Hall counterpart.

\subsection{Ohmic and Hall conductivities at high temperature}

Similar to section \ref{3.2},  we express the angular velocity for $\b=0,-1$ (\ref{angb0})-\eqref{angbn1} in the small magnetic field and high temperature limit, which yields
\begin{align}
    |\tilde\Omega|_{\b=0}=\frac{9 n \cot \left({\theta }{}\right) }{8 \left(3 n^2+1\right)^2 }\left(\frac{T_{min}}{T}\right)^4\left(1-\frac{9 B^2 \csc ^2(\theta ) }{16 \left(3 n^2+1\right)^2 }\left(\frac{T_{min}}{T}\right)^4\right)<<1,\label{othb0}
\end{align}
\begin{align}
    |\tilde\Omega|_{\b=-1}=\frac{9 n \tan \left(\frac{\theta }{2}\right) }{8 \left(3 n^2+1\right)^2 }\left(\frac{T_{min}}{T}\right)^4\left(1-\frac{9 B^2 \csc ^2(\theta ) }{16 \left(3 n^2+1\right)^2 }\left(\frac{T_{min}}{T}\right)^4\right)<<1.\label{othbn}
\end{align}
Notice that similar to $\b=1$, in the high temperature limit ($T>>T_{min}$), the effects due to frame dragging ($\tilde\Omega\sim \frac{1}{T^4}$) are negligible, as shown above (\ref{othb0})-\eqref{othbn}.

The expressions for holographic Ohmic conductivity ($\s_{\phi\phi}$) (\ref{condu1dcT})-\eqref{condthT} and their plots against temperature for $\b=0$, both near and far from the Misner string are given below
\begin{align}
  &\sigma_{\ph\ph,U(1)}\bigg|^{fM}_{T>>T_{min}}\approx\frac{9  J_t}{4 \pi (3+9n^2)}\left(\frac{T_{min}}{T}\right)^2\left(1-\frac{81B^2\csc^2\th}{16(3+9n^2)^2}\left(\frac{T_{min}}{T}\right)^4\right),\label{highdcu1fmb0}\\&\sigma_{\ph\ph,U(1)}\bigg|_{T>>T_{min}}^{nM}\approx\frac{9  J_t}{4 \pi (3+9n^2)}\left(\frac{T_{min}}{T}\right)^2+\frac{81 E^2 n^2  J_t \cot^2\left({\th}\right)}{32(1+3n^2)^3\sin^2\th}\left(\frac{T_{min}}{T}\right)^6,\label{highdcu1nmb0} \\
&\sigma_{\ph\ph,thermal}\Bigg|^{fM}_{T>>T_{min}}\approx1-\frac{81B^2 \csc^2\th}{32(3+9n^2)^2}\left(\frac{T_{min}}{T}\right)^4,\label{highdcthfmb0}\\&\sigma_{\ph\ph,thermal}\Bigg|^{nM}_{T>>T_{min}}\approx1+\frac{9E^2 n^2 \cot^2\left({\th}\right)}{2( 1+ 3 n^2)^2 \sin^2\th }\left(\frac{T_{min}}{T}\right)^4.\label{highdcthb0}
\end{align}
\begin{figure}[H]
     \centering
     \begin{subfigure}[b]{0.495\textwidth}
         \centering
         \includegraphics[width=\textwidth]{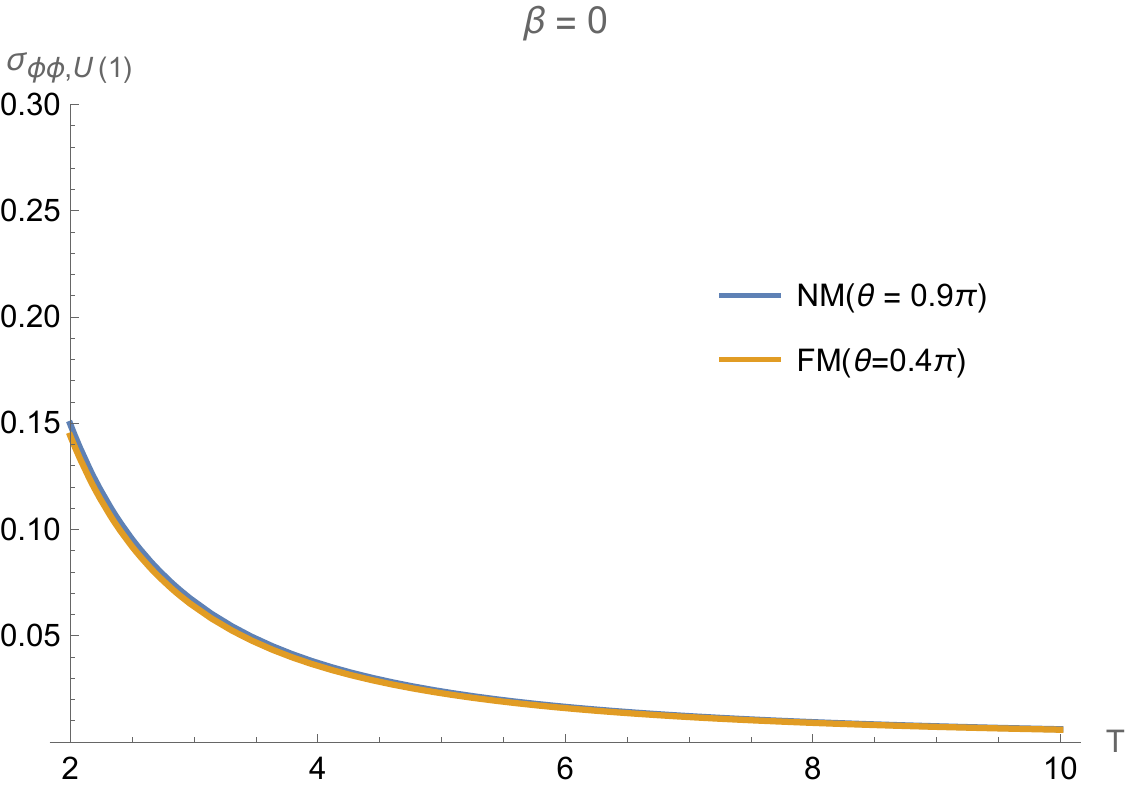}
         \caption{$\s_{\ph\ph,U(1)}$ vs $T$ at $B=0.01$ }\label{fig17a}
              \end{subfigure}
              \hfill
     \begin{subfigure}[b]{0.495\textwidth}
         \centering
         \includegraphics[width=\textwidth]{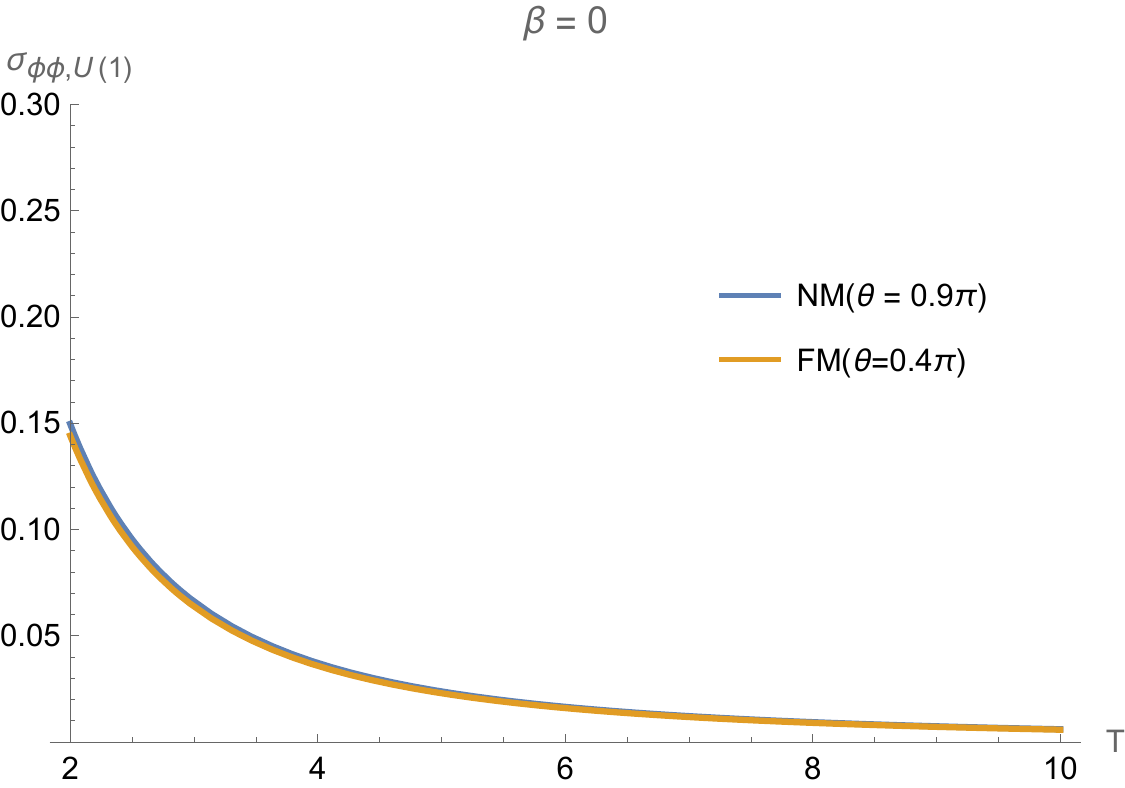}
         \caption{$\s_{\ph\ph,U(1)}$ vs $T$ at $B=0.1$}\label{fig17b}
             \end{subfigure}
        
             \caption{$\s_{\phi\ph,U(1)}$ vs $T$ plot for $\b=0$, $E=0.1$ and $J_t=10$ at small $B$.}
        \label{figure17}
\end{figure}
\begin{figure}[H]
     \centering
     \begin{subfigure}[b]{0.495\textwidth}
         \centering
         \includegraphics[width=\textwidth]{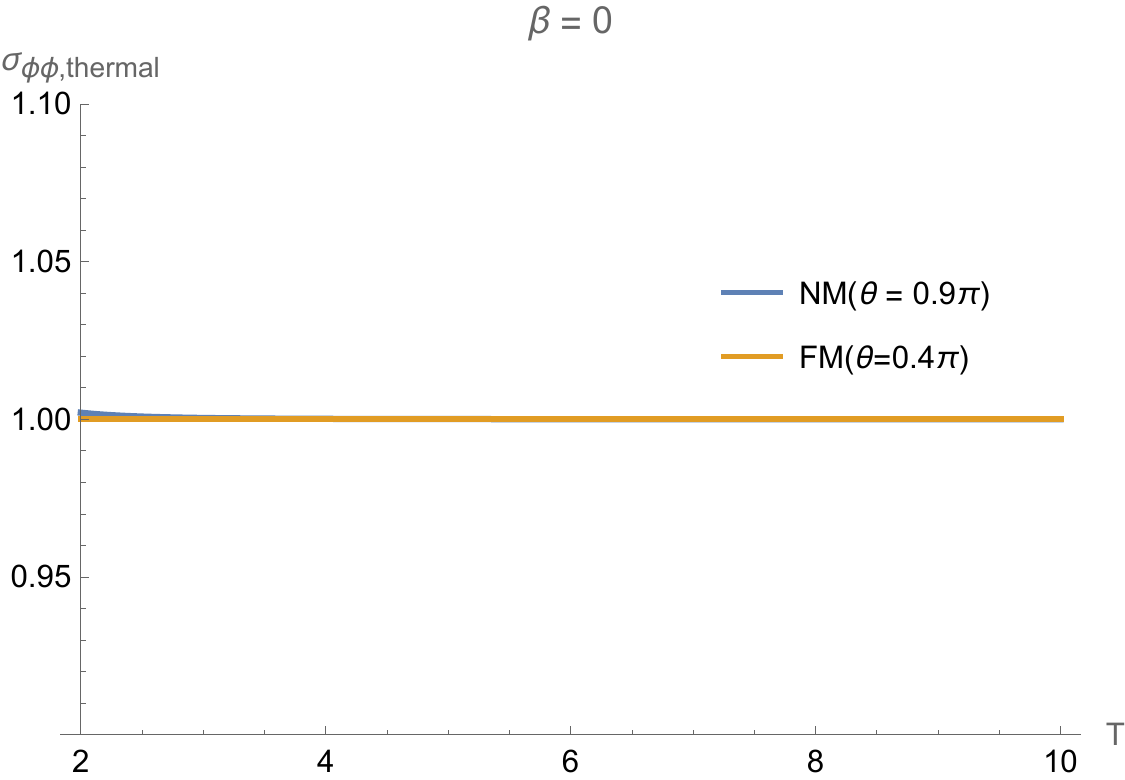}
         \caption{$\s_{\ph\ph,\text{thermal}}$ vs temperature at $B=0.01$ }\label{fig15athb0}
              \end{subfigure}
              \hfill
     \begin{subfigure}[b]{0.495\textwidth}
         \centering
         \includegraphics[width=\textwidth]{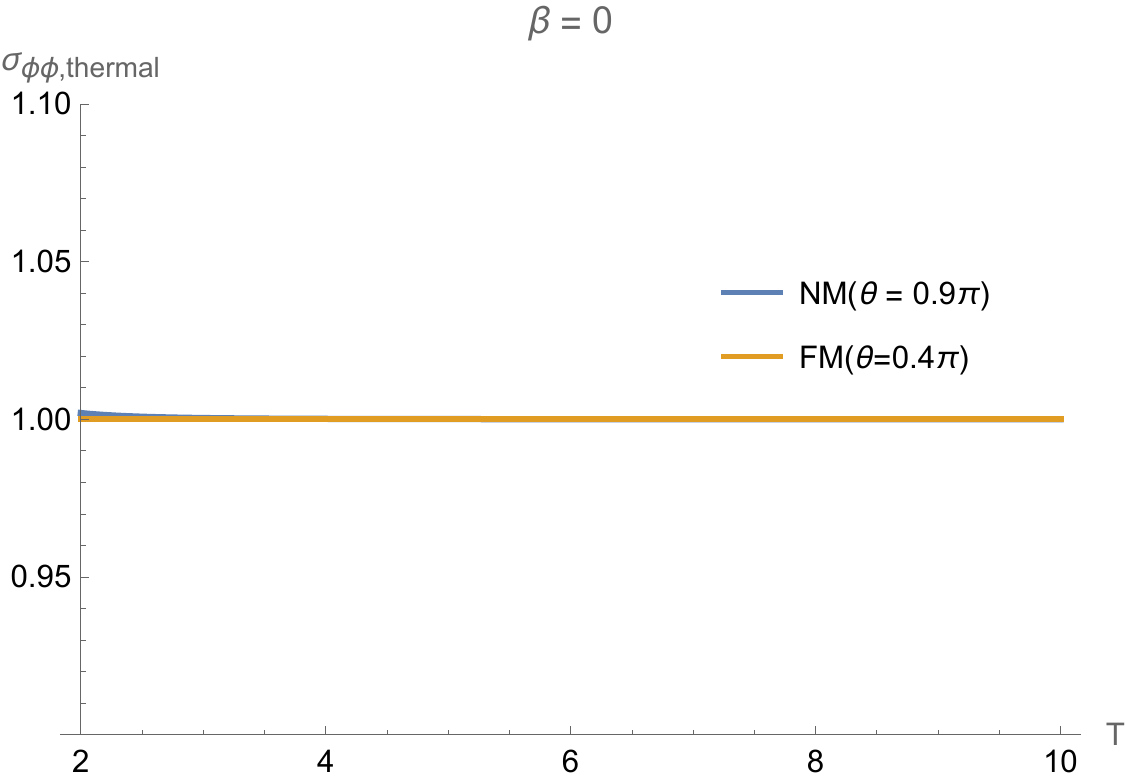}
         \caption{$\s_{\ph \ph ,\text{thermal}}$ vs temperature at $B=0.1$ }\label{fig15bthb0}
             \end{subfigure}
        
             \caption{$\s_{\ph\ph,\text{thermal}}$ vs temperature plot. Here, we set $E=0.1$, $J_t=10$ and $n=0.2$.}
        \label{figure15thb0}
\end{figure}
We can see from the above Figures (\ref{figure17} and \ref{figure15thb0}) that Ohmic conductivity due to $U(1)$ and thermal pairs are identical near far away from Misner string.

 The holographic Hall conductivity $(\s_{\th\ph})$ (see Appendix B \eqref{hallu1T}-\eqref{hallthermal T}) for $\b=0$ and in the high temperature regime take the following forms
\begin{align}&
    \s_{\th\phi,U(1)}\bigg|^{fM}_{T >>T_{min}}=\frac{9 B J_t }{16 \left(3 n^2+1\right)^2 }\left(\frac{T_{\min }}{T}\right)^4, \label{hallu1fmThb0}\\&\s_{\th\ph,U(1)}\bigg|^{nM}_{T >> T_{min}}=\frac{9 B J_t }{16 \left(3 n^2+1\right)^2 }\left(\frac{T_{\min }}{T}\right)^4,\label{hallu1nmthb0}
\\
  &\s_{\th\phi,\text{thermal}}\bigg|^{fM}_{T >> T_{min}}\approx 0,  \\&   \s_{\th\phi,\text{thermal}}\bigg|^{nM}_{T >> T_{min}}=\frac{9 B E n\cot\left({\th}{}\right) }{8 \sin\th \left(3 n^2+1\right)^2}\left(\frac{T_{min}}{T}\right)^4.\label{hallthThb0}
\end{align}

Below, we plot these Hall conductivities against temperature for both near and far from the Misner string.
\begin{figure}[H]
     \centering
     \begin{subfigure}[b]{0.495\textwidth}
         \centering
         \includegraphics[width=\textwidth]{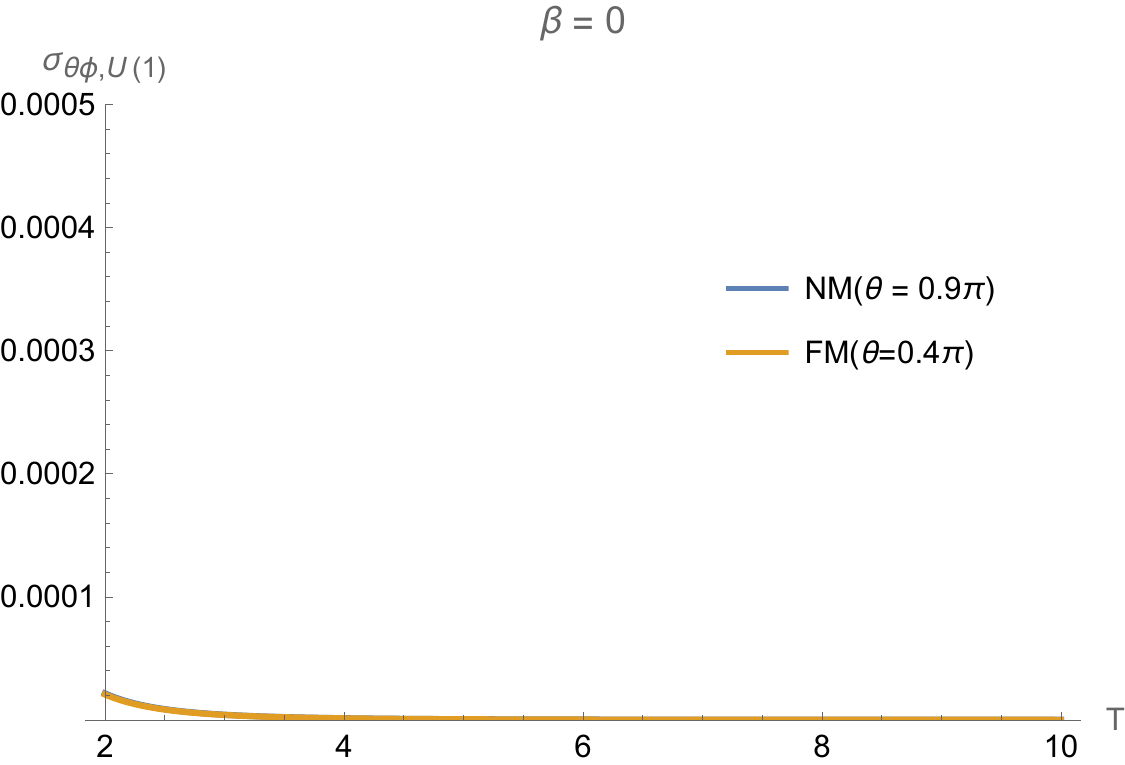}
         \caption{$\s_{\th\ph,U(1)}$ vs $T$ at $B=0.01$ }\label{fig15a}
              \end{subfigure}
              \hfill
     \begin{subfigure}[b]{0.495\textwidth}
         \centering
         \includegraphics[width=\textwidth]{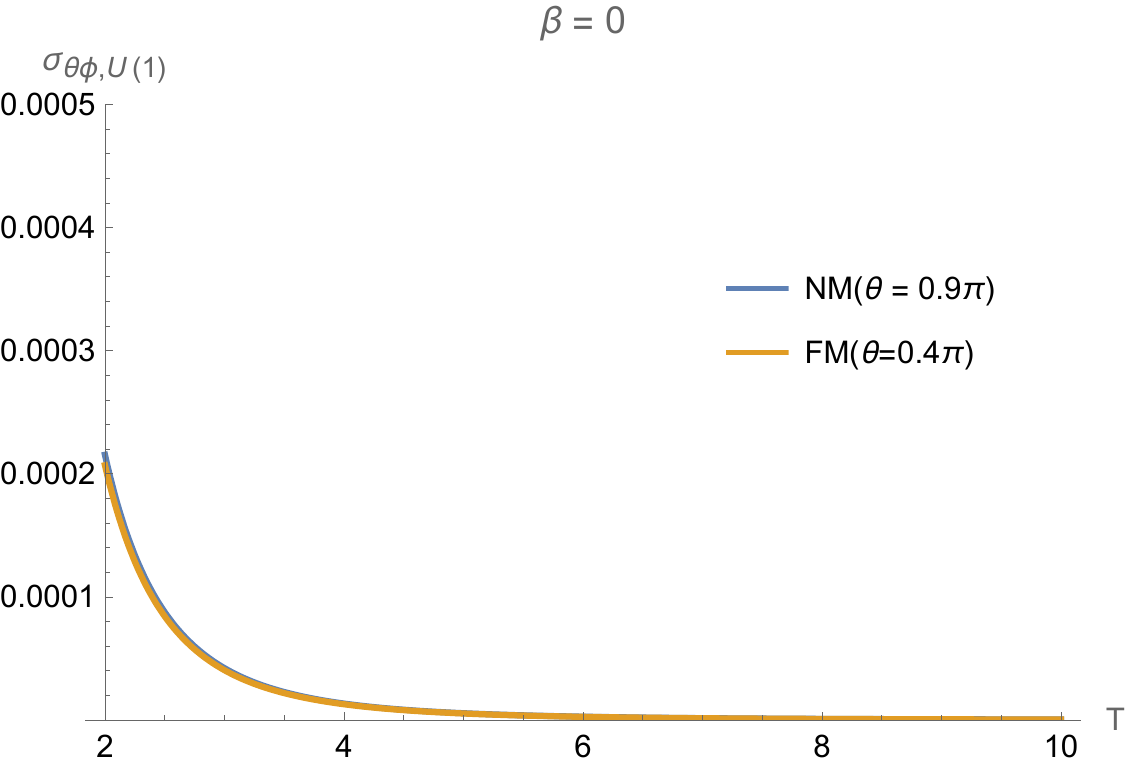}
         \caption{$\s_{\th\ph ,U(1)}$ vs $T$ at $B=0.1$}\label{fig15b}
             \end{subfigure}
        
             \caption{$\s_{\th\ph}$ vs $T$ plot for $\b=0$, $E=0.1$ and $J_t=10$ at small $B$.}
        \label{figure15}
\end{figure}
\begin{figure}[H]
     \centering
     \begin{subfigure}[b]{0.495\textwidth}
         \centering
         \includegraphics[width=\textwidth]{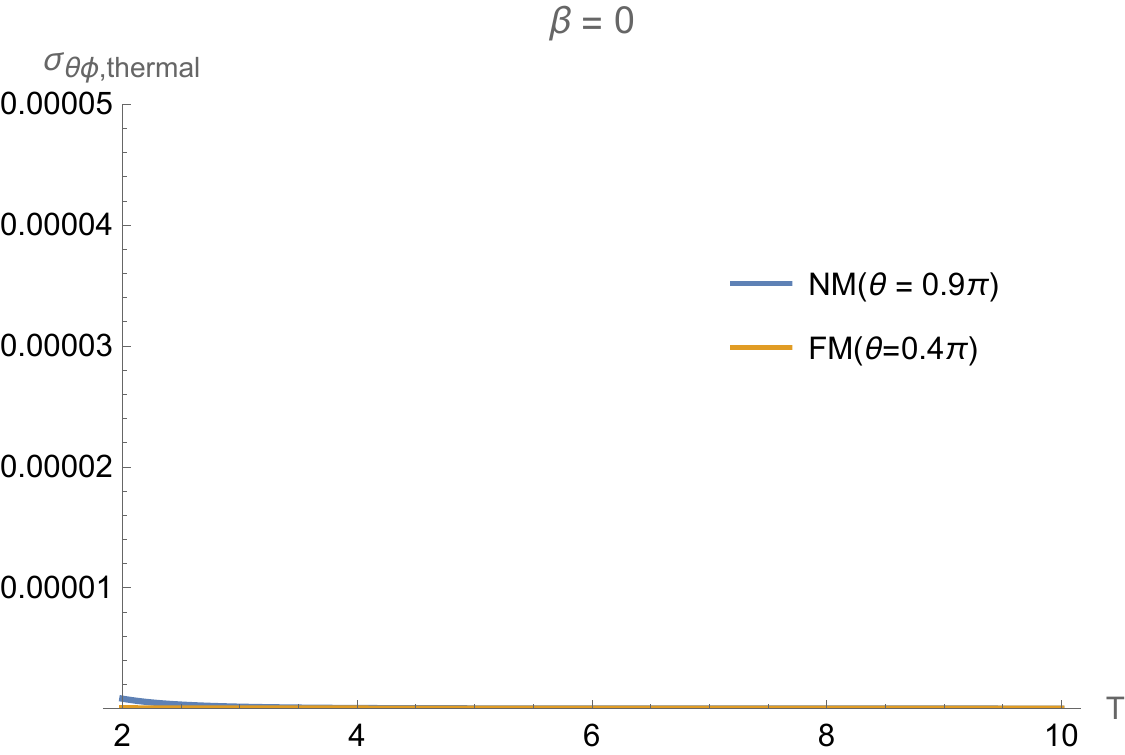}
         \caption{$\s_{\th\ph,\text{thermal}}$ vs temperature at $B=0.01$ }\label{fig5athb0}
              \end{subfigure}
              \hfill
     \begin{subfigure}[b]{0.495\textwidth}
         \centering
         \includegraphics[width=\textwidth]{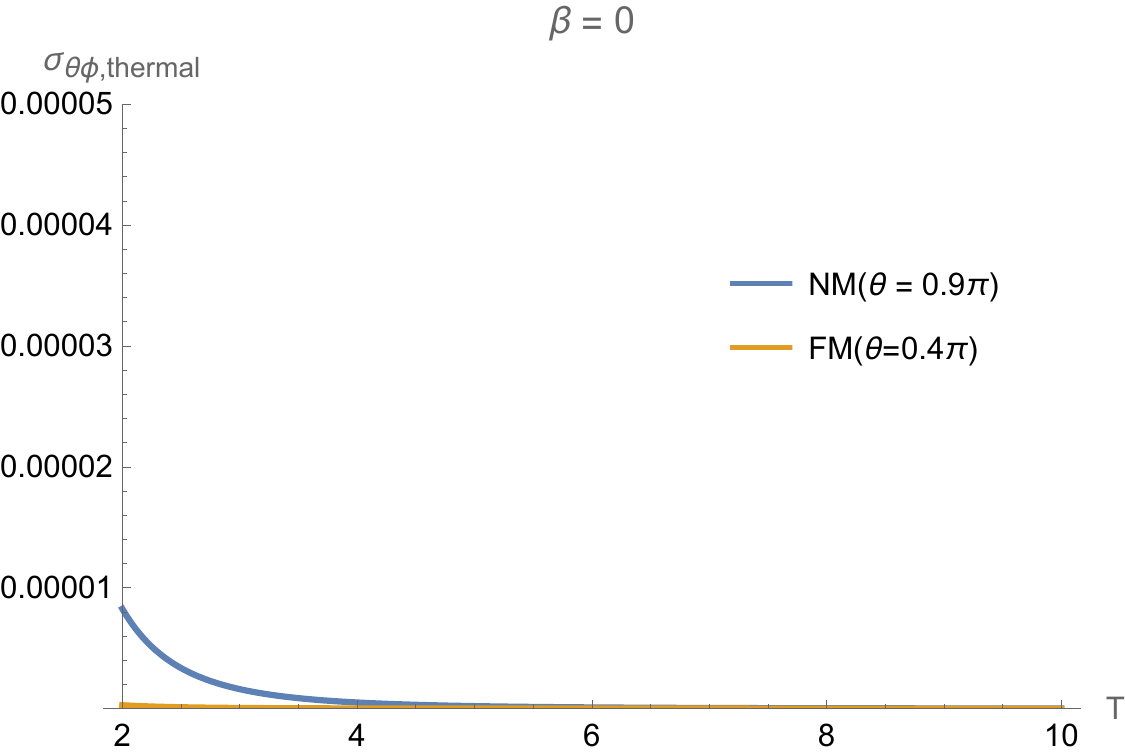}
         \caption{$\s_{\th \ph ,\text{thermal}}$ vs temperature at $B=0.1$ }\label{fig5bthb0}
             \end{subfigure}
        
             \caption{$\s_{\th\ph,\text{thermal}}$ vs temperature plot. Here, we set $E=0.1$, $J_t=10$ and $n=0.2$.}
        \label{figure5thb0}
\end{figure}

Finally, we provide the expressions for Ohmic and Hall conductivity and their plots against temperature for $\b=-1$
\begin{align}
&\sigma_{\ph\ph,U(1)}\bigg|^{fM}_{T>>T_{min}}\approx\frac{9  J_t}{4 \pi (3+9n^2)}\left(\frac{T_{min}}{T}\right)^2\left(1-\frac{81B^2\csc^2\th}{16(3+9n^2)^2}\left(\frac{T_{min}}{T}\right)^4\right),\label{highdcu1fmbn}\\&\sigma_{\ph\ph,U(1)}\bigg|_{T>>T_{min}}^{nM}\approx\frac{9  J_t}{4 \pi (3+9n^2)}\left(\frac{T_{min}}{T}\right)^2+\frac{81 E^2 n^2  J_t \tan^2\left(\frac{\th}{2}\right)}{32(1+3n^2)^3\sin^2\th}\left(\frac{T_{min}}{T}\right)^6,\label{hdcu1nmbn} \\
&\sigma_{\ph\ph,thermal}\Bigg|^{fM}_{T>>T_{min}}\approx1-\frac{81B^2 \csc^2\th}{32(3+9n^2)^2}\left(\frac{T_{min}}{T}\right)^4,\\&\sigma_{\ph\ph,thermal}\Bigg|^{nM}_{T>>T_{min}}\approx1+\frac{9E^2 n^2 \tan^2\left(\frac{\th}{2}\right)}{2( 1+ 3 n^2)^2 \sin^2\th }\left(\frac{T_{min}}{T}\right)^4,\label{highthbn}\\
&\s_{\th\phi,U(1)}\bigg|^{fM}_{T >>T_{min}}=\frac{9 B J_t }{16 \left(3 n^2+1\right)^2 }\left(\frac{T_{\min }}{T}\right)^4, \label{hallu1fmThbn}\\&\s_{\th\phi,U(1)}\bigg|^{nM}_{T >> T_{min}}=\frac{9 B J_t }{16 \left(3 n^2+1\right)^2 }\left(\frac{T_{\min }}{T}\right)^4,\label{hallu1nmthbn}
\\
&\s_{\th\phi,\text{thermal}}\bigg|^{fM}_{T >> T_{min}}\approx 0, \\&   \s_{\th\phi,\text{thermal}}\bigg|^{nM}_{T >> T_{min}}=\frac{9 B E n\tan\left(\frac{\th}{2}\right) }{8 \sin\th \left(3 n^2+1\right)^2}\left(\frac{T_{min}}{T}\right)^4.\label{hallthThbn}
\end{align}

\begin{figure}[H]
     \centering
     \begin{subfigure}[b]{0.495\textwidth}
         \centering
         \includegraphics[width=\textwidth]{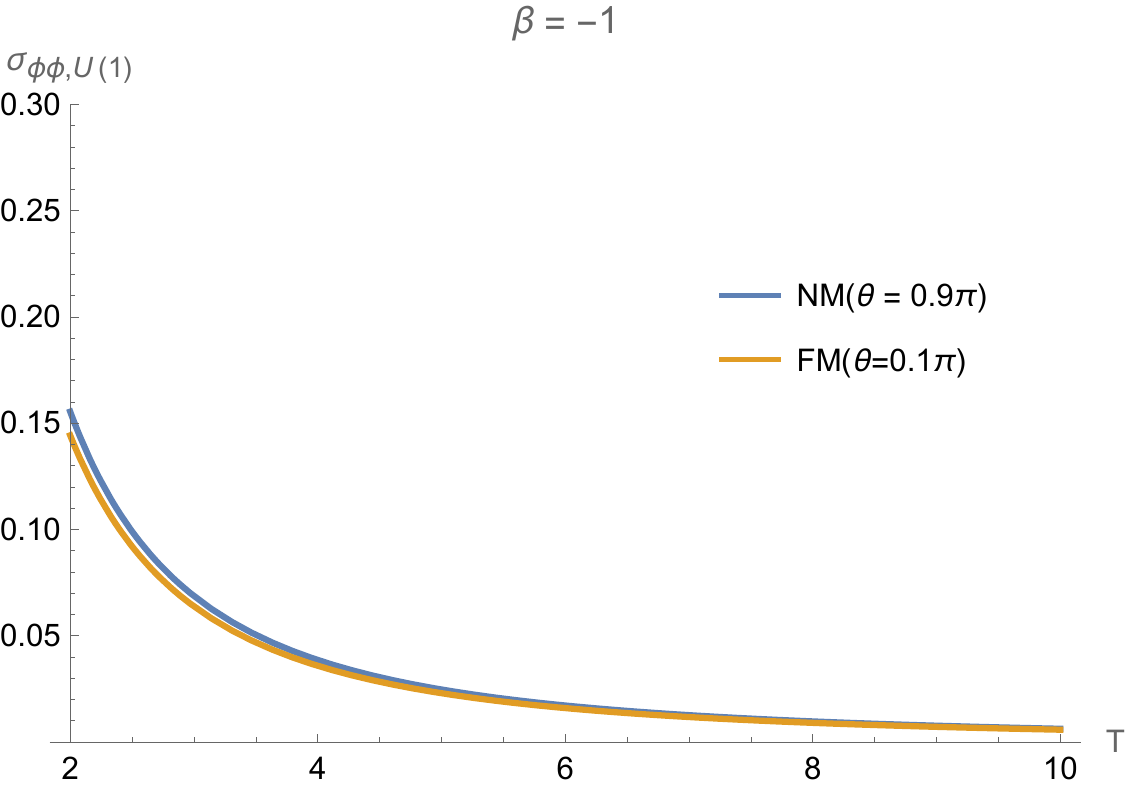}
         \caption{$\s_{\ph\ph,U(1)}$ vs $T$ at $B=0.01$ }\label{fig18a}
              \end{subfigure}
              \hfill
     \begin{subfigure}[b]{0.495\textwidth}
         \centering
         \includegraphics[width=\textwidth]{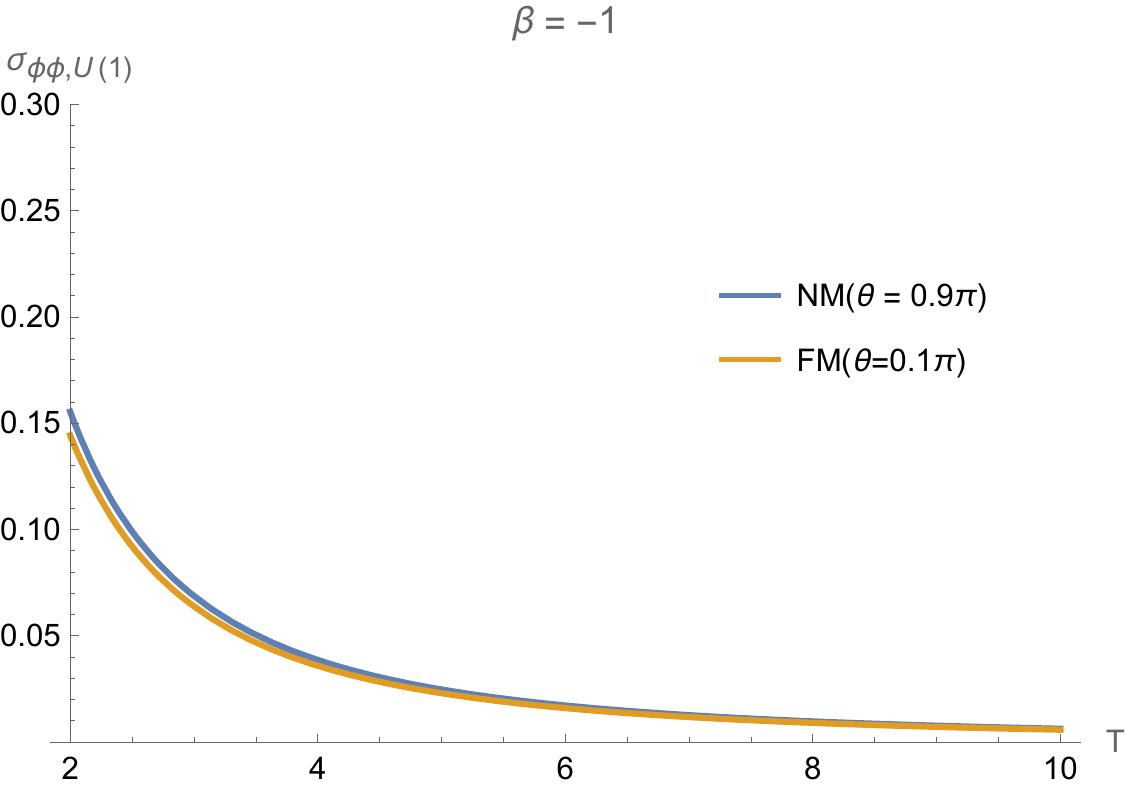}
         \caption{$\s_{\ph\ph,U(1)}$ vs $T$ at $B=0.1$}\label{fig18b}
             \end{subfigure}
        
             \caption{$\s_{\phi\ph,U(1)}$ vs $T$ plot for $\b=-1$, $E=0.1$ and $J_t=10$ at small $B$.}
        \label{figure18}
\end{figure}
\begin{figure}[H]
     \centering
     \begin{subfigure}[b]{0.495\textwidth}
         \centering
         \includegraphics[width=\textwidth]{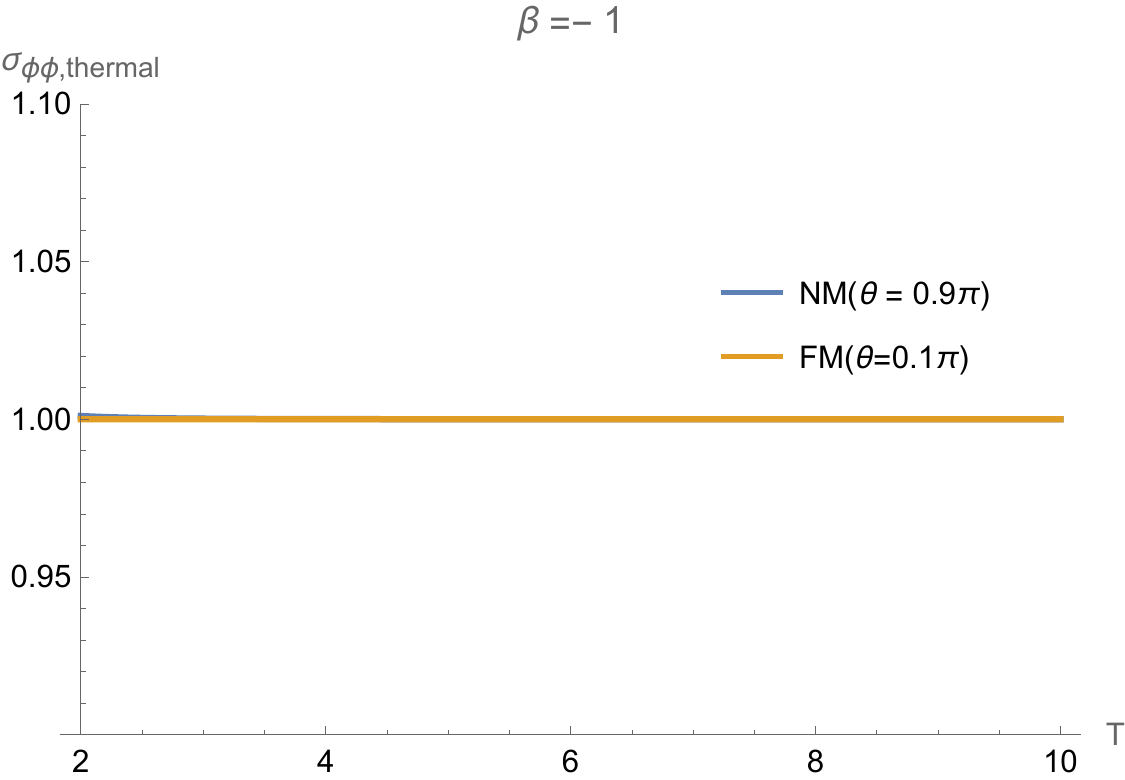}
         \caption{$\s_{\ph\ph,\text{thermal}}$ vs temperature at $B=0.01$ }\label{fig15athbn}
              \end{subfigure}
              \hfill
     \begin{subfigure}[b]{0.495\textwidth}
         \centering
         \includegraphics[width=\textwidth]{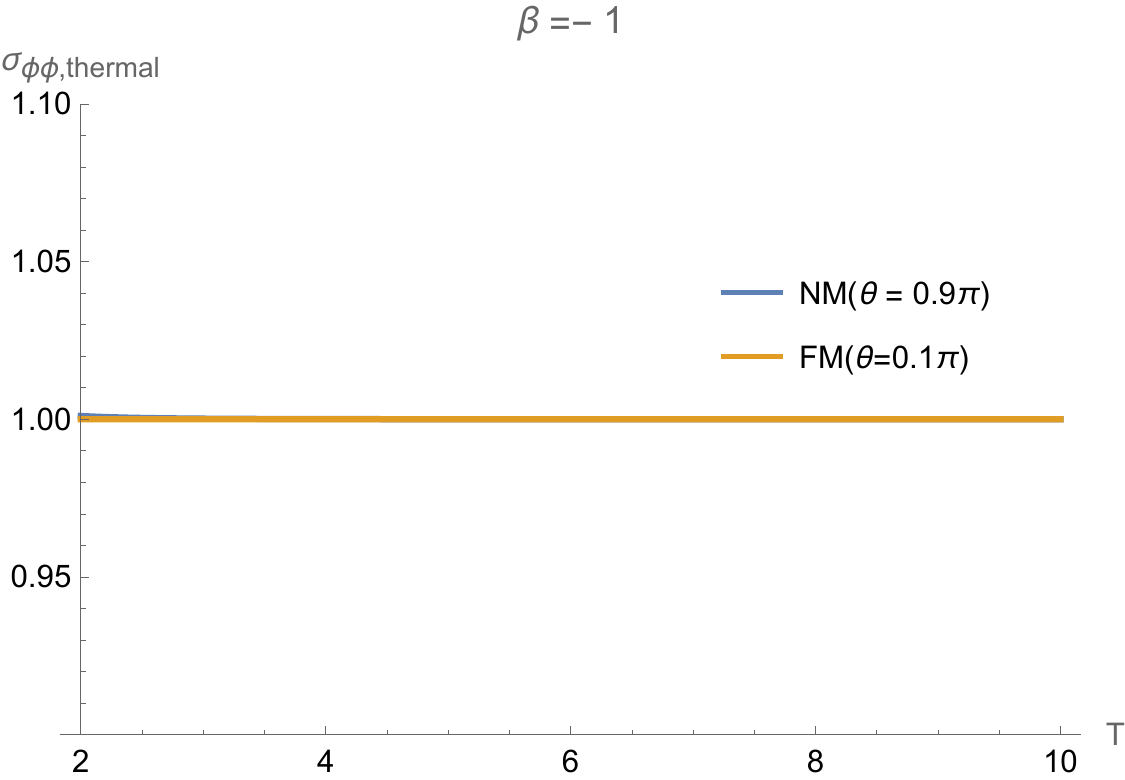}
         \caption{$\s_{\ph \ph ,\text{thermal}}$ vs temperature at $B=0.1$ }\label{fig15bthbn}
             \end{subfigure}
        
             \caption{$\s_{\ph\ph,\text{thermal}}$ vs temperature plot. Here, we set $E=0.1$, $J_t=10$ and $n=0.2$.}
        \label{figure15thbn}
\end{figure}

\begin{figure}[H]
     \centering
     \begin{subfigure}[b]{0.495\textwidth}
         \centering
         \includegraphics[width=\textwidth]{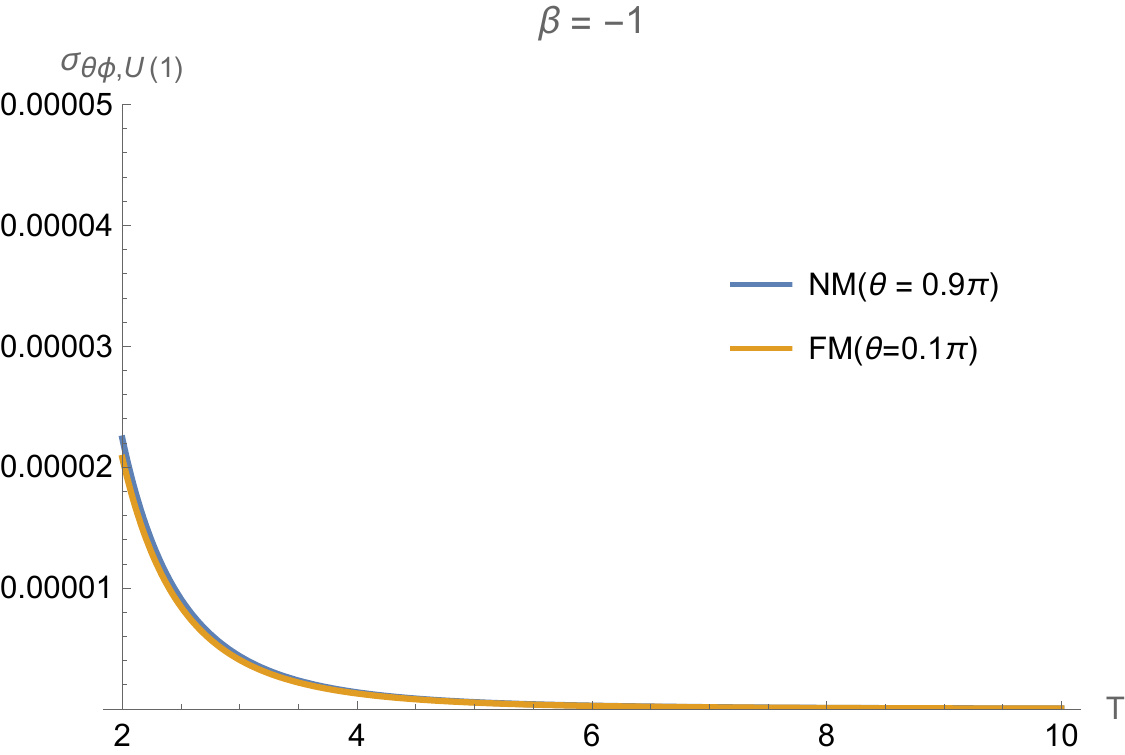}
         \caption{$\s_{\th\ph,U(1)}$ vs $T$ at $B=0.01$ }\label{fig16a}
              \end{subfigure}
              \hfill
     \begin{subfigure}[b]{0.495\textwidth}
         \centering
         \includegraphics[width=\textwidth]{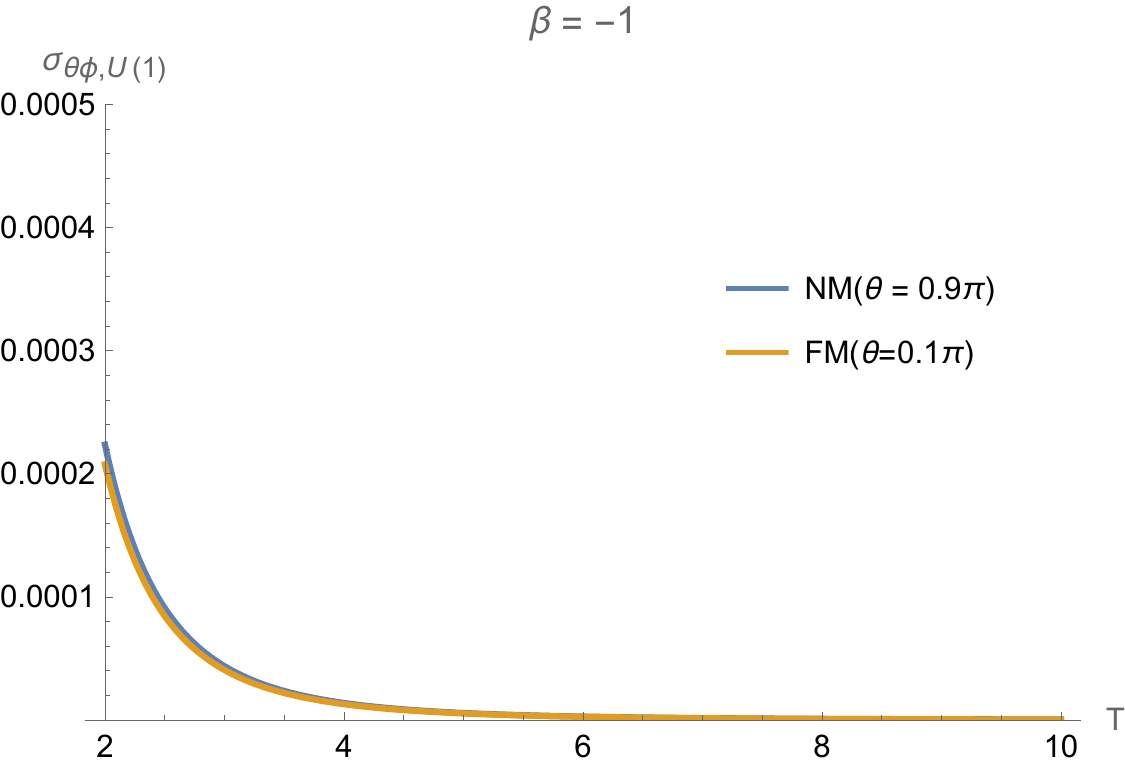}
         \caption{$\s_{\th\ph ,U(1)}$ vs $T$ at $B=0.1$}\label{fig16b}
             \end{subfigure}
        
             \caption{$\s_{\th\ph}$ vs $T$ plot for $\b=-1$, $E=0.1$ and $J_t=10$ at small $B$.}
        \label{figure16}
\end{figure}
\begin{figure}[H]
     \centering
     \begin{subfigure}[b]{0.495\textwidth}
         \centering
         \includegraphics[width=\textwidth]{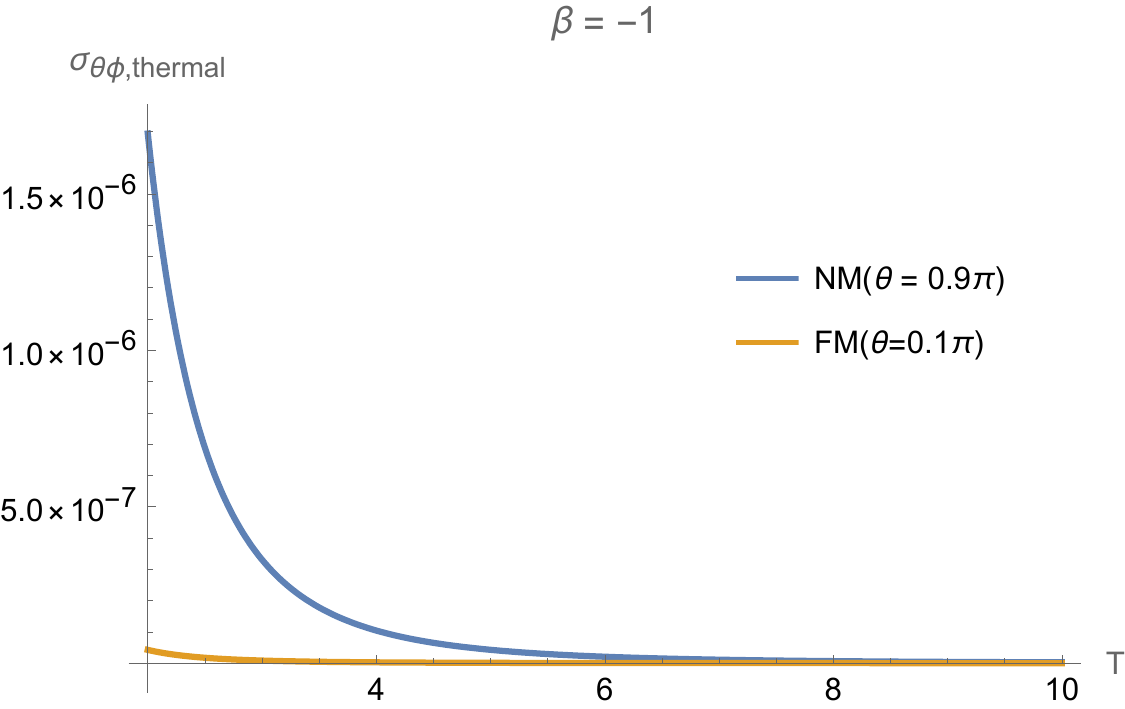}
         \caption{$\s_{\th\ph,\text{thermal}}$ vs temperature at $B=0.01$ }\label{fig5athbn}
              \end{subfigure}
              \hfill
     \begin{subfigure}[b]{0.495\textwidth}
         \centering
         \includegraphics[width=\textwidth]{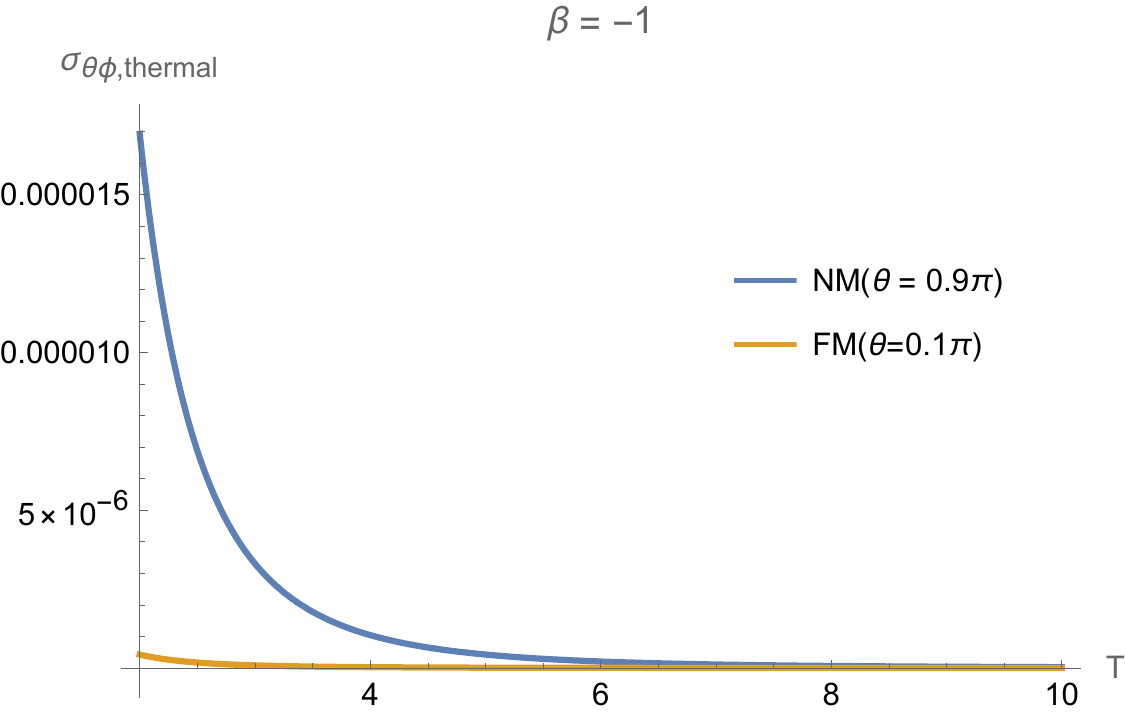}
         \caption{$\s_{\th \ph ,\text{thermal}}$ vs temperature at $B=0.1$ }\label{fig5bthbn}
             \end{subfigure}
        
             \caption{$\s_{\th\ph,\text{thermal}}$ vs temperature plot. Here, we set $E=0.1$, $J_t=10$ and $n=0.2$.}
        \label{figure5thbn}
\end{figure}

\end{document}